\documentclass[prd,nofootinbib,notitlepage,10pt,aps]{revtex4-1}
\usepackage{epsfig,amsmath,amssymb,slashed}
\usepackage[utf8]{inputenc}
\usepackage{graphicx,subfigure}
\usepackage{hyperref}
\usepackage{xcolor}
\def\!{\mskip-\thinmuskip}
\newcommand{\bra}[1]{\langle #1|}
\newcommand{\ket}[1]{|#1\rangle}

\newcommand{\mean}[1]{\langle #1 \rangle}
\def\wa{{\widehat a}}
\def\wad{{\widehat a}^\dagger}
\def\wb{{\widehat b}}
\def\wbd{{\widehat b}^\dagger}
\newcommand{\D}{{\rm d}}
\newcommand{\mc}{\ensuremath{\mathcal}}
\newcommand{\I}{ \mathrm{i}}
\newcommand{\h}[1]{\widehat{#1}}
\newcommand{\tr}{{\rm tr}}  
\newcommand{\E}{{\rm e}}
\newcommand{\de}{\partial}
\newcommand{\subs}[1]{_{\textup{#1}}}
\newcommand{\sups}[1]{^{\textup{#1}}}
\renewcommand{\vec}[1]{\ensuremath{\mathchoice
                     {\mbox{\boldmath$\displaystyle\mathbf{#1}$}}
                     {\mbox{\boldmath$\textstyle\mathbf{#1}$}}
                     {\mbox{\boldmath$\scriptstyle\mathbf{#1}$}}
                     {\mbox{\boldmath$\scriptscriptstyle\mathbf{#1}$}}}}
\hypersetup{pdfinfo={Title={AGM at finite temperature},Author={Matteo Buzzegoli}}}
\begin{document}
%*****************************************************************

%*************************************************************************************
\title{Anomalous gravitomagnetic moment
	and non-universality of the axial vortical effect\texorpdfstring{\\}{} at finite temperature} 
%*************************************************************************************

\author{M. Buzzegoli}
\email{matteo.buzzegoli@unifi.it}
\affiliation{Universit\'a di Firenze and INFN Sezione di Firenze, Via G. Sansone 1, 
	I-50019 Sesto Fiorentino (Firenze), Italy}
\author{Dmitri E. Kharzeev}
\email{dmitri.kharzeev@stonybrook.edu}
\affiliation{Center for Nuclear Theory, Department of Physics and Astronomy, Stony Brook University, New York 11794, USA}
\affiliation{Department of Physics and RIKEN-BNL Research Center, Brookhaven National Laboratory, Upton, New York 11973-5000, USA}

\begin{abstract}
The coupling between the spin of a massive Dirac fermion and the angular momentum 
of the medium, i.e. the gravitomagnetic moment, is shown here to be renormalized by QED interactions at
finite temperature. This means that the anomalous gravitomagnetic moment (AGM) does not vanish, and implies that thermal effects can break the Einstein equivalence principle in quantum
field theory, as argued previously. We also show that the AGM causes radiative corrections to the axial current of massive fermions induced
by vorticity in quantum relativistic fluids, similarly to the previous findings for massless fermions. The radiative QCD effects on the AGM should significantly affect the production of polarized hadrons in heavy-ion collisions.
\end{abstract}
%
%%Old Abstract
%\begin{abstract}
%	It is already known that thermal effects on quantum field theory break the Einstein
%	equivalence principle. We show that, thanks to this violation, the coupling between
%	the spin of a Dirac massive fermion and the rotation of the medium is affected by QED
%	interactions at finite temperature. In analogy with the coupling between spin and
%	magnetic field, this effect is referred to as anomalous gravitomagnetic moment.
%	We also show that this modification of spin-rotation coupling is the cause of
%	the radiative corrections of the axial current induced by vorticity in quantum
%	relativistic fluids. Therefore, the anomalous gravitomagnetic moment may affect the polarization of particles emitted by the quark-gluon plasma produced in heavy-ion
%	collisions and phenomena related to axial charge imbalance.
%\end{abstract}
%%
\maketitle

%************************************************************************************
\section{Introduction}
%************************************************************************************
Recent experiments~\cite{STAR:2017ckg,Adam:2018ivw,Adam:2019srw} at the Relativistic
Heavy Ion Collider (RHIC) have opened the possibility to study the effects of vorticity and magnetic field on the production of polarized hadrons 
in relativistic heavy-ion collisions, see~\cite{Becattini:2020ngo} for a recent review.
The coupling between the spin and vorticity of the quark gluon
plasma~\cite{Liang:2004ph,kharzeev2006parity,Becattini:2007sr} induces a polarization of the hadrons emitted
from the fluid, which can be measured in experiments. The spin and the polarization of
a spin $1/2$ particle, like the $\Lambda$ hyperon, is deeply connected to the axial
current of the particle. In a rotating medium, the coupling between the spin of a fermion and rotation of the medium \cite{Vilenkin:1979ui} also induces a thermal effect on the axial current, known as the Axial Vortical Effect (AVE).

The AVE is a macroscopic quantum effect describing the axial
charge separation along the rotation of the fluid~\cite{Landsteiner:2011iq,Gao:2012ix,Kharzeev:2015znc}.
In the case of a system composed of massless fermions, one can give a simple intuitive picture 
of the phenomenon. Indeed, as a result of the coupling between the fermions' spin and the rotation of the medium, the spins of the
fermions are preferably aligned along the angular momentum vector of the medium. Then, since for
massless fermions chirality coincides with helicity (note that for antifermions, chirality and helicity are opposite), most of the right-handed particles
will move along the direction of the vorticity (pseudo)vector, while most of the left-handed particles will
move in the opposite direction. This flow causes the net separation between the right- and
left-handed particles that we refer to as the AVE. Since the AVE is 
driven by spin-rotation coupling, it can be realized even in a global thermal equilibrium, where the expectation value of the axial current is given by  ~\cite{Buzzegoli:2018wpy}  
\begin{equation*}
\mean{\h{j}^\mu\subs{A}}=n\subs{A} u^\mu+W\sups{A}\frac{\omega^\mu}{T}.
\end{equation*}
where $n_A$ is the density of axial charge, $u$ is the fluid velocity, $\omega$ is the vorticity of the medium, and $T$ is the temperature.
For free massless fermions the AVE conductivity $W\sups{A}$ is found to
be~\cite{Vilenkin:1979ui,Landsteiner:2011iq,Gao:2012ix}
\begin{equation*}
W\sups{A}=\frac{T^3}{6}+\frac{\mu^2+\mu\subs{A}^2}{2\pi^2} T,
\end{equation*}
where $\mu$ is the vector chemical potential and
$\mu\subs{A}$ is the axial chemical potential.

The AVE shares many similarities with other non-dissipative macroscopic quantum effects,
such as the Chiral Magnetic Effect (CME) \cite{kharzeev2006parity,kharzeev2008effects,fukushima2008chiral}   and the Chiral Vortical Effect
(CVE)~\cite{kharzeev2007charge,erdmenger2009fluid}. However, even though both CME and
CVE originate from the chiral anomaly~\cite{Kharzeev:2013ffa}, the link between AVE and anomalies is still under discussion, see ~\cite{Landsteiner:2011cp,Jensen:2012kj,Kalaydzhyan:2014bfa,Golkar:2015oxw,Avkhadiev:2017fxj,Glorioso:2017lcn,Flachi:2017vlp,Buzzegoli:2017cqy,Buzzegoli:2018wpy,Stone:2018zel,Prokhorov:2020npf}.
First, both the axial current
and the vorticity are axial vectors, and therefore AVE does not require parity breaking.
% is allowed even for classical
%symmetries and in principle does . 
 Second, the AVE conductivity is not protected against radiative
corrections: explicit calculations in massless quantum electrodynamics (QED) and
quantum chromodynamics (QCD) show that AVE conductivity is renormalized by interactions~\cite{Hou:2012xg,Golkar:2012kb}.

\begin{figure}[thb!]
	\centering
	\includegraphics[width=0.4\columnwidth]{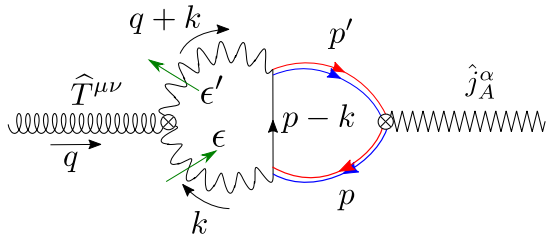}
	\caption{The Feynman diagram describing the radiative corrections to AVE conductivity.
		Red and blue lines represent right and left chiralities of the fermions, and the green arrows represent the photon polarization.}
	\label{fig:AVE_diag}
\end{figure}
This correction only affects the finite-temperature part of the AVE conductivity, while the parts
proportional to the axial chemical potential and the vector chemical potential are
unaltered by interactions. Indeed, at second order in QED coupling constant $e$, the AVE conductivity
for a massless fermion is given by~\cite{Hou:2012xg,Golkar:2012kb}
\begin{equation*}
W\sups{A}=\left(\frac{1}{6}+\frac{e^2}{24\pi^2}\right)T^3.
\end{equation*}
This correction is given solely by the diagram in Fig.~\ref{fig:AVE_diag}.
Higher order corrections are non-vanishing only if they are given by diagrams in which
the axial current is inserted through the anomalous triangle
subdiagram~\cite{Golkar:2012kb}. 
\vskip0.3cm

The radiative corrections to AVE in QED can be
interpreted as driven by the interaction of a graviton with the photon cloud surrounding
the fermion. Indeed, we may understand the coupling between the chirality of fermions and rotation of the medium 
induced by this
diagram as follows. Suppose the lower fermion line inside the loop in
Fig.~\ref{fig:AVE_diag} has a right chirality. Then it couples more effectively to 
photons having a polarization parallel to the fermion spin. Following the loop in the 
figure, those photons then interact with the rotation of the medium (represented by the
insertion of stress-energy tensor in the diagram) forcing their polarization to rotate -- so when 
they couple to the fermion in the loop again, they can induce the chirality flip of the fermion, and transform it into
a left-handed one. Note that this requires the presence of photons in the rotating fluid, which in thermal equlibrium implies a finite temperature. At zero temperature, the photon cloud is part of the coherent state of a charged fermion, and thus rotates together with it. At finite temperature, the thermal bath possesses its own photons, and mixing between these photons and the photons that form the coherent cloud of the fermion can cause the rotation of the fermion's polarization.

%an absorption of a photon with one polarization and its re-emission with another polarization would be impossible. %In general relativity, the effect of the rotation of polarization of light in the gravitational field of a massive rotating body has been proposed long time ago \cite{PhysRev.110.236}. 

%Along the rotation of the
%medium one process that transform right to left fermions is favored compared to the
%other and a separation of chirality is then induced. In this picture, the process in
%Fig.~\ref{fig:AVE_diag} effectively changes the way the spin of the fermions couples
%with the rotation of the medium. This, in turn, modify the AVE conductivity.
\vskip0.3cm

In analogy with the magnetic moment, the quantity that describes the coupling of a fermion's spin to rotation is
a gravitomagnetic moment. The picture discussed above thus suggests that we can
describe the radiative correction to this quantity as an Anomalous Gravitomagnetic Moment (AGM), as Fig.~\ref{fig:AVE_diag} is analogous to the diagram describing the anomalous magnetic moment in QED.
In this work we validate the connection between the AGM and the radiative corrections to the AVE for the case of a massive fermion. We show
that the gravitomagnetic moment of a massive fermion in thermal equilibrium
is indeed anomalous, and that the radiative corrections to the AVE
in the low temperature regime are due to the gravitomagnetic moment.

The emergence of the AGM may seem surprising, as the Einstein equivalence principle
forbids the appearance of an anomalous spin-rotation coupling \cite{kobzarev1962gravitational,Teryaev:2016edw}. 
%This is because the 
%equivalence principle imposes that all angular momentum vectors, including the spin
%of elementary particles, must obey equation of motion of the same form. 
In the
language of quantum field theory, the equivalence principle is translated into
the Lorentz invariance of the local coupling of gravitational fields to the
stress-energy tensor~\cite{Cho:1976de}. This is why the presence of an AGM is excluded
for both elementary~\cite{Cho:1976de} and composite~\cite{Teryaev:2016edw,Teryaev:2006fk} particles.
However, an explicit renormalization of stress-energy tensor for massive QED at
finite temperature has demonstrated that the effects of temperature break the weak equivalence
principle (i.e., gravitational and inertial mass are no longer 
equivalent)~\cite{Donoghue:1984zs,Donoghue:1984ga}. The equivalence principle violation
can be ascribed to the fact that a finite temperature breaks the Lorentz invariance of the
vacuum and that, in the presence of a thermal bath, one can genuinely discern if the
system is under acceleration or under the effect of a gravitational field by making measurements in
reference to the thermal bath. Thus, we can expect, and indeed have found, the emergence of the AGM at finite
temperature.
\vskip0.3cm

This paper is organized as follows. In Sec.~\ref{sec:AVE_Radiative} we evaluate the
AVE conductivity in the low temperature regime for a non-relativistic
particle and show that the radiative corrections to this conductivity can be quantitatively
derived from the anomalous gravitomagnetic moment. In Sec.~\ref{sec:AGM} we introduce
the gravitomagnetic moment of a Dirac fermion and evaluate the radiative correction to it 
at one-loop level in finite temperature QED. In Sec.~\ref{sec:AGMLingrav} we use the scattering
theory in linearized gravity to obtain a formula which connects the gravitomagnetic moment
to the matrix elements of the stress-energy tensor. A summary and a discussion of the
results are given in Sec.~\ref{sec:Discussion}. The details on the renormalization
of the stress-energy tensor at finite temperature and the evaluation of the related
Feynamn diagrams are given respectively in appendix~\ref{sec:SETRen} and~\ref{sec:APPSETRen}.
In App.~\ref{sec:EquivPrinFiniteT} we review how finite temperature effects lead to a
difference between inertial and gravitational mass.

%************************************************************************************
\section{Radiative corrections to the axial vortical effect for a massive field}
%************************************************************************************
\label{sec:AVE_Radiative}
In this section we evaluate the radiative corrections to the axial vortical
effect (AVE) conductivity $W\sups{A}$ for a massive Dirac field in a low temperature
regime. We obtain the AVE conductivity starting from the corresponding Kubo formula that involves
the thermal correlator between the stress-energy tensor and the axial current operators.
As any quantum operator can be written in terms of its matrix elements by expanding it
in multi-particle states~\cite{Weinberg:1995mt}, it then follows that the AVE conductivity
can be obtained by considering the radiative corrections to the following matrix elements:
\begin{equation*}
\bra{p',s'}\h{T}^{\mu\nu}(0)\ket{p,s},\quad  \bra{p',s'}\h{j}\subs{A}^\mu(0)\ket{p,s},
\end{equation*}
where $\h{T}^{\mu\nu}$ is the stress-energy tensor, $\h{j}\subs{A}$ is the axial
current, and $\ket{p,s}$ is the state representing a single fermion with momentum $p$
and spin $s$. The main strategy we adopt to evaluate the radiative corrections of
$W\sups{A}$ is to include the effects of the interactions by renormalazing the 
matrix elements above using finite temperature QED. After renormalization, the formula
for the AVE conductivity is formally identical to the one in the free field case,
the only difference being the values of the renormalized matrix elements.
In particular, the thermal correlator will act solely on the fermionic creation and
annihilation operators and can be readily written down. An equivalent approach to
evaluating statistical averages is described in the textbook~\cite{DeGroot:1980dk},
where the density matrix is written in terms of its ``in-picture'' matrix elements.
In this section we derive the conductivity $W\sups{A}$, while the renormalization of
the matrix elements is performed in App.~\ref{sec:SETRen}.

The AVE conductivity $W\sups{A}$ for a massive Dirac field is given by the thermal
correlator between the axial current and the total angular momentum of the system
and can be written as~\cite{Buzzegoli:2017cqy}
\begin{equation}
\label{eq:WAFormuala}
\begin{split}
W\sups{A}=2 \int_0^\beta \frac{\D\tau}{\beta}\int \D^3 x\, x^1
		\mean{\h{T}^{02}(-\I\tau,\vec{x})\,\h{j}^3\subs{A} (0) }\subs{T,c},
\end{split}
\end{equation}
where the symbol $\mean{\cdots}_T$ denotes thermal averages performed in the rest frame of
thermal bath $u=(1,\vec{0})$ with the familiar homogeneous global equilibrium density operator in the
grand-canonical ensemble
$$
\h{\rho} = \frac{1}{Z} \exp[-\beta(x) \cdot \h{P}]
$$
where $\h{P}^\mu$ is the total four-momentum of the system.
The subscript $c$ on the thermal average in~\eqref{eq:WAFormuala} denotes the
connected part of the correlator, that is, for the simplest case of two operators:
\begin{equation*}
\mean{\h{O}_1 \h{O}_2}_c\equiv \mean{\h{O}_1 \h{O}_2}-\mean{\h{O}_1}\mean{ \h{O}_2}.
\end{equation*}

As mentioned in the beginning of this section, the stress-energy tensor can be
decomposed in the multi-particle Hilbert space basis and can be written as a
combination of creation and annihilation operators. We thus denote with
$\wad_\tau$ and $\wbd_\tau$ the creation operators of, respectively, the particle
and anti-particle state of Dirac field with momentum $p$ and spin $\tau$,
covariantly normalized as:
\begin{equation*}
\{\wa_{\sigma}(q),\wad_{\tau'}(q')\}= 2\, \varepsilon_q
	\,\delta_{\tau\tau'}\delta^3(\vec{q}-\vec{q}'),\qquad
\{\wb_{\sigma}(q),\wbd_{\tau'}(q')\}= 2\, \varepsilon_q
	\,\delta_{\tau\tau'}\delta^3(\vec{q}-\vec{q}'),
\end{equation*}
where $\varepsilon_q=\sqrt{\vec{q}^2+m^2}$. The spinors for particle
$u_\tau(q)$ and for anti-particle $v_\tau(q)$ are normalized as
$$
u^\dagger_\tau(q) u_{\tau'}(q)= \delta_{\tau\tau'}, \qquad
v^\dagger_\tau(q) v_{\tau'}(q)= \delta_{\tau\tau'}.
$$
Furthermore, taking advantage of the fact that $\h{T}^{\mu\nu}$ is an additive operator,
it is uniquely determined by~\cite{Weinberg:1995mt}:
\begin{equation}
\label{eq:SETExpans}
\begin{split}
\h{T}^{\mu\nu}(x)=&\frac{1}{(2\pi)^3}\sum_{\tau,\tau'}\int\D^3 q'\int\D^3 q\,
	\wad_{\tau'}(q') \wa_{\tau}(q) 	\E^{\I(q-q')\cdot x}
	\bar{u}_{\tau'}(q')_A M^{\mu\nu}_{++}(q,q')_{AB}\, u_\tau(q)_B+\\
&+\frac{1}{(2\pi)^3}\sum_{\tau,\tau'}\int\D^3 q'\int\D^3 q\,
	\wad_{\tau'}(q') \wb_{\tau}(q) \E^{-\I(q+q')\cdot x}
	\bar{u}_{\tau'}(q')_A M^{\mu\nu}_{+-}(q,q')_{AB}\, v_\tau(q)_B\\
&+\frac{1}{(2\pi)^3}\sum_{\tau,\tau'}\int\D^3 q'\int\D^3 q\,
	\wbd_{\tau'}(q') \wa_{\tau}(q) \E^{\I(q+q')\cdot x}
	\bar{v}_{\tau'}(q')_A M^{\mu\nu}_{-+}(q,q')_{AB}\, u_\tau(q)_B\\
&+\frac{1}{(2\pi)^3}\sum_{\tau,\tau'}\int\D^3 q'\int\D^3 q\,
	\wbd_{\tau'}(q') \wb_{\tau}(q) \E^{-\I(q-q')\cdot x}
	\bar{v}_{\tau'}(q')_A M^{\mu\nu}_{--}(q,q')_{AB}\, v_\tau(q)_B
+ \h{T}^{\mu\nu}\subs{Photons}(x),
\end{split}
\end{equation}
where $\h{T}^{\mu\nu}\subs{Photons}(x)$ contains only terms in the creation and
annihilation operators for the photons, the sum over the spinor indices $A$ and $B$
is intended, and the matrices $M$ are inferred from the following stress-energy tensor
matrix elements:
\begin{equation}
\label{eq:Mmatrices}
\begin{split}
\bar{u}_{\tau'}(q') M^{\mu\nu}_{++}(q,q')\, u_\tau(q) \equiv &
	\bra{0}\wa_{\tau'}(q')\h{T}^{\mu\nu}(0)\wad_{\tau}(q)\ket{0}
	=\bra{q',\tau'}\h{T}^{\mu\nu}(0)\ket{q,\tau},\\
\bar{u}_{\tau'}(q') M^{\mu\nu}_{+-}(q,q')\, v_\tau(q) \equiv &
	\bra{0}\wa_{\tau'}(q')\h{T}^{\mu\nu}(0)\wbd_{\tau}(q)\ket{0}
	=\bra{q',\tau'}\h{T}^{\mu\nu}(0)\overline{\ket{q,\tau}},\\
\bar{v}_{\tau'}(q') M^{\mu\nu}_{-+}(q,q')\, u_\tau(q) \equiv &
	\bra{0}\wb_{\tau'}(q')\h{T}^{\mu\nu}(0)\wad_{\tau}(q)\ket{0}
	=\overline{\bra{q',\tau'}}\h{T}^{\mu\nu}(0)\ket{q,\tau},\\
\bar{v}_{\tau'}(q') M^{\mu\nu}_{--}(q,q')\, v_\tau(q) \equiv &
	\bra{0}\wb_{\tau'}(q')\h{T}^{\mu\nu}(0)\wbd_{\tau}(q)\ket{0}
	=\overline{\bra{q',\tau'}}\h{T}^{\mu\nu}(0)\overline{\ket{q,\tau}}.
\end{split}
\end{equation}

In the same way the axial current $\h{j}\subs{A}^\mu=\bar{\Psi}\gamma^\mu\gamma^5\Psi$
can be decomposed into
\begin{equation}
\label{eq:jAExpans}
\begin{split}
\h{j}\subs{A}^\mu(0)=&
	\sum_{\sigma,\sigma'}\int\frac{\D^3 k}{(2\pi)^{3/2}} \frac{\D^3 k'}{(2\pi)^{3/2}}
	\left[ \bar{u}_\sigma(k)_{A'} A^\mu_{++}(k,k')_{A'B'}\, u_{\sigma'}(k')_{B'}\wad_\sigma(k)\wa_{\sigma'}(k')\right.+\\
&+\bar{u}_\sigma(k)_{A'} A^\mu_{+-}(k,k')_{A'B'}\, v_{\sigma'}(k')_{B'}\wad_\sigma(k)\wb_{\sigma'}(k')
+\bar{v}_\sigma(k)_{A'} A^\mu_{-+}(k,k')_{A'B'}\, u_{\sigma'}(k')_{B'}\wbd_\sigma(k)\wa_{\sigma'}(k')+\\
&+\left.\bar{v}_\sigma(k)_{A'} A^\mu_{--}(k,k')_{A'B'}\, v_{\sigma'}(k')_{B'}\wbd_\sigma(k)\wb_{\sigma'}(k')\right],
\end{split}
\end{equation}
where
\begin{equation}
\label{eq:Amatrices}
\begin{split}
\bar{u}_{\sigma'}(k') A^\mu_{++}(k,k')\, u_\sigma(k) \equiv &
	\bra{0}\wa_{\sigma'}(k')\h{j}^\mu\subs{A}(0)\wad_{\sigma}(k)\ket{0}
	=\bra{k',\sigma'}\h{j}^\mu\subs{A}(0)\ket{k,\sigma},
\end{split}
\end{equation}
and similarly for the others as in~\eqref{eq:Mmatrices}.
Notice that the pure photon contribution to stress-energy tensor does not affect the AVE
conductivity. Indeed, the thermal correlator in~\eqref{eq:WAFormuala} for
$\h{T}^{\mu\nu}\subs{Photons}(x)$ would result in a combination of thermal averages
between one photonic and one fermionic operator which are vanishing.
Furthermore, the Dirac equation and the Poincaré symmetry constraint the matrix
elements~\eqref{eq:Amatrices} to take the form:
\begin{equation}
\label{eq:AFormfactors}
A^\mu_{s_1 s_2}(k,k')=	F^{A1}_{s_1 s_2}\left((k'-k)^2\right)\gamma^\mu\gamma^5
	+ F^{A2}_{s_1 s_2}\left((k'-k)^2\right)\frac{(k'-k)^\mu}{2m}\gamma^5,
\end{equation}
with $s_1,\,s_2=\pm$ and where $F^{A1}_{s_1 s_2}(0)=1$.

For the sake of clarity we illustrate the calculation only for the particle part, which
we denote with $W\sups{A}_{aa}$. That is we consider only the terms of~\eqref{eq:SETExpans}
and~\eqref{eq:jAExpans} which contain exactly one particle creation operator $\wad$
and one particle annihilation operator $\wa$. The contribution from the other terms
are obtained in a similar fashion and will be added at the end. We can proceed in the
calculation by plugging the expressions~\eqref{eq:SETExpans} and~\eqref{eq:jAExpans}
in the formula~\eqref{eq:WAFormuala}. Then the conductivity $W\sups{A}_{aa}$ involves
the thermal expectation values between four creation and annihilation operators,
which can be reduced by the thermal Wick theorem to the products of two-operator thermal
expectation values as follows:
\begin{equation*}
\begin{split}
\mean{\wad_1 \wa_2 \wad_3 \wa_4}_c=& \mean{\wad_1 \wa_2 \wad_3 \wa_4}
	-\mean{\wad_1 \wa_2}\mean{ \wad_3 \wa_4}
=\mean{\wad_1 \wa_4}\mean{ \wa_2\wad_3 }.
\end{split}
\end{equation*}
The two-operator thermal expectation values for Dirac fields with the
homogeneous grand-canonical ensemble operator $\h{\rho}$ are given by:
\begin{equation}\label{eq:aadagger}
\begin{split}
\mean{\wad_\tau(k) \wa_{\sigma}(q)}_{\beta(x)}=& \delta_{\tau\sigma}
	\delta^3(\vec{k}-\vec{q})n\subs{F}(\varepsilon_k),\quad
\mean{\wa_{\tau'}(k')\wad_{\sigma'}(q')}_{\beta(x)}=\delta_{\tau'\sigma'}
	\delta^3(\vec{k'}-\vec{q'})(1-n\subs{F}(\varepsilon_{k'})),\\
\mean{\wbd_\tau(k) \wb_{\sigma}(q)}_{\beta(x)}=&\delta_{\tau\sigma}
	\delta^3(\vec{k}-\vec{q})n\subs{F}(\varepsilon_k),\quad
\mean{\wb_{\tau'}(k')\wbd_{\sigma'}(q')}_{\beta(x)}=	\delta_{\tau'\sigma'}
	\delta^3(\vec{k'}-\vec{q'})(1-n\subs{F}(\varepsilon_{k'})),
\end{split}
\end{equation}
where other combination are vanishing and $n\subs{F}$ is the Fermi-Dirac distribution function
\begin{equation*}
n\subs{F}(\varepsilon_k) =  \frac{1}{\E^{\varepsilon_k/T } + 1}.
\end{equation*}
After simple calculation, by using the~\eqref{eq:aadagger} and the identity
\begin{equation*}
\sum_\sigma u_\sigma(k) \bar{u}_\sigma(k)=\frac{\slashed{k}+m}{2\varepsilon_p},\quad
\sum_\sigma v_\sigma(k) \bar{v}_\sigma(k)=-\frac{\slashed{k}+m}{2\varepsilon_p},
\end{equation*}
the particle part of the AVE conductivity is
\begin{equation*}
\begin{split}
W\sups{A}_{aa}=&\frac{2}{(2\pi)^6} \int_0^\beta\frac{\D\tau}{\beta}\int\D^3 x \sum_{\sigma,\sigma'}
	\int\frac{\D^3 k}{2\varepsilon_k}\int\frac{\D^3 k'}{2\varepsilon_{k'}}x^1
	\E^{\I(\vec{k}-\vec{k}')\cdot\vec{x}}\E^{-(k_0'-k_0)\tau}\\&\times M^{02}_{++}(k,k')_{AB}
	u_\sigma(k)_B \bar{u}_\sigma(k)_{A'}A^3_{++}(k,k')_{A'B'}
	u_{\sigma'}(k')_{B'} \bar{u}_{\sigma'}(k')_A n\subs{F}(\varepsilon_{k'})(1-n\subs{F}(\varepsilon_k))\\
=&\frac{2}{(2\pi)^6} \int_0^\beta\frac{\D\tau}{\beta}\int\D^3 x 
	\int\frac{\D^3 k}{2\varepsilon_k}\int\frac{\D^3 k'}{2\varepsilon_{k'}}x^1
	\E^{\I(\vec{k}-\vec{k}')\cdot\vec{x}}\E^{-(k_0'-k_0)\tau}\\
	&\times \tr\left[ M^{02}_{++}(k,k')\left(\slashed{k}+m\right)A^3_{++}
	\left(\slashed{k'}+m\right)\right]n\subs{F}(\varepsilon_{k'})(1-n\subs{F}(\varepsilon_k)).
\end{split}
\end{equation*}
Since the matrix elements $M^{02}_{++}(k,k')$ can always be simplified such that they
contain just one gamma matrix, when plugging the form~\eqref{eq:AFormfactors} of $A^3_{++}$
we realize that only the first form factor $F^{A1}_{++}$ can contribute to the trace.
It is now straightforward to integrate over the coordinates by using
\begin{equation*}
\int\D^3 x\, x^1 \E^{\I(\vec{k}'-\vec{k})\cdot\vec{x}}=\I(2\pi)^3
	\frac{\de}{\de k'_x}\delta^3(\vec{k}'-\vec{k}),
\end{equation*}
where $k_x$ is the first component of the spatial momentum $\vec{k}$. Thanks to
the delta function, the result
\begin{equation*}
W\sups{A}_{aa}=\frac{2\I}{(2\pi)^3} \int\frac{\D^3 k}{2\varepsilon_k}
	\int\frac{\D^3 k'}{2\varepsilon_{k'}}\frac{\de}{\de k'_x}\delta^3(\vec{k}'-\vec{k})
	\tr\left[ M^{02}_{++}(k,k')\left(\slashed{k}+m\right)\gamma^3\gamma^5
		\left(\slashed{k'}+m\right)\right] n\subs{F}(\varepsilon_{k'})(1-n\subs{F}(\varepsilon_k)) F^{A1}_{++}\left((k'-k)^2\right)
\end{equation*}
is easily integrated in $k'$:
\begin{equation}
\label{eq:AVE_Maa}
W\sups{A}_{aa}=\frac{\I}{(2\pi)^3}\frac{1}{8} \int\frac{\D^3 k}{\varepsilon_k}
	4(1-n\subs{F}(\varepsilon_k)) \frac{\de}{\de k'_x} \left\{\frac{ F^{A1}_{++}\left((k'-k)^2\right)}{\varepsilon_{k'}}
	\tr\left[ M^{02}_{++}(k,k') \left(\slashed{k}+m\right) \gamma^3\gamma^5
	\left(\slashed{k'}+m\right)\right] n\subs{F}(\varepsilon_{k'})\right\}_{k'=k}.
\end{equation}
Repeating the same steps described above for all the other terms in~\eqref{eq:SETExpans}
and in~\eqref{eq:jAExpans} and including a chemical potential $\mu$ we eventually obtain
\begin{equation}
\label{eq:AVE_M}
W\sups{A}=-2\I\int_0^\beta\frac{\D\tau}{\beta}\int\frac{\D^3 k}{(2\pi)^3}
	\left[\frac{\de}{\de k'_x}S(k,k',\tau)\right]_{k'=k},
\end{equation}
with
\begin{equation*}
\begin{split}
S(k,k',\tau)\equiv & \sum_{s_1,s_2=\pm} F^{A1}_{s_1 s_2}\left((k'-k)^2\right)
	\frac{T_{s_1,s_2}(\varepsilon_k, \varepsilon_{k'},\vec{k},\vec{k}')}
	{4\varepsilon_k \varepsilon_{k'}}
	\E^{-(s_1 \varepsilon_k-s_2\varepsilon_{k'})\tau}\times\\
	&\left(\delta_{s_1,+} - n\subs{F}(\varepsilon_k-s_1\mu)\right)
	\left(\delta_{s_2,-} - n\subs{F}(\varepsilon_{k'}-s_2\mu)\right),\\
T_{s_1,s_2}(\varepsilon_k, \varepsilon_{k'},\vec{k},\vec{k}')\equiv &
	\tr\left[ M^{02}_{s_1 s_2}(k,k')\left(\slashed{k}+m\right)\gamma^3\gamma^5
	\left(\slashed{k'}+m\right)	\right]_{k_0=s_1\varepsilon_k,k'_0=s_2\varepsilon_{k'}}.
\end{split}
\end{equation*}

Before moving on, it is important to check that the proposed method reproduces the known result
for the AVE conductivity in the non-interacting case. For a free theory the matrices
$M^{\mu\nu}_{\pm\pm}$ and the form factor $F^{A1}_{\pm\pm}$ defined in~\eqref{eq:Mmatrices}
and in~\eqref{eq:AFormfactors} do not depend on the signs $s_1$ and $s_2$ and they
are all simply given by
\begin{equation*}
M^{\mu\nu}\subs{Free}(q,q')=\frac{1}{4}\left[\gamma^\mu\left(q^\nu+q^{\prime\nu}\right)
	+\gamma^\nu\left(q^\mu+q^{\prime\mu}\right)\right],\quad
F^{A1}\subs{Free}\left((k'-k)^2\right)=1.
\end{equation*}
Not surprisingly, plugging these forms in~\eqref{eq:AVE_M} we reproduce the same
formula found in~\cite{Buzzegoli:2017cqy} for the AVE conductivity of a free massive
Dirac field. Working out the expression ~\eqref{eq:AVE_M} for a free field we eventually get
\begin{equation*}
W\sups{A}=\frac{1}{(2\pi)^3\beta}\int\D^3 k \left[
	-\frac{\varepsilon_k^2-k^2/3}{\varepsilon_k^3}\left(n\subs{F}(\varepsilon_k-\mu)+n\subs{F}(\varepsilon_k+\mu)\right)
	-\frac{\varepsilon_k^2+k^2/3}{\varepsilon_k^2}\frac{\de}{\de\varepsilon_k}\left(n\subs{F}(\varepsilon_k-\mu)+n\subs{F}(\varepsilon_k+\mu)\right)\right].
\end{equation*}
After integrating by parts we obtain the well-known result:
\begin{equation*}
W\sups{A}=\frac{1}{2\pi^2\beta}\int\D k \frac{\varepsilon_k^2+k^2}{\varepsilon_k}
	\left[n\subs{F}(\varepsilon_k-\mu)+n\subs{F}(\varepsilon_k+\mu)\right].
\end{equation*}
%

%------------------------------------------------------------------------------------
\subsection{Non-relativistic limit}
%------------------------------------------------------------------------------------
We now proceed to evaluate the radiative correction to the axial vortical effect (AVE)
for a non-relativistic particle at low temperatures $T\ll m$. In this regime
the relevant contribution to the AVE conductivity is given by the particle contribution
in Eq.~\eqref{eq:AVE_Maa}, while the other terms in~\eqref{eq:AVE_M} are sub-leading.
We start evaluating the formula~\eqref{eq:AVE_Maa} by expanding the derivative as:
\begin{equation*}
\begin{split}
W\sups{A}\subs{LT}\simeq&W\sups{A}_{aa}=\frac{\I}{(2\pi)^3}\frac{1}{8} \int\frac{\D^3 k}{\varepsilon_k}4(1-n\subs{F}(\varepsilon_k-\mu))
	\left\{	\frac{ F^{A1}_{++}(0)}{\varepsilon_k}  n\subs{F}(\varepsilon_k-\mu) \left[\frac{\de}{\de k'_x}T_{++}(k,k')\right]_{k'=k}+\right.\\
	&\left.+T_{++}(k,k) \frac{\de}{\de k'_x}\left[\frac{F^{A1}_{++}\left((k'-k)^2\right)}{\varepsilon_{k'}} n\subs{F}(\varepsilon_{k'}-\mu)\right]_{k'=k}\right\},
\end{split}
\end{equation*}
where the subscript LT stands for low temperature and
\begin{equation}
\label{eq:TraceTLTpp}
T_{++}(k,k')=\tr\left[ M^{02}_{++}(k,k') \left(\slashed{k}+m\right) \gamma^3\gamma^5
	\left(\slashed{k'}+m\right)\right].
\end{equation}
Taking advantage of Dirac equation in~\eqref{eq:Mmatrices} we see that the matrix
element $M$ can always be written as $M^{02}_{++}(k,k')_{AB}=M^{02}_{++}(k,k')_\lambda \gamma^\lambda_{AB}$.
Then the trace reads
\begin{equation*}
T_{++}(k,k')=M^{02}_{++}(k,k')_\lambda\tr\left[\gamma^\lambda \left(\slashed{k}+m\right) \gamma^3\gamma^5
	\left(\slashed{k'}+m\right)\right]=-4\I M^{02}_{++}(k,k')_\lambda \epsilon^{\lambda 3\mu\nu}k_\mu k'_\nu,
\end{equation*}
from which we see that $T_{++}(k,k')$ is vanishing when $k'=k$. Also, reminding that $F^{A1}_{++}(0)=1$,
the AVE conductivity becomes
\begin{equation*}
\begin{split}
W\sups{A}\subs{LT}=&\frac{\I}{(2\pi)^3}\frac{1}{8} \int\frac{\D^3 k}{\varepsilon_k^2}
	4(1-n\subs{F}(\varepsilon_k-\mu))n\subs{F}(\varepsilon_k-\mu)\left[\frac{\de}{\de k'_x}T_{++}(k,k')\right]_{k'=k}\\
	=& -\frac{\I}{(2\pi)^3}\frac{1}{8\beta} \int\frac{\D^3 k}{\varepsilon_k^2}
	4n\subs{F}'(\varepsilon_k-\mu)\left[\frac{\de}{\de k'_x}T_{++}(k,k')\right]_{k'=k},
\end{split}
\end{equation*}
where we used the identity
\begin{equation*}
(1-n\subs{F}(\varepsilon_k-\mu))n\subs{F}(\varepsilon_k-\mu)=-\frac{1}{\beta}\frac{\de}{\de\epsilon_k}n\subs{F}(\varepsilon_k-\mu)
	=-\frac{1}{\beta}n\subs{F}'(\varepsilon_k-\mu).
\end{equation*}
To evaluate the trace, we notice (see App.~\ref{sec:Selection}) that only the following
form factors of $M$ bring a non-vanishing contribution to $T_{++}(k,k')$:
\begin{equation}
\label{eq:EffSET_M}
\begin{split}
M^{\mu\nu}\subs{Relevant}(k,k')=&I_{P\gamma}(P,q)\left(\gamma^\mu P^\nu+\gamma^\nu P^\mu\right)
	+I_{u\gamma}(P,q)\left(\gamma^\mu u^\nu+\gamma^\nu u^\mu\right)+\\
&+I_{Pl}(P,q)\slashed{\hat{l}}\left(\hat{l}^\mu P^\nu+\hat{l}^\nu P^\mu\right)
	+I_{ul}(P,q)\slashed{\hat{l}}\left(\hat{l}^\mu u^\nu+\hat{l}^\nu u^\mu\right),
\end{split}
\end{equation}
where $u$ is the thermal bath velocity, $P=k+k'$, $q=k'-k$ and
\begin{equation}
\label{eq:Ldef}
l^\mu=\epsilon^{\mu\nu\rho\sigma}u_\nu P_\rho q_\sigma.
\end{equation}
The explicit values of the form factors will be evaluated in Appendix~\ref{sec:SETRen}
and the final results are reported in~\eqref{eq:AGMFormFactors} for $q=0$ and in the
non-relativistic limit of $P$. In the rest frame of thermal bath, where the
formula~\eqref{eq:AVE_Maa} is evaluated, the matrix elements~\eqref{eq:EffSET_M} read
\begin{equation*}
\begin{split}
M^{\mu\nu}(k,k')\subs{Rest}=&I_{P\gamma}(P,q)\left(\gamma^\mu P^\nu+\gamma^\nu P^\mu\right)
	+I_{u\gamma}(P,q)\left(\gamma^\mu \delta^{\nu 0}+\gamma^\nu \delta^{\mu 0}\right)+\\
&+I_{Pl}(P,q)\slashed{\hat{l}}\left(\hat{l}^\mu P^\nu+\hat{l}^\nu P^\mu\right)
	+I_{ul}(P,q)\slashed{\hat{l}}\left(\hat{l}^\mu \delta^{\nu 0}+\hat{l}^\nu \delta^{\mu 0}\right).
\end{split}
\end{equation*}
First, consider the term in $I_{P\gamma}(P,q)$. The calculation is the same as the
free case except for the form factor. The trace reads:
\begin{equation*}
T_{P\gamma}(k,k')=I_{P\gamma}(P,q)\left[\eta^0_\lambda(k_y+k'_y)+\eta^2_\lambda(\varepsilon_k+\varepsilon_{k'})\right]
	(-4\I)\epsilon^{\lambda 3\mu\nu}k_\mu k'_\nu,
\end{equation*}
whose derivative is
\begin{equation*}
\begin{split}
\frac{\de}{\de k'_x}T_{P\gamma}(k,k')\Big|_{k'=k}=&2I_{P\gamma}(P,q=0)\left(\eta^0_\lambda k_y + \eta^2_\lambda \varepsilon_k\right)
	(-4\I)\epsilon^{\lambda 3\mu\nu}k_\mu \eta_{\nu 1}\\
=& -8\I I_{P\gamma}(P,q=0)\left(-\epsilon^{0321}k_y^2+\epsilon^{2301}\varepsilon_k^2\right)
=-8\I I_{P\gamma}(P,q=0)\left(\varepsilon_k^2 + k_y^2 \right).
\end{split}
\end{equation*}
Then this term of matrix elements contributes to the low temperature AVE conductivity as
\begin{equation*}
\begin{split}
W\sups{A}_{P\gamma}=& -\frac{1}{(2\pi)^3\beta} \int\D^3 k\,
	4I_{P\gamma}(P,q=0) \frac{\varepsilon_k^2+k_y^2}{\varepsilon_k^2}n\subs{F}'(\varepsilon_k-\mu)\\
=&-\frac{1}{(2\pi)^3\beta} \int\D^3 k\,
4I_{P\gamma}(P,q=0) \frac{\varepsilon_k^2+k^2/3}{\varepsilon_k^2}n\subs{F}'(\varepsilon_k-\mu).
\end{split}
\end{equation*}
Since the form factor only depend on the spatial modulus of $P$ we can integrate by parts:
\begin{equation*}
\begin{split}
W\sups{A}_{P\gamma}=&\frac{1}{2\pi^2\beta} \int\D k\,\left[
	4I_{P\gamma}(P,q=0) \frac{\varepsilon_k^2+k^2}{\varepsilon_k}
	+4I_{P\gamma}(P,q=0)k^2 \frac{\varepsilon_k^2-k^2/3}{\varepsilon_k^3}+\right.\\
&\left.+4\frac{\de I_{P\gamma}(P,q=0)}{\de\varepsilon_k}\left(k^2+\frac{k^4}{3\varepsilon_k^2}\right)\right]
	n\subs{F}(\varepsilon_k-\mu).
\end{split}
\end{equation*}
At low temperature the second term is sub-leading compared to the first and we can
approximate the coefficient as
\begin{equation}
\label{eq:WALTPGamma}
\begin{split}
W\sups{A}_{P\gamma}=&\frac{1}{2\pi^2\beta} \int\D k\,\left[
	4I_{P\gamma}(P,q=0) \frac{\varepsilon_k^2+k^2}{\varepsilon_k}
	+4\frac{\de I_{P\gamma}(P,q=0)}{\de\varepsilon_k}\left(k^2+\frac{k^4}{3\varepsilon_k^2}\right)\right]
	n\subs{F}(\varepsilon_k-\mu).
\end{split}
\end{equation}

Explicit calculations reveal that the other terms of~\eqref{eq:EffSET_M} when plugged
into~\eqref{eq:AVE_Maa} give contribution of the same form as~\eqref{eq:WALTPGamma}
but with a different form factor. To illustrate this we show that the other terms
of~\eqref{eq:EffSET_M} when traced in~\eqref{eq:TraceTLTpp} give the same result as
the first term, hence $M^{\mu\nu}$ can be effectively by written as proportional to
$\left(\gamma^\mu P^\nu+\gamma^\nu P^\mu\right)$. Firstly, since the form factors
must be evaluated for $q$ going to zero and then integrated over $k$, from
Eq.~\eqref{eq:Ldef} we see that the matrix elements of~\eqref{eq:EffSET_M} proportional
to $\slashed{l}$ are effectively given by
\begin{equation*}
\begin{split}
M^{\mu\nu}(k,k')\subs{Eff}=&I_{P\gamma}(P,q)\left(\gamma^\mu P^\nu+\gamma^\nu P^\mu\right)
	+I_{u\gamma}(P,q)\left(\gamma^\mu \delta^{\nu 0}+\gamma^\nu \delta^{\mu 0}\right)+\\
&-I_{Pl}(P,q)\left(\gamma^\mu P^\nu+\gamma^\nu P^\mu\right)
	-I_{ul}(P,q)\left(\gamma^\mu \delta^{\nu 0}+\gamma^\nu \delta^{\mu 0}\right).
\end{split}
\end{equation*}
Secondly, in the rest frame of thermal bath where $\omega_P=P\cdot u=P^0$
we can write the form factors $I_{u\gamma}(P,q)$ and $I_{ul}(P,q)$ as
\begin{equation*}
\begin{split}
M_{\mu\nu}(k,k')\subs{Eff}=&I_{P\gamma}(P,q)\left(\gamma^\mu P^\nu+\gamma^\nu P^\mu\right)
	+\frac{I_{u\gamma}(P,q)}{\omega_P}\left(\gamma^\mu \delta^{\nu 0}+\gamma^\nu \delta^{\mu 0}\right)P_0+\\
&-I_{Pl}(P,q)\left(\gamma^\mu P^\nu+\gamma^\nu P^\mu\right)
	-\frac{I_{ul}(P,q)}{\omega_P}\left(\gamma^\mu \delta^{\nu 0}+\gamma^\nu \delta^{\mu 0}\right)P_0.
\end{split}
\end{equation*}
Lastly, one can explicitly check that the contribution from
$\left(\gamma^\mu \delta^{\nu 0}+\gamma^\nu \delta^{\mu 0}\right)P_0$ is the
same as $\left(\gamma^\mu P^\nu+\gamma^\nu P^\mu\right)$ in the low temperature limit
and the effective matrix element is
\begin{equation*}
\begin{split}
M^{\mu\nu}(k,k')\subs{Eff}=&\left[I_{P\gamma}(P,q) +\frac{I_{u\gamma}(P,q)}{\omega_P}
	-I_{Pl}(P,q) -\frac{I_{ul}(P,q)}{\omega_P}\right]\left(\gamma^\mu P^\nu+\gamma^\nu P^\mu\right)\\
=&\frac{g_\Omega(P,q)}{4}\,\left(\gamma^\mu P^\nu+\gamma^\nu P^\mu\right).
\end{split}
\end{equation*}
As we will show in the next sections, the quantity $g_\Omega$ defined from the form
factors of the stress-energy tensor above is the
gravitomagnetic moment of a fermion. At finite temperature the interactions renormalize
the angular momentum of the system and the spin-orbit coupling of the fermion, described
by $g_\Omega$, can be different from 1. Following the steps described above, it is now
clear that the AVE conductivity at low temperature is given by
\begin{equation}
\label{eq:WARadCorr}
W\sups{A}\subs{LT}=\frac{1}{2\pi^2\beta}\int\D k \left[g_\Omega(\varepsilon_k)\frac{\varepsilon_k^2+k^2}{\varepsilon_k}
	+g'_\Omega(\varepsilon_k)\left(k^2+\frac{k^4}{3\varepsilon_k^2} \right)\right]n\subs{F}(\varepsilon_k-\mu).
\end{equation}
Eq.~\eqref{eq:WARadCorr} connects the radiative corrections to the AVE to the anomalous
gravitomagnetic moment. If the gravitomagnetic moment is not anomalous, i.e. $g_\Omega=1$,
then the AVE conductivity~\eqref{eq:WARadCorr} remains the one of a free-field, and radiative corrections are not present. On the other hand, if the gravitomagnetic moment is affected by
interactions, the radiative corrections to the AVE can be obtained from the formula above.
Let us stress that the Eq.~\eqref{eq:WARadCorr} holds in the large mass limit, where
we can factorize long- and short-distance contributions. The effects of interactions
above some energy scale (short distances) are contained in the form factors, which we
renormalize at finite temperature, and result in the gravitomagnetic moment.
The low energy contributions (large distance) are collected into the thermal averages
of one-particle states, which correspond to an expansion in $T/m$. We furthermore
considered a non-relativistic particle, such that the contribution coming from the
anti-particle thermal distribution $n\subs{F}(\varepsilon_k+\mu)$ is exponentially
suppressed compared to the particle one $n\subs{F}(\varepsilon_k-\mu)$.
In the following sections we show that the gravitomagnetic moment at finite temperature
is anomalous and we evaluate the first QED radiative correction.

To conclude this section we estimate the integral in Eq.~\eqref{eq:WARadCorr}.
At low temperature the leading contribution is obtained by setting
\begin{equation*}
g_\Omega(\varepsilon_k)\to g_\Omega=\lim_{\vec{k}\to 0}g_\Omega(\varepsilon_k)
,\quad g'_\Omega(\varepsilon_k)=0.
\end{equation*}
At 1-loop in thermal QED we found (see next Section) that for $T\ll m$ the anomalous
gravitomagnetic moment is
\begin{equation*}
g_\Omega-1=-\frac{1}{6}\frac{e^2 T^2}{m^2},
\end{equation*}
where $e$ is the QED coupling constant. The radiative correction to the AVE
then reads
\begin{equation*}
\Delta W\sups{A}\subs{LT}\simeq-\frac{1}{2\pi^2\beta}\frac{1}{6}\frac{e^2 T^2}{m^2}
	\int\D k \frac{\varepsilon_k^2+k^2}{\varepsilon_k} n\subs{F}(\varepsilon_k-\mu)
	=-\frac{1}{6}\frac{e^2 T^2}{m^2} W\sups{A}\subs{free}.
\end{equation*}
Replacing the non-relativistic low temperature limit of the axial vortical effect
conductivity~\cite{Buzzegoli:2017cqy} we finally obtain
\begin{equation*}
\Delta W\sups{A}\subs{LT}\simeq-\frac{1}{6}\frac{e^2 T^2}{m^2}
	\frac{(m T)^{3/2}}{\sqrt{2}\pi^{3/2}}\E^{-(m-\mu)/T}.
\end{equation*}
%

%************************************************************************************
\section{Gravitomagnetic moment}
%************************************************************************************
\label{sec:AGM}
In this section we introduce the gravitomagnetic moment and present the result
for the Anomalous Gravitomagnetic Moment (AGM) in finite temperature QED. The AGM can be 
easily understood in analogy to the anomalous magnetic moment. The Dirac equation
in an external magnetic field and in the presence of rotation can be written using rotating coordinates
and takes the form
\begin{equation*}
\left[\I\gamma^\mu(D_\mu+\Gamma_\mu)-m\right]\psi=0
\end{equation*}
with the covariant derivative $D_\mu=\de_\mu-\I eA_\mu$, the spin connection $\Gamma_\mu=-\frac{\I}{4}\omega_{\mu i j}\sigma^{ij}$
and $\sigma^{ij}=\frac{\I}{2}[\gamma^i,\gamma^j]$. Setting the magnetic field $\vec{B}$ and
the rotation $\vec{\Omega}$ along the $z$ axis we find
\begin{equation*}
\left[\I\gamma^\mu D_\mu+\I\gamma^0\Omega(y\de_x-x\de_y-\I\sigma^{12})-m\right]\psi=
\left[\I\gamma^\mu D_\mu+\gamma^0\Omega(\h{L}_z+\h{S}_z)-m\right]\psi=0.
\end{equation*}
Acting on this equation with $\left(\I\gamma^\mu(D_\mu+\Gamma_\mu)+m\right)$ we obtain
the second order Dirac equation
\begin{equation*}
\left[\de_t^2-\nabla^2-e\vec{B}(\vec{L}+2\vec{S})-\vec\Omega(\vec{L}+\vec{S})+m^2\right]\psi=0.
\end{equation*}
The quantity which couples the magnetic field to the spin $\vec{S}$ is called the magnetic
moment $g_B$, while the quantity $g_\Omega$ that couples the rotation to the spin is the
gravitomagnetic moment. Therefore, Dirac equation alone predicts that the spin couples
to the magnetic moment and rotation with
\begin{equation*}
g_B=2,\qquad g_\Omega=1.
\end{equation*}
This is exactly what we expect for Dirac particles, as discussed 
in~\cite{deOliveira:1962apw,Mashhoon:1988zz,Hehl:1990nf}; spin-rotation coupling
has been reviewed in~\cite{Mashhoon:2000jq,Silenko:2006er}.

However, one of the most precise predictions of QED is that the magnetic moment of
fermions is anomalous, i.e. the value given by the Dirac equation receives radiative corrections. The anomalous magnetic
moment can be found using scattering theory in QED in the non-relativistic limit.
Interactions with magnetic field are described by the Lagrangian $\mathcal{L}=J^\mu A_\mu$,
therefore radiative corrections to magnetic moment are obtained by evaluating the matrix
element of the electromagnetic current, which can be written as:
\begin{equation*}
\bra{p',s'}J^\mu(0)\ket{p,s}=\bar{u}(p',s')\left\{\frac{P^\mu}{2m}F_1(q^2)+\frac{\I\sigma^{\mu\nu}q_\nu}{2m}\left[F_1(q^2)+F_2(q^2)\right]\right\}u(p,s)
\end{equation*}
where $\ket{p,s}$ denotes a state with a fermion of momentum $p$ and polarization
$s$, $u(p,s)$ denotes the Dirac spinor and we defined the momenta $P=p'+p$ and $q=p'-p$;
$F_1,F_2$ are the form factors. At first order in the transferred momentum $q$ the matrix element is
\begin{equation*}
\bra{p',s'}J^\mu(0)\ket{p,s}=\bar{u}(p',s')\left\{\frac{P^\mu}{2m}+\frac{\I\sigma^{\mu\nu}q_\nu}{2m}\left[1+F_2(0)\right]\right\}u(p,s)+\mathcal{O}(q^2).
\end{equation*}
The from factor $F_2(0)$ gives directly the anomalous magnetic moment, which was first evaluated
by Schwinger~\cite{PhysRev.73.416} at order $\alpha=e^2/4\pi$ from the vertex correction.
He obtained the famous result\footnote{Apart from the 
	higher order corrections, the finite temperature effects also give contribution to the
	anomalous magnetic moment, see for instance~\cite{Fujimoto:1982np,Peressutti:1981jg}.}
\begin{equation*}
a_B\equiv\frac{g_B-2}{2}=F_2(0)=\frac{\alpha}{2\pi}.
\end{equation*}

Following a similar argument using scattering theory (as explained in detail in
Section~\ref{sec:AGMLingrav}), we can relate the gravitomagnetic moment and its
radiative corrections to the matrix elements of the stress-energy tensor.
In the weak gravitational field limit, the gravitational coupling is described
to first order in the gravitational field by the Lagrangian $\mathcal{L}=\frac{1}{2}h_{\mu\nu}T^{\mu\nu}$,
where $h$ are small deviations from Minkowski space-time $g_{\mu\nu}=\eta_{\mu\nu}+h_{\mu\nu}$,
and everything else is happening in flat space-time. The rotation of the medium is contained inside $h$ and
to see the effects of interactions we must consider the matrix element of the stress-energy tensor
$\bra{p',s'}T^{\mu\nu}(0)\ket{p,s}$. If we perform this calculation at zero temperature (see
Sec.~\ref{sec:1loopAGM} for the details) we realize that there are no modifications to spin
coupling as dictated by the Einstein equivalence principle~\cite{Cho:1976de}.

We are now interested in finite temperature effects, therefore we consider a Dirac fermion
in thermal equilibrium with the medium. It is possible to use a manifestly Lorentz covariant form of
thermal field theory to write the matrix elements of stress-energy tensor if we also take into
account the thermal bath velocity $u$~\cite{Weldon:1982aq}. We are then using the S-matrix elements
in the QED thermal theory for a non-relativistic fermion. Since the rotation of the medium is
taken in reference to the rest frame of the thermal bath, the S-matrix elements between incoming and
outgoing fermion states of momenta $p$ and $p'$ impose the conservation equation $p\cdot u=p'\cdot u$ or equivalently
$q\cdot u=0$. To detect AGM, it is then convenient to move into the rest frame of the medium $u=(1,\vec{0})$,
which implies
\begin{equation*}
q_0=0,\quad
p_0=p_0'=\sqrt{m^2+(\vec{P}^2+\vec{q}^2)/4},\quad |\vec{p}'|=|\vec{p}|.
\end{equation*}
In a generic frame, we take advantage of orthogonality relation $q\cdot P=0$ and introduce
$\omega_P\equiv P\cdot u$, so that we can define a space-like four-vector $\tilde{P}$ orthogonal to $u$,
and a scalar $P_s$:
\begin{equation*}
\tilde{P}_\mu\equiv(\eta_{\mu\nu}-u_\mu u_\nu)P^\nu=P_\mu -\omega_P u_\mu,\quad
P_s\equiv\sqrt{\omega_P^2-P^2}.
\end{equation*}
To close the tetrad, we define the space-like unit four-vector $\hat{l}$ orthogonal to $u,\,\tilde{P}$
and $q$ (it is also orthogonal to $P$):
\begin{equation*}
l^\mu=\epsilon^{\mu\nu\rho\sigma}u_\nu\tilde{P}_\rho q_\sigma,\quad \hat{l}^\mu=\frac{l^\mu}{\sqrt{-l^2}}.
\end{equation*}
Note that $\tilde{P}\cdot q=0$ and $q\cdot u=0$ also hold true, meaning that
the tetrad $\{ u,\,\tilde{P},\,q,\,\hat{l}\}$ is an orthogonal non-normalized basis.
Using scattering theory (see Sec.~\ref{sec:1loopAGM}), the gravitomagnetic moment is obtained by
\begin{equation}
\label{eq:gOmegaForm}
g_\Omega=\lim_{\substack{q\to 0 \\ P_s\to 0}}4\left(I_{P\gamma}(P,q) + \frac{I_{u\gamma}(P,q)}{\omega_P} -I_{Pl}(P,q) -\frac{I_{ul}(P,q)}{\omega_P} \right),
\end{equation}
where the functions are the following form factors of the stress-energy tensor matrix element
\begin{equation}
\label{eq:SETFormFactors}
\begin{split}
\bra{p',s'}T_{\mu\nu}(0)\ket{p,s}=\bar{u}(p',s')\Big\{&I_{P\gamma}(P,q)\left(P_\mu \gamma_\nu + P_\nu \gamma_\mu\right)
+I_{u\gamma}(P,q)\left(u_\mu \gamma_\nu + u_\nu \gamma_\mu\right)\\
&+I_{Pl}(P,q)\slashed{\hat{l}}\left(P_\mu \hat{l}_\nu + P_\nu \hat{l}_\mu\right)
+I_{ul}(P,q)\slashed{\hat{l}}\left(u_\mu \hat{l}_\nu + u_\nu \hat{l}_\mu\right)+\cdots\Big\}u(p,s)+\mathcal{O}(q^2).
\end{split}
\end{equation}
Notice that in writing eq.~\eqref{eq:SETFormFactors} we have tacitly chosen to decompose the matrix
element with terms that do not contain more than one gamma matrix. At zero temperature the
decomposition~\eqref{eq:SETFormFactors} reads:
\begin{equation}
\label{eq:OurSETFormFactors}
\begin{split}
\bra{p',s'}T_{\mu\nu}(0)\ket{p,s}=\bar{u}(p',s')\Big\{& I_{P\gamma}(q^2) P_{(\mu} \gamma_{\nu)}+
I_{PP}(q^2) \frac{P_\mu P_\nu}{m} \Big\}u(p,s)+\mathcal{O}(q^2).
\end{split}
\end{equation}
However, there is a certain freedom in the choice of the form factors of the
stress-energy tensor stemming from the Gordon identity:
\begin{equation}
\label{eq:GordonId}
\bar{u}(p',s') \gamma^\mu u(p,s) = \bar{u}(p',s') \left[ \frac{P^\mu}{2m}
    + \frac{\I \sigma^{\mu\alpha} q_\alpha}{2m} \right] u(p,s).
\end{equation}
Different forms of this decomposition are used in the literature, for instance
in~\cite{kobzarev1962gravitational} the decomposition is written as:
\begin{equation}
\label{eq:KobzarevFormFactor}
\bra{p',s'}T_{\mu\nu}(0)\ket{p,s}=\bar{u}(p',s')\left[  f_1(q^2) \frac{P_\mu P_\nu}{m}
    + f_2(q^2) \frac{\I \sigma_{(\mu\alpha} q^\alpha P_{\nu)}}{2m} + \cdots \right]u(p,s)+\mathcal{O}(q^2),
\end{equation}
while in~\cite{Teryaev:2016edw} the following decomposition is chosen:
\begin{equation}
\label{eq:FirstOrderFormFactor}
\bra{p',s'}T_{\mu\nu}(0)\ket{p,s}=\bar{u}(p',s')\left[  A(q^2) \gamma_{(\mu} P_{\nu)}
    + B(q^2) \frac{\I \sigma_{(\mu\alpha} q^\alpha P_{\nu)}}{2m} \right]u(p,s)+\mathcal{O}(q^2).
\end{equation}
By using the Gordon identity~\eqref{eq:GordonId} it is straightforward to show that the
form factors in eq.s~\eqref{eq:OurSETFormFactors},~\eqref{eq:KobzarevFormFactor}
and~\eqref{eq:FirstOrderFormFactor} are related to each other by:
\begin{equation*}
I_{P\gamma}(q^2)=A(q^2)+B(q^2)=f_2(q^2),\qquad
I_{PP}(q^2)=-B(q^2)=f_1(q^2)- f_2(q^2).
\end{equation*}
The presence of a finite temperature allows the introduction of additional form factors, which are
also not uniquely defined. However the argument presented in Sec.~\ref{sec:1loopAGM} allows to 
unambiguously identify the gravitomagnetic moment once a consistence choice of the decomposition
is made, in particular the decomposition in eq.~\eqref{eq:SETFormFactors} results in the
gravitomagnetic moment in eq.~\eqref{eq:gOmegaForm}.

In App.~\ref{sec:SETRen} we evaluate the form factors in~\eqref{eq:SETFormFactors} by renormalizing
the stress-energy tensor in finite temperature QED. The full result at one-loop and at first
order in $q$ is reported in App.~\ref{sec:SETrenT}. In the limit $q\to 0$ and $P_s\to 0$, the thermal
contributions to the form factors are
\begin{equation}
\label{eq:AGMFormFactors}
\begin{split}
I_{P\gamma}=&-\frac{10+3\theta\subs{HT}}{72}\frac{e^2 T^2}{4m^2};\\
\frac{I_{u\gamma}}{\omega_P}=& -\frac{1}{9}\frac{e^2 T^2}{4m^2};\\
I_{Pl}=& \frac{2-\theta\subs{HT}}{72}\frac{e^2 T^2}{4m^2};\\
\frac{I_{ul}}{\omega_P}=& -\frac{2+\theta\subs{HT}}{18}\frac{e^2 T^2}{4m^2},\\
\end{split}
\end{equation}
where we introduced the function $\theta\subs{HT}$
\begin{equation*}
\theta\subs{HT}\equiv\begin{cases}
	0 & T \ll m \\
	1 & T \gg m
\end{cases}
\end{equation*}
which turns on the high temperature contributions. It is now straightforward to
compute the AGM from the relation~\eqref{eq:gOmegaForm} which gives
\begin{equation}
\label{eq:ResultAGM}
g_\Omega-1=-\frac{6-\theta\subs{HT}}{36}\frac{e^2 T^2}{m^2}
=\begin{cases}
	-\frac{1}{6}\frac{e^2 T^2}{m^2} & T \ll m \\
	-\frac{5}{36}\frac{e^2 T^2}{m^2} & T \gg m,\,  m > e T
\end{cases}.
\end{equation}
It should also be taken into account that in order to define one particle states and
the form factors of the stress-energy tensor, the mass of the particle must be larger
than $e T$. Then, the high temperature limit can only be taken in the weak coupling limit.
The presence of an anomalous gravitomagnetic moment is to be ascribed to the violation
of Einstein equivalence principle. At finite temperature the vacuum is not Lorentz
invariant. On the contrary, we can always distinguish a preferred reference frame,
which is the one where the thermal bath is at rest. This means that we can distinguish
between acceleration and the genuine effects of gravity. Similar thermal effects
in QED affect modify the values of inertial and gravitational mass of a Dirac fermion
providing an explicit breaking of the weak equivalence
principle~\cite{Donoghue:1984zs,Donoghue:1984ga,Donoghue:1983qx,Mitra:2001ar,Nieves:1998xz}, see
App.~\ref{sec:EquivPrinFiniteT} for details. 

%------------------------------------------------------------------------------------
\section{Gravitomagnetic moment in Linearized gravity}
%------------------------------------------------------------------------------------
\label{sec:AGMLingrav}
In this section we identify the gravitomagnetic moment of a fermion using the scattering theory in linearized gravity. 
The fermion interacts with an external gravitational field $g\subs{Ext}^{\mu\nu}=\eta^{\mu\nu}+h\subs{Ext}^{\mu\nu}$ through
the linearized Hamiltonian
\begin{equation*}
\h{H}\subs{int}=\frac{1}{2}\int\D^3 x\, \h{T}_{\mu\nu}h\subs{Ext}^{\mu\nu}.
\end{equation*}
The corresponding scattering amplitude is
\begin{equation*}
\mc{A} =-\I(2\pi)\delta(p\cdot u-p'\cdot u)\frac{1}{\sqrt{Z_2(p)Z_2(p')}}\frac{1}{2}\bar{u}_\beta(p')M_{\mu\nu}(p,p') u_\beta(p) h\subs{Ext}^{\mu\nu}(p'-p),
\end{equation*}
where $Z_2$ is the wave-function renormalization constant.
To study the effect of rotation, we then just have to use the proper metric accounting for rotation. Using the linear approximation of
gravito-electromagnetism~\cite{Mashhoon:2000jq}, the metric can be written in terms of a gravitational gauge potential
$A_g^\mu=(2\phi_g,\vec{A}_g)$. Let $h_{\mu\nu}$ be the perturbation of metric $g_{\mu\nu}=\eta_{\mu\nu}+h_{\mu\nu}$;
we then define $\bar{h}_{\mu\nu}=h_{\mu\nu}-\frac{1}{2}\eta_{\mu\nu}h^\alpha_\alpha$.
This definition is related to the gravitation gauge potential by $\bar{h}^{00}=4\phi_g$ and $\bar{h}^{0i}=2 A^i_g$.
For the case of rotation around an axis, we have $h^\alpha_\alpha=0$ and therefore $h_{\mu\nu}=\bar{h}_{\mu\nu}$
and $\phi_g=0$. The only non-vanishing components of metric perturbation are $\bar{h}^{0\mu}=2(2\phi_g,\vec{A}_g)$.
We can also define a gravitomagnetic field via $B_g^i=\epsilon^{ijk}\nabla^j A_g^k$, or in terms
of Fourier transform $B_g^i=-\I\epsilon^{ijk}q^j A_g^k(\vec{q})$.
Gravito-electromagnetism is particularly well suited for describing a pure rotation.
For instance, consider a constant rotation around the $z$ axis. In rotating cartesian coordinates, the deviation from
flat space-time is
\begin{equation}
\label{eq:hrotation}
h_{\mu\nu}=\left(\begin{array}{cccc}
-(x^2+y^2)\Omega^2 & y\Omega & -x\Omega & 0\\
y\Omega & 0 & 0 & 0\\
-x\Omega & 0 & 0 & 0\\
0 & 0 & 0 & 0
\end{array}\right)\simeq
\left(\begin{array}{cccc}
0 & y\Omega & -x\Omega & 0\\
y\Omega & 0 & 0 & 0\\
-x\Omega & 0 & 0 & 0\\
0 & 0 & 0 & 0
\end{array}\right)+\mathcal{O}(\Omega^2).
\end{equation}
For the metric in Eq. (\ref{eq:hrotation}), we simply have $\vec{B}_g=\vec{\Omega}$.
For our system at thermal equilibrium, the rotation is taken in reference to the thermal bath velocity $u$
and the metric deviation is therefore given by $h^{\mu\nu}=2u^\mu A^\nu+2u^\nu A^\mu$.
With this metric, the scattering amplitude is
\begin{equation}
\label{eq:QuantumAGMAmpl}
\mc{A} =-\I(2\pi)\delta(p\cdot u-p\cdot u)\frac{2}{\sqrt{Z_2(p)Z_2(p')}}\bar{u}_\beta(p')M_{\mu\nu}(p,p') u_\beta(p) u^\mu A_g^{\nu}(p'-p).
\end{equation}
To identify the gravitomagnetic moment $g_\Omega$ of the fermion, we then compare the previous amplitude to the one obtained
from the potential containing the spin-rotation coupling:
\begin{equation*}
V=-\vec{\mu}_g\cdot \vec{B}_g=-g_\Omega \vec{S}\cdot\vec{\Omega}.
\end{equation*}
This  potential, in the non-relativistic limit, leads to the amplitude
\begin{equation}
\label{eq:PlainAGMAmpl}
\mc{A}=-\I(2\pi)\delta(p_0-p_0')\left[-g_\Omega\,\xi'^\dagger\frac{\vec{\sigma}}{2}\xi\cdot\vec{\Omega}\right],
\end{equation}
where $\xi$ is the two component spinor, normalized such that $\xi^\dagger \xi=1$
and $\vec{\sigma}$ are the Pauli matrices.
By matching the explicit form of the amplitude in Eq.~(\ref{eq:QuantumAGMAmpl}) to the one in Eq.~(\ref{eq:PlainAGMAmpl})
coming from the spin-rotation coupling, we can read off the gravitomagnetic moment obtained in
finite temperature field theory.

%------------------------------------------------------------------------------------
\subsection{Gravitomagnetic moment at leading order}
%------------------------------------------------------------------------------------
\label{sec:AGMLeading}
We now obtain the leading order gravitomagnetic moment by computing
the amplitude~(\ref{eq:QuantumAGMAmpl}).
First, we go to the rest frame $u=(1,\vec{0})$ where the scattering amplitude becomes
\begin{equation*}
\mc{A} =-\I(2\pi)\delta(p_0-p_0')\frac{2}{\sqrt{Z_2(p)Z_2(p')}}\bar{u}(p')M_{0i}(p,p') u(p) A_g^{i}(p'-p).
\end{equation*}
At leading order, $u(p)$ is the usual Dirac spinor and the matrix elements of stress
energy tensor are
\begin{equation*}
M^0_{0i}(p,p')=\frac{1}{4}\left[\gamma_0(p'+p)_i+\gamma_i(p'+p)_0\right],\quad Z_2(p)=Z_2(p')=1.
\end{equation*}
In the limit $\vec{q}\to 0$ and $P_s\to 0$ we take advantage of the spinor identities
\begin{equation*}
\bar{u}(p')\gamma_0 u(p)= \xi'^\dagger \xi,\quad 
\bar{u}(p')\gamma^i u(p)= \xi'^\dagger\left[ \frac{(p'+p)^i}{2m}-\frac{\I\epsilon^{ijk}(p'-p)^j\sigma^k}{2m} \right] \xi
\end{equation*}
and find
\begin{equation*}
\begin{split}
\lim_{\vec{q}\to 0}\mc{A} =&-\I(2\pi)\delta(p_0-p_0')(-1)\frac{1}{2}\xi'^\dagger \left[(p'+p)^i+\frac{(p'+p)_0}{2m}(p'+p)^i-\frac{(p'+p)_0}{2m}\I\epsilon^{ijk}(p'-p)^j\sigma^k  \right]\xi A_g^{i}(\vec{q}).
\end{split}
\end{equation*}
This expression can be simplified using the following approximation valid in the non relativistic limit
\begin{equation*}
\delta(p_0-p_0')(p_0+p'_0)=\delta(p_0-p_0')2p_0=\delta(p_0-p_0')2\sqrt{m^2+(\vec{P}^2+\vec{q}^2)/4}=\delta(p_0-p_0')2m+\mc{O}\left(\vec{q}^2,\vec{P}^2/m^2\right).
\end{equation*}
The amplitude is then
\begin{equation*}
\begin{split}
\lim_{\vec{q}\to 0}\mc{A} =&-\I(2\pi)\delta(p_0-p_0')(-1)\xi'^\dagger \left[(p'+p)^i A_g^{i}(\vec{q})-\frac{1}{2}\I\epsilon^{ijk}q^j\sigma^k A_g^{i}(\vec{q})  \right]\xi \\
=&-\I(2\pi)\delta(p_0-p_0')(-1)\xi'^\dagger \left[(p'+p)^i A_g^{i}(\vec{q})+\frac{1}{2}\vec{\sigma}\cdot\vec{B}_g(\vec{q})   \right]\xi\\
=&-\I(2\pi)\delta(p_0-p_0') \left[-2(p'+p)^i A_g^{i}(\vec{q})\xi'^\dagger\xi- \xi'^\dagger\frac{\vec{\sigma}}{2}\xi\cdot \vec{\Omega}(\vec{q}) \right].
\end{split}
\end{equation*}
Comparing with the amplitude in Eq.~(\ref{eq:PlainAGMAmpl}), the gravitomagnetic moment of a fermion is
\begin{equation*}
g_\Omega=1,
\end{equation*}
which, as expected, is the value predicted by the Dirac equation.

%------------------------------------------------------------------------------------
\subsection{Anomalous gravitomagnetic moment at one loop}
%------------------------------------------------------------------------------------
\label{sec:1loopAGM}
Here we consider the radiative corrections to gravitomagnetic moment.
First, we deal with the zero temperature part.
At zero temperature, the one-loop renormalized stress-energy tensor is~\cite{Milton:1976jr}
\begin{equation*}
M_{\mu\nu}=\bar{u}(p')\left[\left(P_\mu \gamma_\nu+P_\nu \gamma_\mu\right)I_{P\gamma}^0(Q) -P_\mu P_\nu I_{PP}^0(Q)
-\left(q_\mu q_\nu-\eta_{\mu\nu}Q^2\right)I_{qq}^0(Q) \right]u(p)
\end{equation*}
where in the limit of $Q\equiv\sqrt{-q^2}\to 0$ the form factors are
\begin{equation*}
I_{P\gamma}^0(Q)=\frac{1}{4}\left(1-\frac{\alpha\pi}{8}\frac{Q}{m}\right);\quad
I_{PP}^0(Q)=\frac{\alpha\pi}{64m}\frac{Q}{M};\quad
I_{qq}^0(Q)=-\frac{\alpha\pi}{16m}\frac{m}{Q}.
\end{equation*}
Since the spin coupling can only come from a gamma matrix, only the $I_{P\gamma}^0(Q)$ formfactor can contribute
to gravitomagnetic moment. However in the limit of vanishing $\vec{q}$ this term does not affect the gravitomagnetic moment;
therefore there is no AGM for QED at zero temperature. This is what we expect from the Einstein equivalence principle,
which is still valid in interacting quantum field theory at zero temperature.

We now move to temperature modifications. The finite temperature renormalization we adopt is described in App.~\ref{sec:SETRen}.
First, we address the corrections which might come from thermal spinors, see App.~\ref{sec:DiracSpinors}.
We should repeat the previous calculation of the leading order gravitomagnetic moment, except that we 
must now employ thermal on-shell condition and thermal Dirac spinors:
\begin{equation*}
\bar{u}_\beta(p') M^{(0)}_{\mu\nu}u_\beta(p)=\bar{u}_\beta(p')\frac{1}{4}\left(P_\mu \gamma_\nu+P_\nu \gamma_\mu \right)u_\beta(p).
\end{equation*}
In the rest frame where $u=(1,\vec{0})$, the thermal spinors describe the spin interaction, as can be seen by using the identity
\begin{equation*}
\bar{u}_\beta(p')\vec{\gamma}u_\beta(p)=\xi'^\dagger\left[\frac{\vec{P}_T}{2m_p}+\I\frac{\vec{\sigma}\wedge\vec{q}_T}{2m_p}\right]\xi+\mc{O}(e^4),
\end{equation*}
where $P_T,q_T$ and $m_p$ contain temperature modifications according to the notation used in Sec.~\ref{sec:DiracSpinors}.
Here the spin-rotation coupling is divided by the thermal mass, but  this thermal mass is canceled out by the term
$p'_{0T}+p_{0T}\simeq 2m_p$; the gravitomagnetic moment is therefore unaffected.

Now we include the radiative corrections coming from the stress-energy tensor matrix elements
evaluated from the diagrams considered in Sec.~\ref{sec:SETRen}.
First, we select the terms of the stress-energy tensor matrix element $M_{\mu\nu}(p',p)$
that actually contributes to the anomalous gravitomagnetic moment (AGM).
In App.~\ref{sec:Selection} we show that only the following terms can bring contribution to AGM:
\begin{equation*}
\begin{split}
M^{\mu\nu}\subs{Relevant}(p,p')=&I_{P\gamma}(P,q)\left(\gamma^\mu P^\nu+\gamma^\nu P^\mu\right)
	+I_{u\gamma}(P,q)\left(\gamma^\mu u^\nu+\gamma^\nu u^\mu\right)+\\
	&+I_{Pl}(P,q)\slashed{\hat{l}}\left(\hat{l}^\mu P^\nu+\hat{l}^\nu P^\mu\right)
	+I_{ul}(P,q)\slashed{\hat{l}}\left(\hat{l}^\mu u^\nu+\hat{l}^\nu u^\mu\right),
\end{split}
\end{equation*}
which are the same one that contribute to the axial vortical effect.

By comparison with the contribution to AGM from $P_\mu\gamma_\nu+P_\nu\gamma_\mu$ evaluated in Sec.~\ref{sec:AGMLeading},
the formfactor 
\begin{equation*}
I_{P\gamma}(P_s,q^2)\left( P_\mu\gamma_\nu+P_\nu\gamma_\mu\right)
\end{equation*}
leads to the gravitomagnetic moment
\begin{equation*}
g_\Omega=+\lim_{P_s\to 0}\lim_{q\to 0}4I_{P\gamma}(P_s,q^2).
\end{equation*}
Similarly, it is straightforward to show that the other formfactors contribute to AGM as
\begin{equation*}
\begin{split}
I_{u\gamma}(P_s,q^2)&\left(u_\mu\gamma_\nu+u_\nu\gamma_\mu\right) \to g_\Omega=+\lim_{P_s\to 0}\lim_{q\to 0}4\frac{I_{u\gamma}(P_s,q^2)}{\omega_P};\\
I_{ul}(P_s,q^2)&\slashed{\hat{l}}\left(u_\mu \hat{l}_\nu+u_\nu \hat{l}_\mu \right)\to g_\Omega=-\lim_{P_s\to 0}\lim_{q\to 0}4\frac{I_{ul}(P_s,q^2)}{\omega_P};\\
I_{Pl}(P_s,q^2)&\slashed{\hat{l}}\left(P_\mu \hat{l}_\nu+P_\nu \hat{l}_\mu \right)\to g_\Omega=-\lim_{P_s\to 0}\lim_{q\to 0}4I_{Pl}(P_s,q^2).
\end{split}
\end{equation*}
Summing all contributions, the anomalous gravitomagnetic moment is
\begin{equation*}
g_\Omega=\lim_{\substack{q\to 0 \\ P_s\to 0}}4\left(I_{P\gamma}(P,q) + \frac{I_{u\gamma}(P,q)}{\omega_P} -I_{Pl}(P,q) -\frac{I_{ul}(P,q)}{\omega_P} \right).
\end{equation*}
Those form factors are computed in App.~\ref{sec:SETRen} and lead to the result
quoted in Sec.~\ref{sec:AGM}:
\begin{equation*}
g_\Omega-1=-\frac{6-\theta\subs{HT}}{36}\frac{e^2 T^2}{m^2}
=\begin{cases}
	-\frac{1}{6}\frac{e^2 T^2}{m^2} & T \ll m \\
	-\frac{5}{36}\frac{e^2 T^2}{m^2} & T \gg m
\end{cases}.
\end{equation*}
%

%************************************************************************************
\section{Summary and Discussion}
%************************************************************************************
\label{sec:Discussion}
In summary, we showed that in a system at thermal equilibrium the interactions with
photons change the gravitomagnetic moment of a massive fermion, i.e. the coupling
between the spin and the rotation of the medium. Using the scattering theory, in analogy to 
the magnetic moment, we obtained the gravitomagnetic moment from the form factors
of the stress-energy tensor, see Eq.~\eqref{eq:gOmegaForm}. We then renormalized the
stress-energy tensor at one-loop level in the finite temperature QED. The resulting
gravitomagnetic moment, given by ~\eqref{eq:ResultAGM}, receives
radiative corrections only in the presence of thermal effects.  We argued that this is
possible because the thermal bath destroys the Lorentz invariance of stress-energy tensor
and consequently violates the Einstein equivalence principle. To the best of our knowledge, the
possibility of an anomalous gravitomagnetic moment (AGM) in these settings and the
calculation of it are new results.

The effect of spin-rotation coupling has been already observed in the non-vanishing global
polarization of particles emitted by the rotating quark-gluon plasma~\cite{STAR:2017ckg}
and has been investigated in several studies~\cite{Gao:2007bc,Sorin:2016smp,Karpenko:2016jyx,Baznat:2017jfj,Li:2017slc,Xie:2017upb,Kolomeitsev:2018svb,Ivanov:2020qqe,Ivanov:2020wak,Ayala:2020ndx,Karpenko:2021wdm},
see~\cite{Becattini:2020ngo} for a review.
Therefore, in principle, polarization measurements in heavy
ion collisions could reveal the presence of an anomalous gravitomagnetic moment and the
breaking of the Einstein equivalence principle. To give an order of magnitude, we first
extend the result~\eqref{eq:ResultAGM} to QCD. By comparison with the massless QCD
radiative corrections of AVE~\cite{Golkar:2012kb,Hou:2012xg}, we expect that it is
sufficient to replace the QED coupling constant $e^2$ with $(N_c^2-1) g^2/2$:
\begin{equation}
\label{eq:AGMQCD}
g_\Omega\sups{QCD}-1=-\frac{N_c^2-1}{2}\frac{6-\theta\subs{HT}}{36}\frac{g^2 T^2}{m^2}
=\begin{cases}
	-\frac{N_c^2-1}{2}\frac{1}{6}\frac{g^2 T^2}{m^2} & T \ll m \\
	-\frac{N_c^2-1}{2}\frac{5}{36}\frac{g^2 T^2}{m^2} & T \gg m
\end{cases}.
\end{equation}
In a simple recombination picture based on the quark model, the $\Lambda$ polarization is carried predominantly by the
strange quark $s$; therefore the relative importance of the AGM for $\Lambda$ polarization can be inferred from the magnitude of radiative corrections to the gravitomagnetic moment of the $s$ quark. In the quark gluon plasma
phase, due to the high temperature $T=175-300$ MeV and the strong coupling regime, we estimate 
that the constituent mass of the strange quark $m_s\simeq 400$ MeV is larger than $g T \simeq 270- 450$ MeV. Using (\ref{eq:AGMQCD}), we find that the relative contribution of the AGM is quite large, about $40 \%$. Because it depends on the temperature, the AGM contribution may be detected in the data on $\Lambda$ polarization. An anomalous gravimagnetic response in the chirally broken phase of finite density QCD has been discussed in \cite{Aristova:2016wxe} where it was found to contribute significantly to $\Lambda$ polarization. 
Note that the effect of the AGM on polarization has the same sign for fermions and antifermions, so we expect it to contribute equally to the polarization of both $\Lambda$ and ${\bar \Lambda}$ hyperons.

We also established the connection
between the AGM and the axial vortical effect (AVE). In the limit of the large $m/T$ ratio,
we can separate the short distance interactions that renormalize the
angular momentum of the system, and the long distance thermal contributions which result in the
AVE. In this way we obtained the formula in Eq.~\eqref{eq:WARadCorr} which relates the
radiative correction to the AVE for a massive fermion to the momentum-dependent AGM.

While we showed here that the thermal interpretation of the AVE together with
spin-rotation coupling is able to describe the radiative corrections to the AVE, this does
not exclude a connection to the mixed gauge-gravitational anomaly~\cite{Landsteiner:2011cp}.
The presence of radiative corrections itself does not conflict with an interpretation based on the anomaly 
because the AVE is obtained from the matrix elements of the axial current and the
non-renormalization theorems apply to the operators, and not to the matrix elements, as has been established for the case of chiral anomaly in massless QED~\cite{Anselm:1989gi}. However the anomalous
origin of the effect is not yet completely clear; this is particularly true for
massive fermions~\cite{Buzzegoli:2020ycf}.

Radiative corrections to the AVE were presented previously in~\cite{Golkar:2012kb,Hou:2012xg}
for massless fermions and can be linked to the gravitational anomaly of
photons~\cite{Prokhorov:2020npf,Dolgov:1987yp}. Unfortunately we cannot compare these corrections with
those presented above, since our derivation requires massive fermions, and a 
massless limit cannot be performed. Even our definition of the gravitomagnetic moment can not
be applied to a massless particle, as it requires going to the rest frame of the
particle. Therefore, the connection between the AVE and the AGM for a massless Dirac field still has to be clarified. Nevertheless, we believe that the link between the anomalous gravitomagnetic moment and the axial vortical effect for massive fermions established above will help to understand the origin of chiral currents induced by rotation.

\begin{acknowledgments}
We thank Karl Landsteiner for useful comments on the manuscript and stimulating discussions.
M.B. would like to thank the Center for Nuclear Theory at Stony Brook University 
for hospitality and support during his one year visit. The work of M.B. is supported
by Unifi fellowship {\em Polarizzazione nei fluidi relativistici} and
{\em Effetti quantistici nei fluidi relativistici}. The work of D.K.  was supported by the U.S. Department of
Energy under awards DE-FG88ER40388 and DE-SC0012704.

\end{acknowledgments}

\appendix

%************************************************************************************
\section{Equivalence principle at finite temperature}
%************************************************************************************
\label{sec:EquivPrinFiniteT}
It has already been demonstrated~\cite{Donoghue:1984zs,Donoghue:1984ga,Donoghue:1983qx,Mitra:2001ar,Nieves:1998xz}
that radiative corrections at finite temperature lead to a difference between the inertial and
gravitational masses, thus providing  an explicit breaking of weak equivalence principle.
Indeed, one can define three distinct kinds of mass for a particle. The phase-space mass is
the position of the pole in the propagator of the field. The inertial mass is the
response to acceleration caused by an external force, such as an external electric field.
And lastly, the gravitational mass is a measure of how the fermion responds to the gravitational force.
At finite temperature it has been found that the inertial mass and the phase-space masses coincide but the 
gravitational mass is different~\cite{Donoghue:1984zs}.

%------------------------------------------------------------------------------------
\subsection{Phase space mass}
%------------------------------------------------------------------------------------
\label{sec:PhasespaceMass}
Consider a massive Dirac fermion in a QED-like theory at finite temperature.
As in \cite{Donoghue:1984zs}, we refer to phase-space mass as the location of the
pole in the propagator of the fermion field. The full fermion propagator is given by:
\begin{equation*}
	S(p)=\frac{1}{\slashed{p}-m-\Sigma(p)}.
\end{equation*}
The self-energy can be written in covariant form as~\cite{Weldon:1982bn,Mitra:2001ar}
\begin{equation*}
	\Sigma(p)=a\slashed{p}+b\slashed{u}+c,
\end{equation*}
where $a,b,c$ are Lorentz invariant functions. These functions can depend on $m,\,T$ and on the
following Lorentz scalars:
\begin{equation*}
	\omega\equiv p^\alpha u_\alpha,\quad p_s\equiv\sqrt{(p^\alpha u_\alpha)^2-p^2};
\end{equation*}
since $p^2=\omega^2-p_s^2$, one may interpret $\omega$ and $p_s$ as Lorentz invariant energy
and three momentum. It is useful to define a tensor and a vector orthogonal to $u_\mu$ by
\begin{equation*}
	\tilde{\eta}_{\mu\nu}=\eta_{\mu\nu}-u_\mu u_\nu,\quad
	\tilde{p}_\mu=\tilde{\eta}_{\mu\nu}p^\nu=p_\mu-\omega u_\mu.
\end{equation*}
The vector $\tilde{p}$ is automatically space-like:
\begin{equation*}
	\tilde{p}^\mu \tilde{p}_\mu=-p_s^2<0.
\end{equation*}
Inverting the matrices, the full propagator becomes:
\begin{equation*}
	S(p)=\frac{(1-a)\slashed{p}-b\slashed{u}+m+c}{[(1-a)p-b u]^2-(m+c)^2}.
\end{equation*}
Therefore, the location of the pole is determined by the vanishing of the denominator:
\begin{equation*}
	(1-a)^2(\omega^2-p_s^2)+b^2-2(1-a)b\omega=(m+c)^2.
\end{equation*}
The positive solution for the pole is
\begin{equation*}
	\omega=\frac{b}{1-a}+\sqrt{p_s^2+\frac{(m+c)^2}{(1-a)^2}}.
\end{equation*}
Since we are interested in corrections of order $e^2$ in the coupling constant $e$, it suffices to linearize
in the $a,b,c$ functions, which are already $e^2$ order. The phase-space mass is then
\begin{equation}
	\label{eq:PhaseSpaceMass}
	m_p^2\equiv\omega^2-p_s^2=m^2+2\left[a(\omega^2-p_s^2)+b\omega+m c \right].
\end{equation}
At a given a momentum $p$ with scalars $\omega$ and $p_s$, the pole of the propagator is situated at
\begin{equation*}
	\omega^2=E_\beta^2\equiv p_s^2+m_p^2.
\end{equation*}
The functions $a,b,c$ are obtained with the traces over spinor indices
\begin{equation*}
	P_T=\tr\left[\slashed{p}\Sigma^\beta(p)\right],\quad U_T=\tr\left[\slashed{u}\Sigma^\beta(p)\right],\quad T=\tr\left[\Sigma^\beta(p)\right],
\end{equation*}
which give
\begin{equation}
	\label{eq:SelfEnergyScalars}
	a=\frac{(p\cdot u)U_T-P_T}{4[(p\cdot u)^2-p^2]},\quad b=\frac{(p\cdot u) P_T-p^2 U_T}{4[(p\cdot u)^2-p^2]},\quad c=\frac{T}{4}.
\end{equation}
Replacing the Eq.s~(\ref{eq:SelfEnergyScalars}) into Eq.~(\ref{eq:PhaseSpaceMass}) we find
\begin{equation*}
	m_p^2=m^2+\frac{1}{2}\left(P_T+m T\right).
\end{equation*}

In real-time formalism the self energy is given by the one-loop diagram
\begin{equation*}
	\Sigma(p)=\I e^2\int\frac{\D^4 k}{(2\pi)^4}D_{\mu\nu}(k)\gamma^\mu S_F(p+k) \gamma^\nu=\Sigma(T=0)+\Sigma^\beta,
\end{equation*}
where $S_F$ and $D_{\mu\nu}$ are the fermion and the photon propagator in momentum space
(see Sec.~\ref{sec:SETRen}) and we split the zero and the finite temperature part of self-energy.
With standard techniques we can evaluate the self-energy in covariant form and the explicit
form of the functions $a,b$ and $c$.
We find that the phase-space mass can be approximated by~\cite{Weldon:1982bn,Donoghue:1984zs,Mitra:2001ar}
\begin{equation*}
	m^2_p-m^2=\begin{cases}
		\frac{e^2 T^2}{6} & T\ll m,\\
		\frac{e^2 T^2}{8} & T\gg m
	\end{cases},
\end{equation*}
where the different behavior at high temperature arises because the fermion thermal distribution becomes
comparable to the contribution from the photon thermal distribution only when the temperature is much larger
than the mass of the particle.

%------------------------------------------------------------------------------------
\subsection{Inertial mass}
%------------------------------------------------------------------------------------
We refer, as usual, to inertial mass of the particle $m_I$ as the proportionality term between a force and the
acceleration caused by it. To test the inertial mass of a charged Dirac particle, the most natural force
to consider is a constant electric field $\vec{E}$. In this way, it is easy to include it in the Dirac equation as a
minimal coupling with an external gauge field $A^\mu=(\phi,\vec{0})$, where $\vec{E}=-\nabla \phi$.
The corrections to this coupling given by temperature and interactions are the corrections to the
vertex $\Gamma^\mu$. It is found that, when the vertex is contracted between thermal spinors $u_\beta(p)$
(see Sec.~\ref{sec:DiracSpinors}), the modifications exactly compensate each other~\cite{Yee:1984wt}
\begin{equation*}
	e\bar{u}_\beta(p)\Gamma^\mu u_\beta(p)=e \frac{p_\mu}{E_\beta},
\end{equation*}
or in other words, the charge is not renormalized by finite temperature effects.
This suggests that the inertial mass is to be identified with the phase-space mass.

To properly evaluate the inertial mass, we just need to consider the modified Dirac equation
\begin{equation*}
	\left(\slashed{p}_T-m\right)\psi=e\Gamma^\mu A_\mu \psi.
\end{equation*}
In the non-relativistic limit, one can transform the previous equation into a Schrodinger equation
via a Fol\-dy–Wou\-thuy\-sen transformation. In that form, one can easily identify the Hamiltonian $H$
of the system, then the acceleration of the particle is identified via $\vec{a}=-[H,[H,\vec{r}]]$
and hence one can infer the value of inertial mass. It is indeed found~\cite{Donoghue:1984zs} that
inertial mass $m_I$ and phase-space mass $m_p$ coincide.

%------------------------------------------------------------------------------------
\subsection{Gravitational mass}
%------------------------------------------------------------------------------------
\label{sec:gravmass}
Using scattering theory in linearized gravity, we can identify the gravitational mass
of a fermion. We indicate the matrix element of the stress-energy tensor with
\begin{equation*}
	\bra{p',s'}T^{\mu\nu}(0)\ket{p,s}\equiv\bar{u}(p')M_{\mu\nu}(p,p') u(p),
\end{equation*}
where $p$ and $p'$ are the external momenta and $s,\,s'$ the spin of the fermion.
Consider an external gravitational field $g\subs{Ext}^{\mu\nu}=\eta^{\mu\nu}+h\subs{Ext}^{\mu\nu}$;
the interaction Hamiltonian in linearized gravity is therefore given by
$\h{H}\subs{int}=\int\D^3 x \frac{1}{2}\h{T}_{\mu\nu}h\subs{Ext}^{\mu\nu}$.
In the leading order of perturbation theory, the S-matrix element for scattering is:
\begin{equation*}
	\I \mc{A} (2\pi)\delta(p\cdot u-p'\cdot u)=-\I\frac{1}{2} \bar{u}(p')M^0_{\mu\nu}(p,p') u(p)h\subs{Ext}^{\mu\nu}(p'-p),
\end{equation*}
where $h\subs{Ext}^{\mu\nu}(p'-p)$ is the Fourier transform of
$h\subs{Ext}^{\mu\nu}(x)$ and $M^0_{\mu\nu}$ is the tree-level vertex function,
which is given by
\begin{equation*}
	M^0_{\mu\nu}(p,p')=\frac{1}{4}\left[\gamma_\mu(p'+p)_\nu+\gamma_\nu(p'+p)_\mu\right]
	-\frac{1}{2}\eta_{\mu\nu}\left[(\slashed{p}-m)+(\slashed{p}'-m)\right].
\end{equation*}
The radiative corrections modify this expression to
\begin{equation*}
	\mc{A} =-\I(2\pi)\delta(p\cdot u-p'\cdot u)\frac{1}{\sqrt{Z_2(p)Z_2(p')}}\frac{1}{2}\bar{u}_\beta(p')M_{\mu\nu}(p,p') u_\beta(p) h\subs{Ext}^{\mu\nu}(p'-p),
\end{equation*}
where we divided by $\sqrt{Z_2}$ - the wave-function renormalization constant - for each  fermionic leg,
and $u_\beta(p,s)$ is the Dirac thermal spinor which satisfies the Dirac equation including radiative and
thermal corrections and all perturbative diagrams are summed in $M$.

To identify the gravitational mass of the fermion $m_g$, following~\cite{Mitra:2001ar}, we consider the scattering
of a fermion from a static gravitational potential produced by a static mass density $\rho\subs{Ext}(\vec{x})$.
The resulting metric is the linearized solution of Einstein field equations with a matter
stress-energy tensor given by $T^{\mu\nu}=\rho\subs{Ext}u^\mu u^\mu$. Taking advantage
of the Poisson equation $-2\vec{q}^2\phi\subs{Ext}=\rho\subs{Ext}$ the Fourier
transform of Einstein equation solution reads:
\begin{equation*}
	h^{\mu\nu}(\vec{q})=2\phi\subs{Ext}(\vec{q})(2u^\mu u^\nu-\eta^{\mu\nu}).
\end{equation*}
Therefore, inserting this in the scattering amplitude, we find
\begin{equation*}
	\begin{split}
		\mc{A} =&-\I(2\pi)\delta(p_0-p_0')\frac{1}{\sqrt{Z_2(p)Z_2(p')}}\bar{u}_\beta(p')M_{\mu\nu}(p,p') u_\beta(p)(2u^\mu u^\nu -\eta^{\mu\nu})\delta(p_0'-p_0)\phi\subs{Ext}(\vec{q})\\
		=&-\I(2\pi)\delta(p_0'-p_0)\mc{M}(p',p)\phi\subs{Ext}(\vec{q}),
	\end{split}
\end{equation*}
where we have defined
\begin{equation*}
	\mc{M}(p',p) = \frac{1}{\sqrt{Z_2(p)Z_2(p')}}\bar{u}_\beta(p') M_{\mu\nu}(p,p') u_\beta(p)(2u^\mu u^\nu -\eta^{\mu\nu}).
\end{equation*}
If the gravitational field is very slowly varying over a large (macroscopic) region,
$\phi\subs{Ext}(\vec{q})$ will be concentrated around $\vec{q}=0$; then we can take the
limit $\vec{q}\to 0$. In this way, by comparison with the scattering amplitude of a potential
$V(\vec{x})=m_g\phi(\vec{x})$ in the Born approximation, which is
\begin{equation*}
	\mc{A}=-\I m_g \phi(\vec{q}),
\end{equation*}
we can identify the gravitational mass of the fermion. From the previous expression, we see that
the gravitational mass is obtained when the spatial momenta of the fermion are vanishing:
\begin{equation*}
	m_g=\lim_{p_s\to 0}\left[\frac{(2u^\lambda u^\rho-\eta^{\lambda\rho})}{\sqrt{Z_2(p)Z_2(p')}}
	\left(\bar{u}_\beta(p',s)M_{\lambda\rho}(p,p')u_\beta(p,s)\right)\subs{On-shell}\right]\Big|_{p'\to p},
\end{equation*}
where $E_\beta=E_\beta(\vec{p})$ is the on-shell energy of the particle, i.e. the position of the pole
of the self-energy, and $p_s=\sqrt{(p\cdot u)^2-p^2}$.

At leading order, as we are now showing, gravitational mass coincides with inertial and phase space mass.
At leading order $Z_2=1+\mc{O}(e^2)$, the thermal Dirac spinor reduces to the usual free Dirac spinor
and the matrix element is simply the tree-level diagram
\begin{equation*}
	M^0_{\lambda\rho}(p,p')=\frac{1}{4}\left[\gamma_\lambda(p+p')_\rho+\gamma_\rho(p+p')_\lambda\right]
	-\frac{1}{2}\eta_{\lambda\rho}\left(\slashed{p}-m+\slashed{p}'-m\right).
\end{equation*}
Proceeding to evaluate the gravitational mass step by step, we first find
\begin{equation*}
	\left(\bar{u}(p',s)M^0_{\lambda\rho}(p,p')u(p,s)\right)_{p'=p}=
	\bar{u}(p,s)\left\{ \frac{1}{2}\left[\gamma_\lambda p_\rho+\gamma_\rho p_\lambda\right]-\eta_{\lambda\rho}\left(\slashed{p}-m\right)\right\}u(p,s).
\end{equation*}
Then, since $p$ is taken on-shell, we take advantage of spinor properties (see Sec.~\ref{sec:DiracSpinors} for
the conventions used), which in the limit of $p_s$ going to zero give
\begin{equation*}
	\bar{u}(p,s)u(p,s)|_{p_s=0}=1+\mc{O}(e^2),\quad \bar{u}(p,s)\gamma_\mu u(p,s)|_{p_s=0}=\frac{p_\mu}{m}+\mc{O}(e^2),
\end{equation*}
and we obtain
\begin{equation*}
	\begin{split}
		\lim_{p_s\to 0}\left(\bar{u}(p',s)M_{\lambda\rho}(p,p')u(p,s)\right)_{p'=p}=& \lim_{p_s\to 0}\frac{1}{m}\left\{
		\frac{1}{2}\left(p_\lambda p_\rho+p_\rho p_\lambda\right)-\eta_{\lambda\rho}\left(p^2-m^2\right)\right\}\\
		&= \lim_{p_s\to 0}\left\{\frac{1}{2m}\left(p_\lambda p_\rho+p_\rho p_\lambda\right)\right\}.
	\end{split}
\end{equation*}
At last, we find
\begin{equation*}
	m_g=\lim_{p_s\to 0}\left[\frac{1}{2m}(2u^\lambda u^\rho-\eta^{\lambda\rho})\left(p_\lambda p_\rho+p_\rho p_\lambda\right)\right]
	=\lim_{p_s\to 0}\frac{2(u\cdot p)^2 -p^2}{m}=m,
\end{equation*}
where we used the on-shell condition $p\cdot u=E_\beta(p_s=0)=m+\mc{O}(e^2)$.
At leading order, gravitational and inertial mass are indeed equivalent $m_g=m_I=m_p=m$.
Instead, for 1-loop QED it has been proved~\cite{Donoghue:1984zs,Mitra:2001ar} that gravitational mass is different from inertial
mass, in particular for small temperature $T\ll m$ their ratio is
\begin{equation*}
	\frac{m_I}{m_g}=1+\frac{e^2}{3}\frac{T^2}{m^2}.
\end{equation*}
This is a manifest breaking of the weak equivalence principle caused by finite temperature effects.

%************************************************************************************
\section{Selection of the form factors}
%************************************************************************************
\label{sec:Selection}
In this appendix we show that only the following terms
\begin{equation*}
\begin{split}
M^{\mu\nu}\subs{Relevant}(p,p')=&I_{P\gamma}(P,q)\left(\gamma^\mu P^\nu+\gamma^\nu P^\mu\right)
	+I_{u\gamma}(P,q)\left(\gamma^\mu u^\nu+\gamma^\nu u^\mu\right)+\\
&+I_{Pl}(P,q)\slashed{\hat{l}}\left(\hat{l}^\mu P^\nu+\hat{l}^\nu P^\mu\right)
	+I_{ul}(P,q)\slashed{\hat{l}}\left(\hat{l}^\mu u^\nu+\hat{l}^\nu u^\mu\right),
\end{split}
\end{equation*}
with $P=p'+p$ and $q=p'-p$, can contribute to the axial vortical effect (AVE) or to
the gravitomagnetic moment. First, notice that $M$ reproduces the stress-energy tensor
matrix elements when evaluated between the two Dirac spinors $\bar{u}(p')$ and $u(p)$:
\begin{equation*}
\bar{u}_{s'}(p') M^{\mu\nu}(p,p')\, u_s(p) 
		=\bra{p',s'}\h{T}^{\mu\nu}(0)\ket{p,s}.
\end{equation*}
Therefore, taking advantage of the equation of motion and the gamma matrices algebra,
we can write each term of $M$ in terms of the tetrad $\{u,P,q,\hat{l}\}$ (defined in
Sec.~\ref{sec:AGM}), the metric $\eta$ and at maximum one $\gamma$ matrix.
Comparing the spin-rotation amplitude~\eqref{eq:PlainAGMAmpl} with the thermal spinor
identities
\begin{equation*}
	\bar{u}_\beta(p')\gamma_0 u_\beta(p)= \xi'^\dagger \xi+\mc{O}(e^2),\quad 
	\bar{u}_\beta(p')\vec{\gamma}u_\beta(p)=\xi'^\dagger\left[\frac{\vec{P}}{2m_p}+\I\frac{\vec{\sigma}\wedge\vec{q}}{2m_p}\right]\xi+\mc{O}(e^4),
\end{equation*}
we conclude that only the terms of $M$ which contain exactly one gamma matrix can bring
contribution to the gravitomagnetic moment. We come to the same conclusion for the AVE
by looking at the trace in Eq.~\eqref{eq:TraceTLTpp}. Among the terms of $M$
which contains one gamma matrix, the ones that contains $\slashed{P}$ and $\slashed{q}$ give
\begin{equation*}
	\bar{u}(p')\slashed{P}u(p)=\bar{u}(p')\,2m\,u(p),\quad \bar{u}(p')\slashed{q}u(p)=0,
\end{equation*}
and hence do not contribute to AVE or to gravitomagnetic moment.

Furthermore, from Eq.~\eqref{eq:TraceTLTpp} the AVE is evaluated with the components
$M^{02}$ in the rest frame of the thermal bath. Then, only the terms which have a
non-vanishing time-space component can contribute. Similarly, the gravitomagnetic
moment is evaluated with the contraction $M_{\mu\nu}(p',p)u^\mu A^\nu_g$, therefore a 
relevant term must not vanish when contracted with $u$ and $A_g$ (notice that $A_g\cdot u=0$).
Therefore between the terms
\begin{equation*}
P_\mu\gamma_\nu+P_\nu\gamma_\mu,\quad
u_\mu\gamma_\nu+u_\nu\gamma_\mu,\quad
q_\mu\gamma_\nu+q_\nu\gamma_\mu,\quad
\hat{l}_\mu\gamma_\nu+\hat{l}_\nu\gamma_\mu,
\end{equation*}
since $u\cdot q=u\cdot\hat{l}=0$ only the first two are relevant. The other terms left
that contain a gamma matrix and that satisfy the conditions stated above are the
following
\begin{equation}
\label{eq:Lterms}
\slashed{\hat{l}} P_\mu P_\nu,\quad
\slashed{\hat{l}}\left(u_\mu P_\nu+u_\nu P_\mu \right),\quad
\slashed{\hat{l}}\left(u_\mu q_\nu+u_\nu q_\mu \right),\quad
\slashed{\hat{l}}\left(u_\mu \hat{l}_\nu+u_\nu \hat{l}_\mu \right),\quad
\slashed{\hat{l}}\left(P_\mu q_\nu+P_\nu q_\mu \right),\quad
\slashed{\hat{l}}\left(P_\mu \hat{l}_\nu+P_\nu \hat{l}_\mu \right).
\end{equation}
For the AVE we see that to obtain the thermal coefficient we must evaluate the terms
in~\eqref{eq:Lterms} for $q=0$, then the terms proportional to $q_\mu$ or $q_\nu$ can
not bring contribution. Moreover, by plugging the term $\slashed{\hat{l}} P_\mu P_\nu$
in Eq.~\eqref{eq:AVE_Maa} and after evaluating the trace and the derivatives, we see that it is
odd under the transformation $\vec{k}\to -\vec{k}$ and therefore vanish after momentum
integration. For the gravitomagnetic moment we can write each terms in~\eqref{eq:Lterms} 
as $\slashed{\hat{l}}(w_\mu v_\nu+w_\nu v_\mu)$, where $w$ can be either $u$ or $P$ and
$v$ can be $P,q$ or $\hat{l}$. To evaluate the gravitomagnetic moment we should consider
it at thermal bath rest frame:
\begin{equation*}
\bar{u}(p')\slashed{\hat{l}} u(p)(w_\mu v_\nu+w_\nu v_\mu )u^\mu A_g^\nu
	=\bar{u}(p')(\vec{l}\cdot\vec{\gamma}) u(p)w_0 (\vec{v}\cdot \vec{A}_g).
\end{equation*}
Since the gravitomagnetic moment is the coupling of spin and rotation, i.e.
$\vec{\gamma}\cdot\vec{A}_g$, we can take advantage of three vector properties and
write the scalar products as
\begin{equation*}
(\vec{\gamma}\cdot\vec{l})(\vec{v}\cdot\vec{A}_g)
	=(\vec{\gamma}\cdot\vec{A}_g)(\vec{v}\cdot\vec{l})-(\vec{\gamma}\wedge\vec{v})(\vec{l}\wedge\vec{A}_g).
\end{equation*}
By definition of $l$, $(\vec{v}\cdot\vec{l})$ is non vanishing if and only if $v=\hat{l}$.
Then, only the terms with $\hat{l}_\mu$ can contribute to gravitomagnetic moment.
At the end, we found that only the following terms can bring contribution to the
AVE or to the gravitomagnetic moment:
\begin{equation*}
\begin{split}
	u_\mu\gamma_\nu+u_\nu\gamma_\mu,&\quad
	P_\mu\gamma_\nu+P_\nu\gamma_\mu,\quad
	\slashed{\hat{l}}\left(u_\mu \hat{l}_\nu+u_\nu \hat{l}_\mu \right),\quad
	\slashed{\hat{l}}\left(P_\mu \hat{l}_\nu+P_\nu \hat{l}_\mu \right).
\end{split}
\end{equation*}
%

%************************************************************************************
\section{Renormalization of stress-energy tensor}
%************************************************************************************
\label{sec:SETRen}
\begin{figure}[thb!]
	\centering
	\subfigure[Tree Level]{\label{fig:M_Tree}\includegraphics[width=0.3\columnwidth]{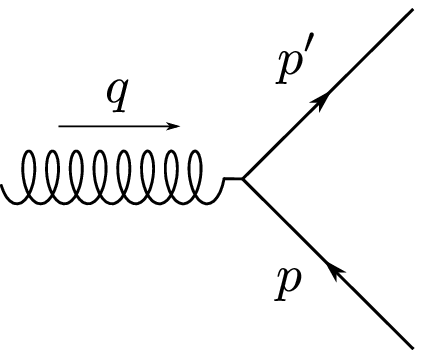}}
	\subfigure[Self Energy]{\label{fig:M_SelfEnergy}\includegraphics[width=0.3\columnwidth]{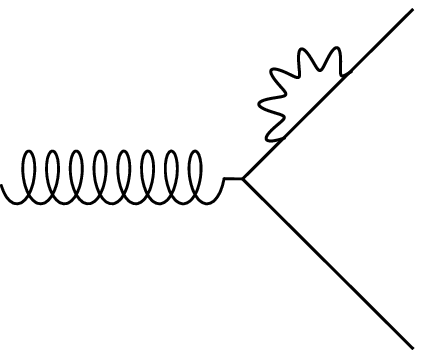}}
	\subfigure[Counter Term]{\label{fig:M_Counter}\includegraphics[width=0.3\columnwidth]{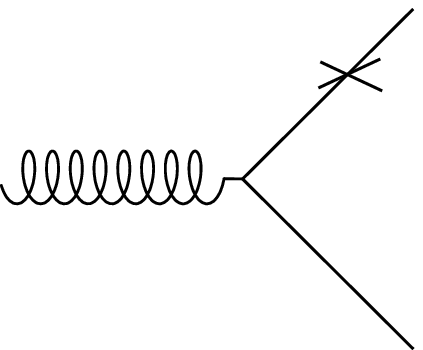}}\\
	\subfigure[Electromagnetic Vertex]{\label{fig:M_EMVertex}\includegraphics[width=0.3\columnwidth]{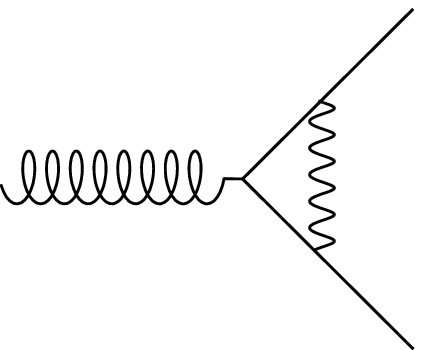}}
	\subfigure[Contact]{\label{fig:M_Contact}\includegraphics[width=0.3\columnwidth]{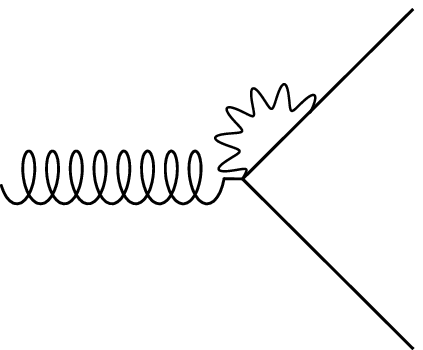}}
	\subfigure[Photon Polarization]{\label{fig:M_Polarization}\includegraphics[width=0.3\columnwidth]{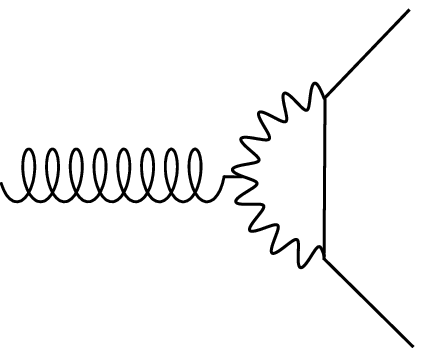}}
	\caption{The diagrams contributing to the radiative corrections to the stress-energy tensor.
	The curly line is the stress-energy tensor,	the wavy line is the photon and the solid line is the fermion.}
	\label{fig:SETRen}
\end{figure}
In order to evaluate the radiative corrections to gravitomagnetic moment at finite temperature
we have to consider the renormalization of the stress-energy tensor at finite temperature and
at first order on momentum transfer $q$. The zero temperature renormalization is performed with
usual techniques and we do not discuss it here. To address thermal corrections to the
matrix-element of stress-energy tensor we use the real-time formalism of thermal field theory.
The major modification of QFT at finite temperature is the value of the vacuum of the theory,
which is not empty but contain a number of bosons and fermions given
respectively by the Bose-Einstein and the Fermi-Dirac distribution functions:
\begin{equation*}
	\h{a}\subs{B}^\dagger\h{a}\subs{B}(p)|0\rangle =n\subs{B}(E)|0\rangle,\quad
	\h{a}\subs{F}^\dagger\h{a}\subs{F}(p)|0\rangle =n\subs{F}(E)|0\rangle,
\end{equation*}
where as usual we indicate
\begin{equation*}
	n\subs{B}(x)=\frac{1}{\exp(\beta x)-1},\quad n\subs{F}(x)=\frac{1}{\exp(\beta x)+1}.
\end{equation*}
The resulting propagators of the gauge field and of the fermionic field in real-time formalism are
\begin{equation*}
	\begin{split}
		\I S_F(p)&=\left(\slashed{p}+m\right)\left[\frac{\I}{p^2-m^2}-2\pi\delta\left(p^2-m^2\right)n_F(p)\right],\\
		\I D_{\mu\nu}(k)&=-\eta_{\mu\nu}\left[\frac{\I}{k^2}+2\pi\delta\left(k^2\right)n_B(k)\right].
	\end{split}
\end{equation*}
Perturbation theory and Feynman diagrams at finite temperature are unmodified compared to usual quantum field
theory except for the previous propagators. We see that the propagators in real-time formalism are naturally
separated into a temperature part and into a $T=0$ part. The $T=0$ part has been addressed as usual
and from now on we just consider the thermal part. The diagrams responsible for radiative
corrections to the stress-energy tensor matrix element
\begin{equation*}
	\bra{p',s'}T^{\mu\nu}(0)\ket{p,s}=\frac{1}{\sqrt{Z_2(p)Z_2(p')}}\bar{u}_\beta(p') M_{\mu\nu}(p,p') u_\beta(p),
\end{equation*}
are reported in Fig.~\ref{fig:SETRen}. Furthermore, since we are only interested in the thermal part,
each integrals corresponding to a Feynman diagram is weighted with a thermal function and it is therefore
ultraviolet convergent and finite. All the divergent part are already been taken care of with the $T=0$
renormalization.

Note that we can distinguish between two different regimes of temperature. Indeed, finite-temperature
modification may arise both from the boson or the photon propagators. However, for low temperatures
such that $T\ll m$, the fermion distribution function is suppressed by a factor $\exp(-m/T)$, which
is negligible compared to contributions of order $(T/m)^2$ coming from the photon distribution.
In the opposite regime, when $T\gg m$, the photon contribution is still the same as low temperatures,
while the fermion distribution can now also contribute with terms of order $(T/m)^2$. Therefore, as
in the case of the pole of the propagator, we can expect two different values of AGM valid in the
two regimes of low and high temperatures.

As first step, we need to identify the mass shift and the wave-function renormalization constant.
These quantities are obtained starting from the self-energy of the fermion, which is discussed
in Sec.~\ref{sec:PhasespaceMass}. We found that the self-energy can be written as $\Sigma(p)=a\slashed{p}+b\slashed{u}+c$
and that the full fermion propagator becomes
\begin{equation}
	\label{eq:newprop}
	S(p)=\frac{(1-a)\slashed{p}-b\slashed{u}+m+c}{[(1-a)p-b u]^2-(m+c)^2},
\end{equation}
which has a pole in $\omega^2=E_\beta^2=p_s^2+m_p^2$, with $m_p$ given by Eq.~(\ref{eq:PhaseSpaceMass}).

%------------------------------------------------------------------------------------
\subsection{Thermal Dirac spinor}
%------------------------------------------------------------------------------------
\label{sec:DiracSpinors}
The Dirac equation in momentum space is modified according to the self-energy, which also include thermal modifications.
The thermal Dirac spinors $u_\beta(p)$ satisfy the modified Dirac equation corresponding to the
new propagator~(\ref{eq:newprop}):
\begin{equation*}
	\left[\slashed{p}-m_R-\Sigma^\beta(p)\right]u_\beta(p)=\left[(1-a)\slashed{p}-b\slashed{u}-m_R-c\right]u_\beta(p)=0,
\end{equation*}
where $m_R$ indicates the zero temperature renormalized mass. The thermal spinor $u_\beta(p)$ satisfies
the previous equation when $p$ is the pole of the propagator, i.e. such that $p\cdot u=E_\beta(p_s)$.
The thermal Dirac spinors are actually required to properly account for stress-energy tensor
renormalization at finite temperature~\cite{Donoghue:1984zs,Weldon:1982bn}.

In thermal bath rest frame, we choose the normalization
\begin{equation*}
	u^\dagger_\beta(p)u_\beta(p)=1.
\end{equation*}
For convenience, we furthermore define a four vector and a scalar
\begin{equation*}
	p_T\equiv (1-a)p-b u,\quad m_T\equiv m_R+c,
\end{equation*}
so that the modified thermal Dirac equation is written as
\begin{equation*}
	\left[\slashed{p}_T-m_T\right]u_\beta(p)=0.
\end{equation*}
Therefore, it follows from the modified thermal Dirac equation that the thermal spinors satisfy the following identities:
\begin{equation*}
%	\label{eq:thermalspinor}
	\bar{u}_\beta(p,s)u_\beta(p,s)=\frac{m_T}{(1-a)E_\beta-b},\quad
	\bar{u}_\beta(p,s)\gamma_\mu u_\beta(p,s)=\frac{{p_T}_\mu}{(1-a)E_\beta-b},\quad
	\sum_s u_\beta(p,s)\bar{u}_\beta(p,s)=\frac{\slashed{p}_T+m_T}{2(1-a)E_\beta-2b}.
\end{equation*}
In the non relativistic limit and in the frame where $u=(1,\vec{0})$, we can also show the
validity of the following identities:
\begin{equation*}
	\bar{u}_\beta(p')\gamma_0 u_\beta(p)= \xi'^\dagger \xi+\mc{O}(e^2),\quad 
	\bar{u}_\beta(p')\vec{\gamma}u_\beta(p)=\xi'^\dagger\left[\frac{\vec{P}}{2m_p}+\I\frac{\vec{\sigma}\wedge\vec{q}}{2m_p}\right]\xi+\mc{O}(e^4),
\end{equation*}
where $\xi$ is the two component spinor, normalized such that $\xi^\dagger \xi=1$.
We used the identities above to compute the gravitational mass in Sec.~\ref{sec:gravmass}
and the gravitomagnetic moment in Sec.~\ref{sec:AGMLingrav}.

%------------------------------------------------------------------------------------
\subsection{Wave-function renormalization constant \texorpdfstring{$Z_2$}{Z2}}
%------------------------------------------------------------------------------------
The wave-function renormalization constant $Z_2$ is obtained by requesting that
the fermion field is properly renormalized
\begin{equation*}
	\psi\subs{R}(x)=\frac{1}{\sqrt{Z_2}}\psi_0(x).
\end{equation*}
The temperature part of the wave-function renormalization constant is~\cite{Donoghue:1983qx,Mitra:2001ar}
\begin{equation*}
	^\beta Z_2(\omega,p_s)=\frac{2\left[(1-a)E_\beta-b\right]}{\frac{\de D}{\de\omega}\Big|_{\omega=E_\beta}},
\end{equation*}
where $E_\beta$ is the pole of the propagator and $D$ is the denominator of the propagator~(\ref{eq:newprop})
\begin{equation*}
	D(\omega,p_s)=(1-a)^2(\omega^2-p_s^2)+b^2-2(1-a)b\omega-(m+c)^2.
\end{equation*}
At order $e^2$ we find
\begin{equation}
	\label{eq:Z2Formula}
	^\beta Z_2=1+\left( a +\frac{m^2}{\omega}\frac{\de a}{\de\omega} +\frac{\de b}{\de\omega} +\frac{m}{\omega} \frac{\de c}{\de\omega}\right)\Big|_{\omega=\sqrt{p_s^2+m^2}}.
\end{equation}
For the computation of AGM we are interested to the quantity $(\sqrt{Z_2(p)Z_2(p')})^{-1/2}$ in the limits of $q\to 0$
and of $P_s=\sqrt{(P\cdot u)^2-P^2}\to 0$. Therefore, we just have to evaluate $Z_2(p)$ with $p$ an on-shell momentum and then perform the $p_s\to 0$ limit.
Using the prescription in Eq.~(\ref{eq:Z2Formula}) and the behavior of the functions $a,b,c$ we find:
\begin{equation*}
	\lim_{p_s\to 0}\, ^\beta Z_2(p)=1-\frac{e^2 T^2}{12 m^2}+\frac{e^2}{2\pi^2}\int_0^\infty\D k\,\frac{n\subs{B}(k)}{k}
	-\theta\subs{HT}\frac{e^2 T^2}{24 m^2},
\end{equation*}
where the $\theta\subs{HT}$ is defined such that it turns on the high temperatures contribution:
\begin{equation*}
	\theta\subs{HT}\equiv\begin{cases}
		0 & T \ll m \\
		1 & T \gg m
	\end{cases}.
\end{equation*}
The low-temperature part is in agreement with~\cite{Donoghue:1983qx,Donoghue:1984zs,Mitra:2001ar}.
Therefore we have
\begin{equation*}
	\lim_{p_s\to 0}\frac{1}{\sqrt{\,^\beta Z_2(p)\,^\beta Z_2(p')}}=1+\frac{e^2 T^2}{12 m^2}-\frac{e^2}{16\pi^3}I\subs{A}
	+\theta\subs{HT}\,\frac{e^2 T^2}{24 m^2},
\end{equation*}
where we denoted
\begin{equation*}
	I\subs{A}\equiv 8\pi \int_0^\infty\D k\,\frac{n\subs{B}(k)}{k}.
\end{equation*}
For future convenience, noticing that for $p_s=0$ we have $\omega_P=2m$, we can write the factor coming
from wave-function renormalization constants using $\omega_P\equiv u\cdot P$ instead of the mass $m$:
\begin{equation*}
	\lim_{P_s\to 0}\lim_{q\to 0}\frac{1}{\sqrt{\,^\beta Z_2(p)\,^\beta Z_2(p')}}=1+\frac{e^2 T^2}{3 \omega_P^2}-\frac{e^2}{16\pi^3}I\subs{A}
	+\theta\subs{HT}\,\frac{e^2 T^2}{6 \omega_P^2}.
\end{equation*}
%

%------------------------------------------------------------------------------------
\subsection{Renormalization of stress-energy tensor at finite temperature}
%------------------------------------------------------------------------------------
\label{sec:SETrenT}
As last step to renormalize the stress-energy tensor we have to calculate the temperature contribution of the diagrams in Fig.~\ref{fig:SETRen}.
Here we write the general procedure and we leave the details in Appendix~\ref{sec:APPSETRen}.

First, we recap the notation used. We indicate with $q$ and $P$ the momenta
\begin{equation*}
	P=p'+p,\quad q=p'-p
\end{equation*}
where $p$ and $p'$ are the momenta of external legs of the diagrams in Fig.~\ref{fig:SETRen}.
Notice that $q\cdot P=0$. Moreover, scattering theory imposes the conservation of the time component
of $p$ and $p'$ in the thermal bath rest frame, meaning $q\cdot u=0$. We are using this constraint
when evaluating all the diagrams. It is then convenient to define the following scalar and four-vectors:
\begin{equation*}
	\omega_P=P\cdot u,\quad P_s=\sqrt{\omega_P^2-P^2},\quad \tilde{P}_\mu=(\eta_{\mu\nu}-u_\mu u_\nu)P^\nu=P_\mu-\omega_P u_\mu,\quad
	l^\mu=\epsilon^{\mu\nu\rho\sigma}u_\nu\tilde{P}_\rho q_\sigma,\quad \hat{l}^\mu=\frac{l^\mu}{\sqrt{-l^2}},
\end{equation*}
we also denote with $a$ the ratio
\begin{equation*}
	a\equiv\frac{\sqrt{\omega_P^2-P^2}}{\omega_P}=\frac{P_s}{\omega_P},\quad 0<a<1.
\end{equation*}

Since we are considering 1-loop corrections, a generic diagrams of Fig.~\ref{fig:SETRen}, which we label with $X$,
can be written as
\begin{equation}
\label{eq:genericDiagram}
	M_{\mu\nu}^X(p,p')=\int\frac{\D^4 k}{(2\pi)^4}\frac{f^X(p,p',k)}{D^X(p,p',k)}\bar{u}(p')N^X_{\mu\nu}u(p),
\end{equation}
where all the spinorial structure is contained inside the numerator $N^X$.
Therefore $N^X$ can be simplified using Dirac equation and it can be decomposed in the following terms:
\begin{equation}
	\label{eq:DiracSimplify}
	\begin{split}
		\bar{u}(p')N^X_{\mu\nu}u(p)=&\bar{u}(p')\Big[ N_{kk}^X \slashed{k}\, k_\mu k_\nu + N_{Pk}^X \slashed{k}\, (P_\mu k_\nu +P_\nu k_\mu)+\\
		&+(N_{P\gamma}^{X(s)} +N_{P\gamma}^{X(k)}\slashed{k} )(P_\mu \gamma_\nu +P_\nu \gamma_\mu) 
		+ (N_{k\gamma}^{X(s)} +N_{k\gamma}^{X(k)}\slashed{k} ) (k_\mu \gamma_\nu +k_\nu \gamma_\mu)+\cdots\Big]u(p)
	\end{split}
\end{equation}
where each term $N_{yy}^X$ can depend on the scalars $\{k^2,\,P^2,\,q^2,k\cdot P,\, k\cdot q,\, P\cdot q,\,m\}$
and the dots stands for terms that to do not contribute to AGM. Indeed, we show in App.~\ref{sec:Selection} that 
the only terms that are relevant for AGM are
\begin{equation*}
	\begin{split}
		u_\mu\gamma_\nu+u_\nu\gamma_\mu,&\quad
		P_\mu\gamma_\nu+P_\nu\gamma_\mu,\quad
		\slashed{\hat{l}}\left(u_\mu \hat{l}_\nu+u_\nu \hat{l}_\mu \right),\quad
		\slashed{\hat{l}}\left(P_\mu \hat{l}_\nu+P_\nu \hat{l}_\mu \right).
	\end{split}
\end{equation*}
Using the orthogonal non-normalized basis $\{ u,\,\tilde{P},\,q,\,\hat{l}\}$, the integration variable $k$ can be decomposed into:
\begin{equation*}
	\begin{split}
		k_\mu=&(k\cdot u) u_\mu+\frac{(k\cdot \tilde{P})}{\tilde{P}^2}\tilde{P}_\mu+\frac{(k\cdot q)}{q^2}q_\mu-(k\cdot\hat{l})\hat{l}_\mu\\
		=&\left[(k\cdot u)-\frac{\omega_P(k\cdot \tilde{P})}{\tilde{P}^2}\right]u_\mu +\frac{(k\cdot \tilde{P})}{\tilde{P}^2}P_\mu+\frac{(k\cdot q)}{q^2}q_\mu-(k\cdot\hat{l})\hat{l}_\mu\\
		\equiv&A_u\,u_\mu +A_P\,P_\mu+A_q\,q_\mu+A_l\,\hat{l}_\mu.
	\end{split}
\end{equation*}
This decomposition is used to write the diagrams in a covariant form.
Consider the term $\bar{u}(p')\slashed{k} u(p)v_\mu w_\mu$, where $v$ and $w$ are any vector between $\{u,P,q,l\}$,
the term is decomposed into
\begin{equation*}
	\bar{u}(p')\slashed{k} u(p)v_\mu w_\nu= \bar{u}(p')\left[A_u\,\slashed{u}+A_l\,\slashed{\hat{l}}+2mA_P\right]u(p)v_\mu w_\nu,
\end{equation*}
and for what we show in Sec.~\ref{sec:1loopAGM} only the term in $\slashed{l}$ can contribute to AGM:
\begin{equation*}
	\bar{u}(p')\slashed{k} u(p)v_\mu w_\nu\to A_l\,\bar{u}(p')\,\slashed{\hat{l}}\,v_\mu w_\nu\,u(p)
	\quad\text{ with }(v,w)=(u,l)\text{ or }(v,w)=(P,l).
\end{equation*}
With the same argument, we can select only the parts relevant to AGM of all the possible terms:
\begin{equation*}
	\begin{split}
		\bar{u}(p')\slashed{k} u(p)\left(k_\mu v_\nu + k_\nu v_\mu\right)\to& \bar{u}(p')\Big[A_u A_l \slashed{\hat{l}} (u_\mu v_\nu + u_\nu v_\mu)
		+ A_P A_l \slashed{\hat{l}} (P_\mu v_\nu + P_\nu v_\mu)\Big] u(p) \text{ if }v=l\\
		\to& \bar{u}(p')\Big[A_l^2 \slashed{\hat{l}} (\hat{l}_\mu v_\nu + \hat{l}_\nu v_\mu)\Big] u(p) \text{ if }v=u\text{ or }v=P\\
		\bar{u}(p')\slashed{k}k_\mu k_\nu u(p)\to &\bar{u}(p')\Big[A_u A_l^2\,\slashed{\hat{l}}(u_\mu \hat{l}_\nu+u_\nu \hat{l}_\mu)
		+A_P A_l^2\,\slashed{\hat{l}}(P_\mu \hat{l}_\nu+P_\nu \hat{l}_\mu) \Big]u(p).
	\end{split}
\end{equation*}
We then approximate the integrand to first order in $q$ and we perform the loop integral in $k$
decomposing its component along the tetrad  $\{u,\tilde{P},q,l\}$.
The results at first order in $q$ are (see Appendix~\ref{sec:APPSETRen}):
\begin{equation*}
	\begin{split}
		(Z_2(p)Z_2(p'))^{-1/2}M_{\mu\nu}^{(0)}=& I^{Z}_{P\gamma} \left( P_\mu\gamma_\nu + P_\nu \gamma_\mu\right),\\
		M_{\mu\nu}^{SE}+M_{\mu\nu}^{CT}=&I^{S}_{P\gamma}\left( P_\mu\gamma_\nu + P_\nu \gamma_\mu\right),\\
		M_{\mu\nu}^{V}=&\left(I^V_{P\gamma}+\theta\subs{HT}I^V_{P\gamma,f}\right)\left( P_\mu\gamma_\nu + P_\nu \gamma_\mu\right)
		+\left(I^V_{u\gamma}+\theta\subs{HT}I^V_{u\gamma,f}\right)\left( u_\mu \gamma_\nu + u_\nu \gamma_\mu \right)\\
		&+\left(I^V_{Pl}+\theta\subs{HT}I^V_{Pl,f}\right) \slashed{\hat{l}}\left( P_\mu \hat{l}_\nu + P_\nu \hat{l}_\mu \right)
		+\theta\subs{HT}I^V_{ul,f}\slashed{\hat{l}}\left( u_\mu \hat{l}_\nu + u_\nu \hat{l}_\mu \right),\\
		M_{\mu\nu}^{C}=& \left(I^C_{P\gamma}+\theta\subs{HT}I^C_{P\gamma,f}\right)\left( P_\mu\gamma_\nu + P_\nu \gamma_\mu\right)
		+\left(I^C_{u\gamma}+\theta\subs{HT}I^C_{u\gamma,f}\right) \left( u_\mu \gamma_\nu + u_\nu \gamma_\mu \right),\\
		M_{\mu\nu}^{P}=&  I^P_{P\gamma}\left( P_\mu\gamma_\nu + P_\nu \gamma_\mu\right) + I^P_{u\gamma} \left( u_\mu \gamma_\nu + u_\nu \gamma_\mu \right)
		+\left(I^P_{Pl}+\theta\subs{HT}I^P_{Pl,f}\right) \slashed{\hat{l}}\left( P_\mu \hat{l}_\nu + P_\nu \hat{l}_\mu \right)\\
		&+\left(I^P_{ul}+\theta\subs{HT}I^P_{ul,f}\right) \slashed{\hat{l}}\left( u_\mu \hat{l}_\nu + u_\nu \hat{l}_\mu \right).
	\end{split}
\end{equation*}
Here, the function
\begin{equation*}
	\theta\subs{HT}\equiv\begin{cases}
		0 & T \ll m \\
		1 & T \gg m
	\end{cases}
\end{equation*}
turns on the fermionic contributions at high temperature and removes them for low temperatures;
we also introduce the IR divergent integral
\begin{equation*}
	I\subs{A}\equiv 8\pi\int_0^\infty \D k \frac{n\subs{B}(k)}{k}.
\end{equation*}
The form factors of $P_\mu\gamma_\nu + P_\nu \gamma_\mu$ are
\begin{equation*}
	\begin{split}
		\lim_{P_s\to 0}\lim_{q\to 0}I^{Z}_{P\gamma}=&\frac{1}{4}+\frac{2+\theta\subs{HT}}{24}\frac{e^2 T^2}{\omega_P^2}-\frac{1}{4}\frac{e^2}{16\pi^3}I\subs{A};\\
		\lim_{P_s\to 0}\lim_{q\to 0}I^{S}_{P\gamma}=&-\frac{2+\theta\subs{HT}}{12}\frac{e^2 T^2}{\omega_P^2}+\frac{1}{2}\frac{e^2}{16\pi^3}I\subs{A};\\
		I^V_{P\gamma} =&-\frac{1}{18}\frac{e^2 T^2}{\omega_P^2}\left[\frac{3}{2a^3}\log\left(\frac{1+a}{1-a}\right)-\frac{3}{a^2}\right]
		-\frac{1}{4}\frac{e^2}{16\pi^3}\frac{4m^2}{\omega_P^2}\frac{I_A}{1-a^2};\\
		I^V_{P\gamma,f} =&-\frac{1}{36}\frac{e^2 T^2}{\omega_P^2}\left[\frac{3}{2a^3}\log\left(\frac{1+a}{1-a}\right)-\frac{3}{a^2}\right];\\
		I^C_{P\gamma} =& \frac{1}{18}\frac{e^2 T^2}{\omega_P^2}\left[\frac{3}{2a^3}\log\left(\frac{1+a}{1-a}\right)-\frac{3}{a^2}\right];\\
		I^C_{P\gamma,f} =& \frac{1}{36}\frac{e^2 T^2}{\omega_P^2}\left[\frac{3}{2a^3}\log\left(\frac{1+a}{1-a}\right)-\frac{3}{a^2} \right];\\
		I^P_{P\gamma} =& -\frac{1}{18}\frac{e^2 T^2}{\omega_P^2}\left[\frac{3}{2a^3}\log\left(\frac{1+a}{1-a}\right)-\frac{3}{a^2}\right].
	\end{split}
\end{equation*}
The form factors of $u_\mu\gamma_\nu + u_\nu \gamma_\mu$ are
\begin{equation*}
	\begin{split}
		I^V_{u\gamma} =& -\frac{1}{9}\frac{e^2 T^2}{\omega_P}\left[\frac{3}{2a^3}\log\left(\frac{1+a}{1-a}\right)-\frac{3}{a^2}\right];\\
		I^V_{u\gamma,f} =& -\frac{1}{18}\frac{e^2 T^2}{\omega_P}\left[\frac{3}{2a^2}-\frac{3(1-a^2)}{4a^3}\log\left(\frac{1+a}{1-a}\right) \right];\\
		I^C_{u\gamma} =& \frac{1}{9}\frac{e^2T^2}{\omega_P}\left[\frac{3}{2a^2}-\frac{3(1-a^2)}{4a^3}\log\left(\frac{1+a}{1-a}\right) \right];\\
		I^C_{u\gamma,f} =& \frac{1}{18}\frac{e^2 T^2}{\omega_P}\left[\frac{3}{2a^2}-\frac{3(1-a^2)}{4a^3}\log\left(\frac{1+a}{1-a}\right) \right];\\
		I^P_{u\gamma} =& -\frac{1}{9}\frac{e^2 T^2}{\omega_P}\left[\frac{3}{2a^2}-\frac{3(1-a^2)}{4 a^3}\log\left(\frac{1+a}{1-a}\right)\right].
	\end{split}
\end{equation*}
The form factors of $\slashed{\hat{l}}(P_\mu\hat{l}_\nu + P_\nu \hat{l}_\mu)$ are
\begin{equation*}
	\begin{split}
		I^V_{Pl} =& \frac{1}{12}\frac{e^2 T^2}{\omega_P^2}\left[\frac{3}{2 a^3}\log\left(\frac{1+a}{1-a}\right)-\frac{3}{a^2}\right];\\
		I^V_{Pl,f} =& \frac{7}{360}\frac{e^2 T^2}{\omega_P^2}\left[\frac{12}{7}\frac{4m^2}{\omega_P^2}\left(\frac{5(3-2a^2)}{2a^4(1-a^2)}-\frac{15}{4a^5}\log\left(\frac{1+a}{1-a}\right) \right)
		-\frac{5}{7}\left(\frac{3}{2a^3}\log\left(\frac{1+a}{1-a}\right) -\frac{3}{a^2}\right) \right];\\
		I^P_{Pl} =&I^P_{P\gamma}= -\frac{1}{18}\frac{e^2 T^2}{\omega_P^2}\left[\frac{3}{2a^3}\log\left(\frac{1+a}{1-a}\right)-\frac{3}{a^2}\right];\\
		I^P_{Pl,f} =&-\frac{1}{30}\frac{e^2 T^2}{\omega_P^2}\frac{4m^2}{\omega_P^2}\left[\frac{15(3-a^2)}{8 a^5}\log\left(\frac{1+a}{1-a}\right)-\frac{45}{4 a^4}\right].
	\end{split}
\end{equation*}
The form factors of $\slashed{\hat{l}}(u_\mu\hat{l}_\nu + u_\nu \hat{l}_\mu)$ are
\begin{equation*}
	\begin{split}
		I^V_{ul,f} =& -\frac{1}{30}\frac{e^2 T^2}{\omega_P}\left[-\frac{2}{3}\frac{4m^2}{\omega_P^2}\left(\frac{15(3-2a^2)}{8a^5}\log\left(\frac{1+a}{1-a}\right)-\frac{45}{4a^4} \right)
		+\frac{5}{3}\left(\frac{3}{2a^2}-\frac{3(1-a^2)}{4a^3}\log\left(\frac{1+a}{1-a}\right)\right) \right];\\
		I^P_{ul} =&I^P_{u\gamma}= -\frac{1}{9}\frac{e^2 T^2}{\omega_P}\left[\frac{3}{2a^2}-\frac{3(1-a^2)}{4 a^3}\log\left(\frac{1+a}{1-a}\right)\right];\\
		I^P_{ul,f} =&-\frac{1}{45}\frac{e^2 T^2}{\omega_P}\frac{4m^2}{\omega_P^2}\left[\frac{15(3-a^2)}{8 a^5}\log\left(\frac{1+a}{1-a}\right)-\frac{45}{4 a^4}\right].
	\end{split}
\end{equation*}
These functions are written such that the quantity inside the square brackets is 1 for $a=0$,
which correspond to the non relativistic particle limit ($P_s\to 0$).
 It is straightforward
to check that by summing all these terms in the non-relativistic limit, one get the form
factors quoted in Eq.~(\ref{eq:AGMFormFactors}).

%************************************************************************************
\section{Radiative correction to the stress-energy tensor}
%************************************************************************************
\label{sec:APPSETRen}
In this appendix we evaluate the temperature modification of the diagrams in Fig.~\ref{fig:SETRen}.
The general strategy and the final results are written in Sec.~\ref{sec:SETrenT}, in what follows
we provide a detailed calculations of all the diagrams.

To perform the loop integration on the momentum $k$ in the generic diagram~\eqref{eq:genericDiagram},
since the set $\{ u,\,\tilde{P},\,q,\,\hat{l}\}$ is a basis,
we choose $k_0$ along $u$, $k_z$ along $q_\mu$ and $k_x$ along $\tilde{P}$
and $k_y$ along $l$; thus, defining
\begin{equation*}
	\epsilon=\sqrt{-q^2},\quad 
	a=\frac{\sqrt{-\tilde{P}^2}}{\omega_P}=\frac{\sqrt{\omega_P^2-P^2}}{\omega_P},\quad 0<a<1,
\end{equation*}
we have
\begin{equation*}
	\begin{split}
		(k\cdot u)&=k_0,\quad (k\cdot q)=-\epsilon\,k_z,\quad
		(k\cdot \tilde{P})=-a\omega_P \, k_x\quad (k\cdot P)=\omega_P(k_0-a k_x),\quad(k\cdot \hat{l})=- k_y\\
		A_u&=k_0-\frac{k_x}{a},\quad A_P=\frac{k_x}{a\omega_P},\quad A_q=\frac{k_z}{\epsilon},\quad A_l=k_y.
	\end{split}
\end{equation*}
At last we define the following unit vectors inside the $k$ integration:
\begin{equation*}
	\hat{k}_x=\frac{k_x}{|\vec{k}|}=\cos\phi\sin\theta,\quad \hat{k}_y=\frac{k_y}{|\vec{k}|}=\sin\phi\sin\theta,\quad \hat{k}_z=\frac{k_z}{|\vec{k}|}=\cos\theta.
\end{equation*}

For the fermionic part we only consider the high temperature limit, $T\gg m,P_s,\omega_p,q$.
Every fermionic form factor at first order in $\epsilon$ can be written as
\begin{equation*}
	I=C\int\frac{\D k}{E_k} k^l n\subs{F}(E_k)\sum_{s=\pm}\int\D\Omega_k f_s(\hat{k},m,\omega_p,a),
\end{equation*}
where $C$ is a numerical constant and $l$ an integer.
Because of the Fermi-Dirac distribution function, the major contribution of the integral comes from $k\sim T$.
Then in the integrand we can consider $m$ and $\omega_P$ to be small compared to $k$. To consider the relevant
part of the integral we first replace
\begin{equation*}
	m=m_r k y,\quad \omega_P=\omega_{Pr} k y,\quad E_k=k\sqrt{1+m_r^2 y^2},
\end{equation*}
where $y$ is a dummy variable to perform a Taylor series. Then we expand the integrand in series of $y$ around zero
and we keep only the first terms. Then we replace back
\begin{equation*}
	y=1,\quad m_r=\frac{m}{k},\quad \omega_{Pr}=\frac{\omega_P}{k}.
\end{equation*}
In this way we can select the relevant contribution, which has the form
\begin{equation*}
	I\simeq C\int\D k\, n\subs{F}(E_k)\sum_{s=\pm}\int\D\Omega_k \tilde{f}_s(\hat{k},m,\omega_p,a)
	=\frac{C \pi^2 T^2}{12}\sum_{s=\pm}\int\D\Omega_k \tilde{f}_s(\hat{k},m,\omega_p,a).
\end{equation*}
The last factor can be obtained by summing and integrating over the angles. The angular integrals
have the form
\begin{equation*}
	I_\Omega(n_x,n_y,n_z,n_d)=\int\D\Omega_k \frac{2\hat{k}_x^{n_x}\hat{k}_y^{n_y}\hat{k}_z^{n_z}}{(1-a^2 \hat{k}_x^2)^{n_d}}
\end{equation*}
and their results are quoted in Sec.~\ref{sec:AngularIntegrals}.

%------------------------------------------------------------------------------------
\subsection{Self-energy and Counter-terms}
%------------------------------------------------------------------------------------
The fermion-self energy diagram, Fig.~\ref{fig:SETRen}(b), is 
\begin{equation*}
	M_{\mu\nu}^{SE}= \bar{u}(p')\, ^\beta\Sigma(p') \I S_F(p') V_{\mu\nu}(p',p)u(p),
\end{equation*}
where $V_{\mu\nu}$ is the stress-energy tensor fermion coupling:
\begin{equation*}
	V_{\mu\nu}(p,p')=\frac{1}{4}\left[\gamma_\mu(p+p')_\nu+\gamma_\nu(p+p')_\mu\right]
	-\frac{1}{2}\eta_{\mu\nu}\left[\slashed{p}-m+\slashed{p}'-m\right].
\end{equation*}
Near the pole the self-energy can be written as
\begin{equation*}
	^\beta\Sigma(p)=\left(1-\, ^\beta Z_2(p)^{-1}\right)\left(\slashed{p}-m\right)+a\slashed{p}+b\slashed{u}+\delta m.
\end{equation*}
Most of this diagram is canceled by the counter-term in Fig.~\ref{fig:SETRen}(c).
Since the temperature dependent Dirac equation is
\begin{equation*}
	\left((1-a)\slashed{p}-b\slashed{u}-m\subs{T}\right)u_\beta(p)=0,
\end{equation*}
we must use the finite temperature counter-term of the momentum Lagrangian
\begin{equation*}
	\delta \mc{L}=a\slashed{p}+b\slashed{u}+\delta m.
\end{equation*}
Therefore, the counter term is
\begin{equation*}
	\begin{split}
		M_{\mu\nu}^{CT}=&-\bar{u}(p')\, \left(a\slashed{p'}+b\slashed{u}+\delta m\right) \I S_F(p') V_{\mu\nu}(p',p)u(p).
	\end{split}
\end{equation*}

Considering both legs we have in total
\begin{equation*}
	\begin{split}
		M_{\mu\nu}^{SE}+M_{\mu\nu}^{CT}=&\left(2-\,^\beta Z_2(p)^{-1}-\, ^\beta Z_2(p')^{-1}\right)\bar{u}(p') V_{\mu\nu}(p',p)u(p),
	\end{split}
\end{equation*}
whose contribution to AGM is
\begin{equation*}
	\begin{split}
		M_{\mu\nu}^{SE}+M_{\mu\nu}^{CT}=&\frac{1}{4}\left(2-\,^\beta Z_2(p)^{-1}-\, ^\beta Z_2(p')^{-1}\right)\bar{u}(p')\left(P_\mu\gamma_\nu+P_\nu\gamma_\mu \right)u(p)\\
		=& I^{S}_{P\gamma}\bar{u}(p')\left(P_\mu\gamma_\nu+P_\nu\gamma_\mu \right)u(p).
	\end{split}
\end{equation*}
In the limit $q\to 0$ and $P_s\to 0$ we have
\begin{equation*}
	\lim_{P_s\to 0}I^{S}_{P\gamma}=-\frac{e^2 T^2}{6 \omega_P^2}+\frac{1}{2}\frac{e^2}{16\pi^3}I\subs{A}-\theta\subs{HT}\,\frac{e^2 T^2}{12 \omega_P^2},
\end{equation*}
where $\theta\subs{HT}$ is vanishing for $T\ll m$ and it goes to one for $T\gg m$.

%------------------------------------------------------------------------------------
\subsection{Photon polarization diagram}
%------------------------------------------------------------------------------------
\label{subsec:PolDiag}
Now we want to check if gravitomagnetic moment gets finite temperature corrections from the diagram
of photon polarization Fig.~\ref{fig:SETRen}(f):
\begin{equation*}
	M^P_{\mu\nu}(p,p')= \int\frac{\D^4k}{(2\pi)^4}
	\bar{u}(p')(-\I e\gamma_\rho)\I S_F(p-k)(-\I e\gamma_\sigma)u(p)T^{\alpha\beta}_{\mu\nu}(k,q+k)\I D^\rho_\beta(q+k)\I D^\sigma_\alpha(k)
\end{equation*}
where the stress-energy tensor-photon coupling vertex is
\begin{equation*}
	\begin{split}
		T^{\alpha\beta}_{\mu\nu}(k,p)=&
		-(\delta^\alpha_\mu\delta^\beta_\nu+\delta^\alpha_\nu\delta^\beta_\mu)k\cdot p
		-g^{\alpha\beta}(k_\mu p_\nu+k_\nu p_\mu)+(k_\mu\delta^\beta_\nu+k_\nu\delta^\beta_\mu)p^\alpha+\\
		&+(p_\mu\delta^\alpha_\nu+p_\nu\delta^\alpha_\mu)k^\beta+g_{\mu\nu}(\delta^{\alpha\beta}k\cdot p-k^\alpha p^\beta).
	\end{split}
\end{equation*}
At low temperature $T\ll m$  we can neglect the part coming from the fermionic thermal distribution.
The real part of the thermal contribution of the diagram is then
\begin{equation*}
	\begin{split}
		\text{Re}^\beta M^P_{\mu\nu}(p,p')=& e^2 \int\frac{\D^4k}{(2\pi)^3}
		\bar{u}(p')\gamma_\alpha\,\frac{\slashed{p}-\slashed{k}+m}{(k-p)^2-m^2}\gamma_\beta u(p)T^{\alpha\beta}_{\mu\nu}(k,q+k)\times\\
		&\times\left[\frac{\delta\left((q+k)^2\right)n_B(q+k)}{k^2}+\frac{\delta\left(k^2\right)n_B(k)}{(q+k)^2}\right].
	\end{split}
\end{equation*}
Making the changing of variables $k\to k-q$ in the first term we have
\begin{equation}
	\label{eq:ReMpLowT}
	\begin{split}
		\text{Re}^\beta M^P_{\mu\nu}(p,p')=& e^2 \int\frac{\D^4k}{(2\pi)^3}
		\bar{u}(p')\gamma_\alpha\Bigg[\frac{\slashed{p}-\slashed{k}+\slashed{q}+m}{(k-q-p)^2-m^2}\frac{T^{\alpha\beta}_{\mu\nu}(k-q,k)}{(k-q)^2}+\\
		&+\frac{\slashed{p}-\slashed{k}+m}{(k-p)^2-m^2}\frac{T^{\alpha\beta}_{\mu\nu}(k,q+k)}{(q+k)^2}\Bigg]\gamma_\beta u(p)\delta\left(k^2\right)n_B(k)\\
		\equiv&e^2 \int\frac{\D^4k}{(2\pi)^3}\bar{u}(p')\left[\frac{N_1}{D_1}+\frac{N_2}{D_2}\right]u(p)\delta\left(k^2\right)n_B(k).
	\end{split}
\end{equation}
We are interested in linear order of $q$, therefore we are using the momenta $P$ and $q$:
\begin{equation*}
	\begin{cases}
		P=p'+p, & p=\frac{1}{2}\left(P-q\right)\\
		q=p'-p, & p'=\frac{1}{2}\left(P+q\right).
	\end{cases}
\end{equation*}

We can use gamma algebra to simplify the expressions of the numerators of the diagram in Eq. (\ref{eq:ReMpLowT}).
Taking advantage of the Dirac equation and setting $k^2=0$ from the Dirac delta, we find:
\begin{equation*}
	\begin{split}
		\bar{u}(p')N_1 u(p)=\bar{u}(p')\Big\{& \left(-4\slashed{k}-4m\right) k_\mu k_\nu+(k\cdot q)(P_\mu\gamma_\nu+P_\nu\gamma_\mu)
		+2(k\cdot P)(k_\mu\gamma_\nu+k_\nu\gamma_\mu)+\left(2m+2\slashed{k}\right)(k_\mu q_\nu+k_\nu q_\mu)\\
		&-(k\cdot P)(q_\mu\gamma_\nu+q_\nu\gamma_\mu)+\left(-2m(k\cdot q)-(k\cdot P)\slashed{k}-(k\cdot q)\slashed{k}+q^2\slashed{k}\right)\eta_{\mu\nu}\Big\}u(p);\\
		\bar{u}(p')N_2 u(p)=\bar{u}(p')\Big\{& \left(-4\slashed{k}-4m\right) k_\mu k_\nu-(k\cdot q)(P_\mu\gamma_\nu+P_\nu\gamma_\mu)
		+2(k\cdot P)(k_\mu\gamma_\nu+k_\nu\gamma_\mu)-\left(2m+2\slashed{k}\right)(k_\mu q_\nu+k_\nu q_\mu)\\
		&+(k\cdot P)(q_\mu\gamma_\nu+q_\nu\gamma_\mu)+\left(2m(k\cdot q)-(k\cdot P)\slashed{k}+(k\cdot q)\slashed{k}+q^2\slashed{k}\right)\eta_{\mu\nu}\Big\}u(p).
	\end{split}
\end{equation*}
Decomposing the four-vector $k$ with the tetrad $\{u,\tilde{P},q,\hat{l}\}$ we obtain:
\begin{equation*}
	\begin{split}
		\bar{u}(p')N_1 u(p)=&\bar{u}(p')\Big\{ \left(A_u\,\slashed{u}+A_l\,\slashed{\hat{l}}+m+2A_p\,m\right) \Big[-4A_u^2 u_\mu u_\nu -4A_P^2 P_\mu P_\nu
		-4A_l^2 \hat{l}_\mu \hat{l}_\nu -4A_u A_P (u_\mu P_\nu+u_\nu P_\mu)\\
		&-4A_u A_l (u_\mu \hat{l}_\nu+u_\nu \hat{l}_\mu)-4A_P A_l (P_\mu \hat{l}_\nu+P_\nu \hat{l}_\mu)+(4A_q-4A_q^2)q_\mu q_\nu+(2A_u-4A_u A_q)(u_\mu q_\nu+u_\nu q_\mu)\\
		&+(2A_P-4A_P A_q)(P_\mu q_\nu+P_\nu q_\mu)+(2A_l-4A_l A_q)(\hat{l}_\mu q_\nu+\hat{l}_\nu q_\mu)\Big]+2(k\cdot P)A_u(u_\mu\gamma_\nu+u_\nu\gamma_\mu)\\
		&+\left[2(k\cdot P)A_P+(k\cdot q)\right](P_\mu\gamma_\nu+P_\nu\gamma_\mu)+2(k\cdot P)A_l(\hat{l}_\mu\gamma_\nu+\hat{l}_\nu\gamma_\mu)
		+\left[2(k\cdot P)A_q-(k\cdot P)\right](q_\mu\gamma_\nu+q_\nu\gamma_\mu)\\
		&-\left[2m(k\cdot q)+\left((k\cdot q)+(k\cdot P)-q^2\right)\left(A_u\,\slashed{u}+A_l\,\slashed{\hat{l}}+2mA_P\right)\right]\eta_{\mu\nu}\Big\}u(p);
	\end{split}
\end{equation*}
and
\begin{equation*}
	\begin{split}
		\bar{u}(p')N_2 u(p)=&\bar{u}(p')\Big\{ \left(A_u\,\slashed{u}+A_l\,\slashed{\hat{l}}+m+2A_p\,m\right) \Big[-4A_u^2 u_\mu u_\nu -4A_P^2 P_\mu P_\nu
		-4A_l^2 \hat{l}_\mu \hat{l}_\nu -4A_u A_P (u_\mu P_\nu+u_\nu P_\mu)\\
		&-4A_u A_l (u_\mu \hat{l}_\nu+u_\nu \hat{l}_\mu)-4A_P A_l (P_\mu \hat{l}_\nu+P_\nu \hat{l}_\mu)+(4A_q-4A_q^2)q_\mu q_\nu+(2A_u-4A_u A_q)(u_\mu q_\nu+u_\nu q_\mu)\\
		&+(2A_P-4A_P A_q)(P_\mu q_\nu+P_\nu q_\mu)+(2A_l-4A_l A_q)(\hat{l}_\mu q_\nu+\hat{l}_\nu q_\mu)\Big]+2(k\cdot P)A_u(u_\mu\gamma_\nu+u_\nu\gamma_\mu)\\
		&+\left[2(k\cdot P)A_P-(k\cdot q)\right](P_\mu\gamma_\nu+P_\nu\gamma_\mu)+2(k\cdot P)A_l(\hat{l}_\mu\gamma_\nu+\hat{l}_\nu\gamma_\mu)
		+\left[2(k\cdot P)A_q+(k\cdot P)\right](q_\mu\gamma_\nu+q_\nu\gamma_\mu)\\
		&+\left[2m(k\cdot q)+\left((k\cdot q)-(k\cdot P)+q^2\right)\left(A_u\,\slashed{u}+A_l\,\slashed{\hat{l}}+2mA_P\right)\right]\eta_{\mu\nu}\Big\}u(p).
	\end{split}
\end{equation*}
For the denominators, using $k^2=0,\,P\cdot q=0,\,P^2=4m^2-q^2$, we find
\begin{equation*}
	\begin{split}
		D_1=&\left(P\cdot k+q\cdot k\right)\left(2q\cdot k-q^2\right)\\
		D_2=&-\left(P\cdot k-q\cdot k\right)\left(2q\cdot k+q^2\right).
	\end{split}
\end{equation*}
Choosing the frame described at the beginning of this section to decompose the momentum $k$ we can write the denominators as:
\begin{equation*}
	\begin{split}
		D_1=&-2\omega_P k_z\epsilon\left(k_0-k_x a\right)+\epsilon^2\left(\omega_P(k_0-k_x a)+2k_z^2\right)-k_z\epsilon^3\\
		D_2=&2\omega_P k_z\epsilon\left(k_0-k_x a\right)+\epsilon^2\left(\omega_P(k_0-k_x a)+2k_z^2\right)+k_z\epsilon^3.
	\end{split}
\end{equation*}
Since the denominators does not contains $k_y$ every term that is odd in $A_l=k_y$ is vanishing.
Moreover, we can now select only the pieces that could give contribution to AGM, they are the following
\begin{equation*}
	\begin{split}
		\bar{u}(p')N_1 u(p)=\bar{u}(p')\Big\{& -4A_u A_l^2\,\slashed{\hat{l}}(u_\mu \hat{l}_\nu+u_\nu \hat{l}_\mu)
		-4A_P A_l^2\,\slashed{\hat{l}}(P_\mu \hat{l}_\nu+P_\nu \hat{l}_\mu)+2(k\cdot P)A_u(u_\mu\gamma_\nu+u_\nu\gamma_\mu)\\
		&+\left[2(k\cdot P)A_P+(k\cdot q)\right](P_\mu\gamma_\nu+P_\nu\gamma_\mu)\Big\}u(p),
	\end{split}
\end{equation*}
and
\begin{equation*}
	\begin{split}
		\bar{u}(p')N_2 u(p)=\bar{u}(p')\Big\{& -4A_u A_l^2\,\slashed{\hat{l}}(u_\mu \hat{l}_\nu+u_\nu \hat{l}_\mu)
		-4A_P A_l^2\,\slashed{\hat{l}}(P_\mu \hat{l}_\nu+P_\nu \hat{l}_\mu)+2(k\cdot P)A_u(u_\mu\gamma_\nu+u_\nu\gamma_\mu)\\
		&+\left[2(k\cdot P)A_P-(k\cdot q)\right](P_\mu\gamma_\nu+P_\nu\gamma_\mu)\Big\}u(p).
	\end{split}
\end{equation*}

For the term in $\slashed{l}(u_\mu \hat{l}_\nu +u_\nu \hat{l}_\mu )$ at first order in $\epsilon$ we have
\begin{equation*}
	\begin{split}
		I^P_{ul}=&-e^2 \int\frac{\D^4k}{(2\pi)^3}\left[\frac{4}{D_1}+\frac{4}{D_2}\right]A_u A_l^2 \delta\left(k^2\right)n_B(k)\\
		\simeq& \lim_{\epsilon\to 0}\frac{2e^2}{\omega_P^2} \int\frac{\D^4k}{(2\pi)^3}\left[\frac{\omega_P}{(k_0-ak_x)(k_z^2-\epsilon^2/4)}
		+\frac{2}{(k_0-a k_x)^2}\right](k_0-k_x/a) k_y^2 \delta\left(k^2\right)n_B(k),
	\end{split}
\end{equation*}
integrating $k_0$ with the delta we find that the second term in square bracket is odd on $k_x$ and so vanishing; the first term becomes
\begin{equation*}
	\begin{split}
		I^P_{ul}=& \lim_{\epsilon\to 0}\frac{e^2}{\omega_P} \int\frac{\D\,k}{(2\pi)^3} k\, n_B(k)
		\int\D\Omega_k \frac{2(1-\hat{k}_x^2)\hat{k}_y^2}{(1-a^2\hat{k}_x^2)(\hat{k}_z^2-\epsilon^2/(4k^2))}.
	\end{split}
\end{equation*}
The angular integrals do not converge for $\epsilon=0$ but they have a finite result in the principal value sense:
\begin{equation*}
	\begin{split}
		\lim_{\tau\to 0}\int\D\Omega_k \frac{2 \hat{k}_y^2}{(1-a^2\hat{k}_x^2)(\hat{k}_z^2-\tau^2)}=&-\frac{8\pi}{2a}\log\left(\frac{1+a}{1-a}\right),\\
		\lim_{\tau\to 0}\int\D\Omega_k \frac{2 \hat{k}_x^2 \hat{k}_y^2}{(1-a^2\hat{k}_x^2)(\hat{k}_z^2-\tau^2)}=&-\frac{8\pi}{3}\left(\frac{3}{2a^3}\log\left(\frac{1+a}{1-a}\right)-\frac{3}{a^2}\right)
	\end{split}
\end{equation*}
and therefore
\begin{equation*}
	I^P_{ul}\equiv I^P_u= -\frac{e^2 T^2}{9\omega_P}\left[\frac{3}{2a^2}-\frac{3(1-a^2)}{4a^3}\log\left(\frac{1+a}{1-a}\right)\right].
\end{equation*}

For the term in $\slashed{\hat{l}}(P_\mu \hat{l}_\nu +P_\nu \hat{l}_\mu )$ we have
\begin{equation*}
	\begin{split}
		I^P_{Pl}=&-e^2 \int\frac{\D^4k}{(2\pi)^3}\left[\frac{4}{D_1}+\frac{4}{D_2}\right]A_P A_l^2 \delta\left(k^2\right)n_B(k)\\
		\simeq& \lim_{\epsilon\to 0}\frac{2e^2}{a\omega_P^3} \int\frac{\D^4k}{(2\pi)^3}\left[\frac{\omega_P}{(k_0-ak_x)(k_z^2-\epsilon^2/4)}
		+\frac{2}{(k_0-a k_x)^2}\right]k_x k_y^2 \delta\left(k^2\right)n_B(k),
	\end{split}
\end{equation*}
integrating $k_0$ with the delta we find that the second term in square bracket is odd on $k_x$ and so vanishing; the first term becomes
\begin{equation*}
	\begin{split}
		I^P_{Pl}=& \lim_{\tau\to 0}\frac{e^2}{\omega_P^2} \int\frac{\D\,k}{(2\pi)^3} k\, n_B(k)\int\D\Omega_k \frac{2\hat{k}_x^2 \hat{k}_y^2}{(1-a^2\hat{k}_x^2)(\hat{k}_z^2-\tau^2)}.
	\end{split}
\end{equation*}
After integration we obtain
\begin{equation*}
	I^P_{Pl}\equiv I^P_P= -\frac{e^2 T^2}{18\omega_P^2}\left[\frac{3}{2a^3}\log\left(\frac{1+a}{1-a}\right)-\frac{3}{a^2}\right].
\end{equation*}

Consider now the part in $(u_\mu \gamma_\nu +u_\nu \gamma_\mu )$. At linear order of $\epsilon$ the scalar part in front of it is
\begin{equation*}
	\begin{split}
		I_{u\gamma}^P=& e^2 \int\frac{\D^4k}{(2\pi)^3}\left[\frac{1}{D_1}+\frac{1}{D_2}\right]2(k\cdot P)A_u\delta\left(k^2\right)n_B(k)
		\simeq e^2 \int\frac{\D^4k}{(2\pi)^3}\left[\frac{2}{a\omega_P}\frac{k_x-a k_0}{k_0-a k_x}+\frac{k_x-ak_0}{ak_z^2}\right]\delta\left(k^2\right)n_B(k)\\
		=& \frac{e^2}{2(2\pi)^3} \sum_{s=\pm}\int \D k\,k \,n_B(k)\int\D\Omega_k \left[\frac{2}{a\omega_P}\frac{\hat{k}_x-a s}{s-a \hat{k}_x}
		+\frac{1}{k}\frac{\hat{k}_x-a s}{a\hat{k}_z^2}\right]\\
		=&\frac{e^2}{2(2\pi)^3} \int \D k \,k\,n_B(k)\int\D\Omega_k  \left[-\frac{4}{\omega_P}\frac{1-\hat{k}_x^2}{1-a^2 \hat{k}_x^2}
		+\frac{2}{k}\frac{\hat{k}_x}{a\hat{k}_z^2}\right]\\
		=&-\frac{e^2}{\omega_P}\frac{1}{(2\pi)^3} \int \D k \,k\,n_B(k)\int\D\Omega_k \frac{2(1-\hat{k}_x^2)}{(1-a^2 \hat{k}_x^2)}.
	\end{split}
\end{equation*}
Using the angular integrals
\begin{equation*}
	\begin{split}
		\int\D\Omega_k \frac{2}{1-a^2 \hat{k}_x^2}=\frac{8\pi}{2a}\log\left(\frac{1+a}{1-a}\right),\quad
		\int\D\Omega_k \frac{2\hat{k}_x^2}{1-a^2 \hat{k}_x^2}=\frac{8\pi}{3}\left(\frac{3}{2a^3}\log\left(\frac{1+a}{1-a}\right)-\frac{3}{a^2}\right)
	\end{split}
\end{equation*}
we obtain
\begin{equation*}
	I_{u\gamma}^P=I^P_u=-\frac{e^2 T^2}{9\omega_P}\left(\frac{3}{2a^2}-\frac{3(1-a^2)}{4 a^3}\log\left(\frac{1+a}{1-a}\right)\right).
\end{equation*}

The part proportional to $(P_\mu\gamma_\nu+P_\nu\gamma_\mu)$ at first order in $\epsilon$ is the integral
\begin{equation*}
	\begin{split}
		I_{P\gamma}^P=& e^2 \int\frac{\D^4k}{(2\pi)^3}\left[\frac{\left(2(k\cdot\tilde{P})A_P+(k\cdot q)\right)}{D_1}
		+\frac{\left(2(k\cdot\tilde{P})A_P-(k\cdot q)\right)}{D_2}\right]\delta\left(k^2\right)n_B(k)\\
		\simeq& e^2 \int\frac{\D^4k}{(2\pi)^3}\left[\frac{1}{\omega_P}\frac{1}{k_0-a k_x}
		-\frac{2k_x}{a\omega_P^2(k_0-a k_x)}-\frac{k_x}{a\omega_P k_z^2}\right]\delta\left(k^2\right)n_B(k)\\
		=& \frac{e^2}{2\omega_P}\frac{1}{(2\pi)^3} \sum_{s=\pm}\int \D k\,k \,n_B(k)\int\D\Omega_k \left[\frac{1}{k}\frac{1}{s-a \hat{k}_x}
		-\frac{2\hat{k}_x}{a\omega_P(s-a \hat{k}_x)}-\frac{1}{k}\frac{\hat{k}_x}{a \hat{k}_z^2}\right]\\
		=&\frac{e^2}{2\omega_P}\frac{1}{(2\pi)^3} \int \D k \,k\,n_B(k)\int\D\Omega_k  \left[\frac{1}{k}\frac{2a\hat{k}_x}{1-a^2 \hat{k}_x^2}
		-\frac{4\hat{k}_x^2}{\omega_P(1-a^2 \hat{k}_x^2)}-\frac{2}{k}\frac{\hat{k}_x}{a \hat{k}_z^2}\right]\\
		=&-\frac{e^2}{\omega_P^2}\frac{1}{(2\pi)^3} \int \D k \,k\,n_B(k)\int\D\Omega_k \frac{2\hat{k}_x^2}{(1-a^2 \hat{k}_x^2)},
	\end{split}
\end{equation*}
where we first integrated $k_0$ with the delta and then we performed the angular integration and removed the manifestly vanishing angular integrations.
After integration we obtain:
\begin{equation*}
	I_{P\gamma}^P=I^P_P=-\frac{e^2 T^2}{18\omega_P^2}\left(\frac{3}{2a^3}\log\left(\frac{1+a}{1-a}\right)-\frac{3}{a^2}\right).
\end{equation*}

Summing all the relevant terms we found that the diagram can be written as:
\begin{equation*}
	M^P_{\mu\nu}=  I^P_u \left( u_\mu \gamma_\nu + u_\nu \gamma_\mu \right) + I^P_P\left( P_\mu\gamma_\nu + P_\nu \gamma_\mu\right)
	+I^P_u \slashed{\hat{l}}\left( u_\mu \hat{l}_\nu + u_\nu \hat{l}_\mu \right)+I^P_P \slashed{\hat{l}}\left( P_\mu \hat{l}_\nu + P_\nu \hat{l}_\mu \right).
\end{equation*}
Using this result, the contribution to gravitomagnetic moment from the photon polarization diagram
at low temperatures is vanishing
\begin{equation*}
	g_\Omega^P=\lim_{a\to 0}4\left(I^P_{P\gamma} + \frac{I^P_{u\gamma}}{\omega_P}
	-I^P_{Pl} -\frac{I^P_{ul}}{\omega_P} \right)=0.
\end{equation*}
%

%------------------------------------------------------------------------------------
\subsubsection{Fermionic part: High temperature}
%------------------------------------------------------------------------------------
The fermionic part is negligible at low temperature but it is comparable to the bosonic part at high temperatures.
We obtain the fermionic part from
\begin{equation*}
	\begin{split}
		\text{Re}^\beta M^P_{\mu\nu}(p,p')=&- e^2 \int\frac{\D^4k}{(2\pi)^3}
		\bar{u}(p')\gamma_\beta\,\left(\slashed{p}-\slashed{k}+m\right)\gamma_\alpha u(p)T^{\alpha\beta}_{\mu\nu}(k,q+k)
		\frac{\delta\left((p-k)^2-m^2\right)n_F(p-k)}{(q+k)^2 k^2}.
	\end{split}
\end{equation*}
Changing variables into $k\to p-k$ we have
\begin{equation*}
	\begin{split}
		\text{Re}^\beta M^P_{\mu\nu}(p,p')=& - e^2 \int\frac{\D^4k}{(2\pi)^3}
		\bar{u}(p')\gamma_\beta\,\left(\slashed{k}+m\right)\gamma_\alpha u(p)T^{\alpha\beta}_{\mu\nu}(p-k,q+p-k)
		\frac{\delta\left(k^2-m^2\right)n_F(k)}{(q+p-k)^2 (p-k)^2}\\
		=&- e^2 \int\frac{\D^4k}{(2\pi)^3}\frac{\bar{u}(p') N^P u(p)}{D^P}\delta\left(k^2-m^2\right)n_F(k).
	\end{split}
\end{equation*}
The numerator can be simplified into:
\begin{equation*}
	\begin{split}
		\bar{u}(p')N^P u(p)=\bar{u}(p')\Big\{& \left(4\slashed{k}-8m\right) k_\mu k_\nu+\slashed{k}\,q_\mu q_\nu
		+\left[(k\cdot P)-2m^2\right](P_\mu\gamma_\nu+P_\nu\gamma_\mu)-\slashed{k} P_\mu P_\nu+q^2 (k_\mu\gamma_\nu+k_\nu\gamma_\mu)\\
		&-(k\cdot q)(q_\mu\gamma_\nu+q_\nu\gamma_\mu)+2m(k_\mu P_\nu+k_\nu P_\mu)+\left(4m^3-2(k\cdot P)m-mq^2\right)\eta_{\mu\nu}\Big\}u(p).
	\end{split}
\end{equation*}
and the part relevant to AGM at first order in $q$ is
\begin{equation*}
	\begin{split}
		\bar{u}(p')N^P u(p)=\bar{u}(p')\Big\{& 4\slashed{k}\, k_\mu k_\nu
		+\left[(k\cdot P)-2m^2\right](P_\mu\gamma_\nu+P_\nu\gamma_\mu) \Big\}u(p).
	\end{split}
\end{equation*}
and after $k$ decomposition the relevant part is
\begin{equation*}
	\begin{split}
		\bar{u}(p')N^P u(p)=\bar{u}(p')\Big\{& 4A_u A_l^2\slashed{\hat{l}}(u_\mu \hat{l}_\nu+u_\nu \hat{l}_\mu)
		+4A_P A_l^2\slashed{\hat{l}} (P_\mu\hat{l}_\nu+ P_\nu\hat{l}_\mu)
		+\left(k\cdot P-2m^2\right)(P_\mu\gamma_\nu+P_\nu\gamma_\mu)\Big\}u(p).
	\end{split}
\end{equation*}
The denominator is
\begin{equation*}
	D^P=(q+p-k)^2 (p-k)^2=\left(2m^2-(k\cdot P)\right)^2-(k\cdot q)^2
	=(2m^2-\omega_P(k_0 -a k_x))^2-\epsilon^2 k_z^2.
\end{equation*}

For the term in $\slashed{\hat{l}}(u_\mu \hat{l}_\nu +u_\nu \hat{l}_\mu )$ we have
\begin{equation*}
	\begin{split}
		I^P_{ul,f}=&-e^2 \int\frac{\D^4k}{(2\pi)^3}\frac{4A_u A_l^2}{D} \delta\left(k^2-m^2\right)n_F(k)\\
		\simeq& -4e^2 \int\frac{\D^4k}{(2\pi)^3} \frac{(k_0-k_x/a) k_y^2}{[\omega_P(k_0-a k_x)-2m^2]^2} \delta\left(k^2-m^2\right)n_F(k)\\
		=&-\frac{e^2}{4\pi^3\omega_P^2} \int \D k\frac{k^3}{E_k} \,n_F(E_k)\sum_{s=\pm}\int\D\Omega_k \frac{(sE_k/k-\hat{k}_x/a) \hat{k}_y^2}{[(sE_k/k-a \hat{k}_x)-\frac{2m^2}{k\omega_P}]^2}
	\end{split}
\end{equation*}
Using the high temperature expansion described at the beginning of this appendix we find:
\begin{equation*}
	\begin{split}
		I^P_{ul,f}\simeq&-\frac{e^2 T^2 m^2}{12\pi\omega_P^3} \left(I_\Omega(0,2,0,3)+3(a^2-1)I_\Omega(2,2,0,3)-a^2 I_\Omega(4,2,0,3)\right)
	\end{split}
\end{equation*}
and at the end
\begin{equation*}
	I^P_{ul,f}\simeq -\frac{1}{45}\frac{e^2 T^2}{\omega_P}\frac{4m^2}{\omega_P^2}\left[\frac{15(3-a^2)}{8 a^5}\log\left(\frac{1+a}{1-a}\right)-\frac{45}{4 a^4}\right].
\end{equation*}

For the term in $(P_\mu \hat{l}_\nu +P_\nu \hat{l}_\mu )$ we have
\begin{equation*}
	\begin{split}
		I^P_{Pl,f}=&-e^2 \int\frac{\D^4k}{(2\pi)^3}\frac{4A_P A_l^2}{D} \delta\left(k^2-m^2\right)n_F(k)\\
		\simeq& -\frac{4e^2}{a\omega_P} \int\frac{\D^4k}{(2\pi)^3} \frac{k_x k_y^2}{[\omega_P(k_0-a k_x)-2m^2]^2} \delta\left(k^2-m^2\right)n_F(k)\\
		=&-\frac{e^2}{4\pi^3 a\omega_P^3} \int \D k\,\frac{k^3}{E_k} \,n_F(E_k)\sum_{s=\pm}\int\D\Omega_k \frac{\hat{k}_x \hat{k}_y^2}{\left[(sE_k/k-a \hat{k}_x)-\frac{2m^2}{k\omega_P}\right]^2}.
	\end{split}
\end{equation*}
At high temperatures it becomes
\begin{equation*}
	I^P_{Pl,f}=-\frac{e^2m^2T^2}{12\pi\omega_P^4}\left(3I_\Omega(2,2,0,3) +a^2 I_\Omega(4,2,0,3)\right)
\end{equation*}
and hence
\begin{equation*}
	\begin{split}
		I^P_{Pl,f}=-\frac{1}{30}\frac{e^2 T^2}{\omega_P^2}\frac{4m^2}{\omega_P^2}\left[\frac{15(3-a^2)}{8 a^5}\log\left(\frac{1+a}{1-a}\right)-\frac{45}{4 a^4}\right].
	\end{split}
\end{equation*}

The part proportional to $(P_\mu\gamma_\nu+P_\nu\gamma_\mu)$ at first order in $\epsilon$ is the integral
\begin{equation*}
	\begin{split}
		I_{P\gamma,f}^P=& -e^2 \int\frac{\D^4k}{(2\pi)^3}\frac{(k\cdot P)-2m^2+A_P q^2}{D}\delta\left(k^2-m^2\right)n_F(k)\\
		\simeq& -e^2 \int\frac{\D^4k}{(2\pi)^3}\frac{1}{\omega_P(k_0-ak_x)-2m^2}\delta\left(k^2-m^2\right)n_F(k)\\
		=& -\frac{e^2}{16\pi^3\omega_P} \int \D k\frac{k}{E_k} \,n_F(E_k)\sum_{s=\pm}\int\D\Omega_k \frac{1}{sE_k/k-a\hat{k}_x-\frac{2m^2}{k\omega_P}}.
	\end{split}
\end{equation*}
The high temperature expansion give a contribution of the form
\begin{equation*}
	I_{P\gamma,f}^P=\int \D k \,n_F(k)\sum_{n=1}k^{-n}f_n(a),
\end{equation*}
which gives logarithmic and sub-leading contributions in temperature that we can neglect. Notice that there is no IR divergence, because the previous
integral is just an approximation for high $k$, at low $k$ when the divergence would occur the mass of the particle prevent the divergence.

Summing all the relevant terms we found that the diagram can be written as:
\begin{equation*}
	M^P_{\mu\nu,f}=  I^P_{u,f} \slashed{\hat{l}}\left( u_\mu \hat{l}_\nu + u_\nu \hat{l}_\mu \right)+I^P_{P,f} \slashed{\hat{l}}\left( P_\mu \hat{l}_\nu + P_\nu \hat{l}_\mu \right),
\end{equation*}
with
\begin{equation*}
	\begin{split}
		I^P_{u,f}=& -\frac{4e^2 T^2 m^2}{45\omega_P^3}\left[\frac{15(3-a^2)}{8 a^5}\log\left(\frac{1+a}{1-a}\right)-\frac{45}{4 a^4}\right],\\
		I^P_{P,f}=& -\frac{2e^2m^2T^2}{15\omega_P^4}\left[\frac{5(3-2a^2)}{2a^4(1-a^2)}-\frac{15}{4 a^5}\log\left(\frac{1+a}{1-a}\right)\right].
	\end{split}
\end{equation*}
Using this result to evaluate the amplitude $\mc{A}\propto M^P_{\mu\nu}u^\mu A_g^\nu$ we find
that the gravitomagnetic moment coming from photon polarization diagram is
\begin{equation*}
	g_\Omega^P=-\lim_{P_s\to 0}\lim_{q\to 0}4 \left(\frac{I^P_{u,f}}{\omega_P}+ I^P_{P,f}\right)\theta\subs{HT}=\frac{e^2 T^2}{18 m^2}\theta\subs{HT}.
\end{equation*}
Only at high temperatures the polarization diagram contribute to AGM.

%------------------------------------------------------------------------------------
\subsection{Electromagnetic vertex}
%------------------------------------------------------------------------------------
From electromagnetic vertex correction, Fig.~\ref{fig:SETRen}(d), we have
\begin{equation*}
	M_{\mu\nu}^V=\int\frac{\D^4 k}{(2\pi)^4}\bar{u}(p')(-\I e\gamma_\alpha) \I S_F(p'-k) V_{\mu\nu}(p-k,p'-k)\I S_F(p-k)(-\I e\gamma_\beta) \I D^{\alpha\beta}(k)u(p),
\end{equation*}
where $V_{\mu\nu}$ is the stress-energy tensor fermion coupling:
\begin{equation*}
	V_{\mu\nu}(p,p')=\frac{1}{4}\left[\gamma_\mu(p+p')_\nu+\gamma_\nu(p+p')_\mu\right]
	-\frac{1}{2}\eta_{\mu\nu}\left[\slashed{p}-m+\slashed{p}'-m\right].
\end{equation*}
As before, we evaluate the temperature part and the relevant part at low temperature is only given by the Bose distribution term:
\begin{equation*}
	\begin{split}
		^\beta M_{\mu\nu}^V=&-e^2\int\frac{\D^4 k}{(2\pi)^3}
		\frac{\bar{u}(p')\gamma_\alpha(\slashed{p}'-\slashed{k}+m) V_{\mu\nu}(p-k,p'-k)(\slashed{p}-\slashed{k}+m)\gamma^\alpha u(p)}{[(p'-k)^2-m^2][(p-k)^2-m^2]}\delta\left(k^2\right)n_B(k)\\
		\equiv& -e^2\int\frac{\D^4 k}{(2\pi)^3}\frac{\bar{u}(p')N^V u(p)}{D^V}.
	\end{split}
\end{equation*}
After simplification we obtain
\begin{equation*}
	\begin{split}
		\bar{u}(p')N^V u(p)=-\bar{u}(p')\Big\{& \left(4\slashed{k}+4m\right)\, k_\mu k_\nu+2\slashed{k} P_\mu P_\nu
		-\left(m+3\slashed{k}\right)(k_\mu P_\nu+k_\nu P_\mu)\\
		&+\left(m^2-q^2/2-(k\cdot P)\right)(P_\mu\gamma_\nu+P_\nu\gamma_\mu)+\left(-2m^2+q^2+2(k\cdot P)\right)(k_\mu\gamma_\nu+k_\nu\gamma_\mu)\\
		&+2\left(m+\slashed{k}\right)(k\cdot P)\eta_{\mu\nu}\Big\}u(p).
	\end{split}
\end{equation*}
The denominator is
\begin{equation*}
	D^V=\left[(k\cdot P)+(k\cdot q)\right]\left[(k\cdot P)-(k\cdot q)\right]=\omega_P^2(k_0-a k_x)^2 - k_z^2\epsilon^2.
\end{equation*}
The part of the nominator that brings contribution to AGM is then
\begin{equation*}
	\begin{split}
		\bar{u}(p')N^V u(p)=\bar{u}(p')\Big\{& 4\slashed{k}\, k_\mu k_\nu-3\slashed{k}(k_\mu P_\nu+k_\nu P_\mu)
		+\left(m^2-q^2/2-(k\cdot P)\right)(P_\mu\gamma_\nu+P_\nu\gamma_\mu)\\
		&+\left(-2m^2+q^2+2(k\cdot P)\right)(k_\mu\gamma_\nu+k_\nu\gamma_\mu)\Big\}u(p).
	\end{split}
\end{equation*}
After decomposing $k$ and removing odd terms in $A_l$, we have to consider:
\begin{equation*}
	\begin{split}
		\bar{u}(p')N^V u(p)=\bar{u}(p')\Big\{& 4A_u A_l^2 \slashed{\hat{l}}(u_\mu\hat{l}_\nu+u_\nu\hat{l}_\mu)+(4A_P-3)A_l^2\slashed{\hat{l}}(P_\mu\hat{l}_\nu+P_\nu\hat{l}_\mu)\\
		&+\left[\left(m^2-q^2/2-(k\cdot P)\right)+\left(-2m^2+q^2+2(k\cdot P)\right)A_P\right](P_\mu\gamma_\nu+P_\nu\gamma_\mu)\\
		&+\left(-2m^2+q^2+2(k\cdot P)\right)A_u(u_\mu\gamma_\nu+u_\nu\gamma_\mu)\Big\}u(p).
	\end{split}
\end{equation*}

The part in $\slashed{\hat{l}}(u_\mu \hat{l}_\nu+u_\nu \hat{l}_\mu)$ is
\begin{equation*}
	\begin{split}
		I^V_{ul}=&-e^2 \int\frac{\D^4k}{(2\pi)^3}\frac{4A_u A_l^2}{D^V}  \delta\left(k^2\right)n_B(k)
		\simeq -4e^2\int\frac{\D^4k}{(2\pi)^3}\frac{(k_0-k_x/a) k_y^2}{\omega_P^2(k_0-ak_x)^2} \delta\left(k^2\right)n_B(k)\\
		=&-\frac{2e^2}{(2\pi)^3\omega_P^2}\int \D k\,k^2\,n_B(k)\sum_{s=\pm}\int\D\Omega_k\frac{(s-\hat{k}_x/a) \hat{k}_y^2}{(s-a\hat{k}_x)^2}\\
		=&-\frac{2e^2}{(2\pi)^3\omega_P^2}\int \D k\,k^2\,n_B(k)\int\D\Omega_k\frac{2 \left(a^2 \hat{k}_x^3-2 a^2 \hat{k}_x+\hat{k}_x\right)\hat{k}_y^2}{a(1-a^2\hat{k}_x^2)^2}=0.
	\end{split}
\end{equation*}
The part in $\slashed{\hat{l}}(P_\mu \hat{l}_\nu+P_\nu \hat{l}_\mu)$ is
\begin{equation*}
	\begin{split}
		I^V_{Pl}=&-e^2 \int\frac{\D^4k}{(2\pi)^3}\frac{(4A_P-3)A_l^2}{D^V}  \delta\left(k^2\right)n_B(k)
		\simeq e^2\int\frac{\D^4k}{(2\pi)^3}\frac{(3-4k_x/a\omega_P) k_y^2}{\omega_P^2(k_0-ak_x)^2} \delta\left(k^2\right)n_B(k)\\
		=&\frac{e^2}{2(2\pi)^3\omega_P^2}\int \D k\,k\,n_B(k)\sum_{s=\pm}\int\D\Omega_k\frac{(3-4k\hat{k}_x/a\omega_P)\hat{k}_y^2}{(s-a\hat{k}_x)^2}
		=\frac{3e^2}{2(2\pi)^3\omega_P^2}\int \D k\,k\,n_B(k)\int\D\Omega_k\frac{2(1+a^2 \hat{k}_x^2)\hat{k}_y^2}{(1-a^2\hat{k}_x^2)^2}\\
		=&\frac{e^2 T^2}{12 \omega_P^2}\left[\frac{3}{2 a^3}\log\left(\frac{1+a}{1-a}\right)-\frac{3}{a^2}\right].
	\end{split}
\end{equation*}
The term in  $P_\mu \gamma_\nu+P_\nu \gamma_\mu$ gives
\begin{equation*}
\begin{split}
		I^V_{P\gamma}\simeq&- e^2\int\frac{\D^4k}{(2\pi)^3}\left[\frac{m^2}{(k_0-ak_x)^2}-\frac{\omega_P}{k_0-ak_x}\right]\frac{a\omega_p-2k_x}{a\omega_P^3} \delta\left(k^2\right)n_B(k)\\
		=&-\frac{e^2}{2(2\pi)^3a\omega_P^3}\int \D k\,k\,n_B(k)\sum_{s=\pm}\int\D\Omega_k\left[\frac{m^2}{k^2}\frac{a\omega_p-2k\,\hat{k}_x}{(s-a\hat{k}_x)^2}
		-\frac{\omega_P}{k}\frac{a\omega_p-2k\,\hat{k}_x}{s-a\hat{k}_x}\right]\\
		=&-\frac{e^2}{16\pi^3a\omega_P^3}\int \D k\,k\,n_B(k)\int\D\Omega_k\left[\frac{m^2}{k^2}\frac{2(a\omega_p-2k\,\hat{k}_x)(1+a^2\hat{k}_x^2)}{(1-a^2\hat{k}_x^2)^2}
		-\frac{\omega_P}{k}\frac{(a\omega_p-2k\,\hat{k}_x)2a\hat{k}_x}{1-a^2\hat{k}_x^2}\right]\\
		=&-\frac{e^2}{16\pi^3\omega_P^2}\int \D k\,k\,n_B(k)\int\D\Omega_k\left[\frac{m^2}{k^2}\frac{2(1+a^2\hat{k}_x^2)}{(1-a^2\hat{k}_x^2)^2}
		+2\frac{2\hat{k}_x^2}{1-a^2\hat{k}_x^2}\right]\\
		=&-\frac{e^2}{16\pi^3\omega_P^2}\int\D\Omega_k\left[m^2\frac{I_A}{8\pi}\frac{2(1+a^2\hat{k}_x^2)}{(1-a^2\hat{k}_x^2)^2}
		+\frac{2\pi^2 T^2}{6}\frac{2\hat{k}_x^2}{1-a^2\hat{k}_x^2}\right]\\
		=&-\frac{e^2m^2}{16\pi^3\omega_P^2}\frac{I_A}{1-a^2}
		-\frac{e^2 T^2}{18\omega_P^2}\left(\frac{3}{2a^3}\log\left(\frac{1+a}{1-a}\right)-\frac{3}{a^2}\right).
	\end{split}
\end{equation*}
The term in  $u_\mu \gamma_\nu+u_\nu \gamma_\mu$ gives
\begin{equation*}
	\begin{split}
		I^V_{u\gamma}\simeq& 2e^2\int\frac{\D^4k}{(2\pi)^3}\left[\frac{m^2}{(k_0-ak_x)^2}-\frac{\omega_P}{k_0-ak_x}\right]\frac{k_0-k_x/a}{\omega_P^2} \delta\left(k^2\right)n_B(k)\\
		=&\frac{e^2}{(2\pi)^3\omega_P^2}\int \D k\,k\,n_B(k)\sum_{s=\pm}\int\D\Omega_k\left[\frac{m^2}{k}\frac{(s-\hat{k}/a)}{(s-a\hat{k}_x)^2}
		-\frac{\omega_p(s-\hat{k}_x/a)}{s-a\hat{k}_x}\right]\\
		=&\frac{e^2}{(2\pi)^3\omega_P^2}\int \D k\,k\,n_B(k)\int\D\Omega_k\left[\frac{m^2}{k}\frac{2(\hat{k}_x-2a^2\hat{k}_x^2+a^2\hat{k}_x^3))}{a(1-a^2\hat{k}_x^2)^2}
		-\omega_P \frac{2(1-\hat{k}_x^2)}{1-a^2\hat{k}_x^2}\right]\\
		=&-\frac{e^2}{(2\pi)^3\omega_P}\frac{\pi^2T^2}{6}\int\D\Omega_k\frac{2(1-\hat{k}_x^2)}{1-a^2\hat{k}_x^2}\\
		=&-\frac{e^2 T^2}{9\omega_P}\left(\frac{3}{2a^3}\log\left(\frac{1+a}{1-a}\right)-\frac{3}{a^2}\right).
	\end{split}
\end{equation*}
Summing all the relevant terms we found that the diagram can be written as:
\begin{equation*}
	M^V_{\mu\nu}=  I^V_{u\gamma} \left( u_\mu \gamma_\nu + u_\nu \gamma_\mu \right) + I^V_{P\gamma}\left( P_\mu\gamma_\nu + P_\nu \gamma_\mu\right)
	+I^V_{Pl} \slashed{\hat{l}}\left( P_\mu \hat{l}_\nu + P_\nu \hat{l}_\mu \right).
\end{equation*}

%------------------------------------------------------------------------------------
\subsubsection{Fermionic part: High temperature}
%------------------------------------------------------------------------------------
The fermionic part is:
\begin{equation*}
	\begin{split}
		\text{Re}^\beta M^V_{\mu\nu}(p,p')=&e^2\int\frac{\D^4 k}{(2\pi)^3}
		\frac{\bar{u}(p')\gamma_\alpha(\slashed{p}'-\slashed{k}+m) V_{\mu\nu}(p-k,p'-k)(\slashed{p}-\slashed{k}+m)\gamma^\alpha u(p)}{k^2}\times\\
		&\times\left[\frac{\delta\left((p-k)^2-m^2\right)}{(p'-k)^2-m^2}n_F(p-k)+\frac{\delta\left((p'-k)^2-m^2\right)}{(p-k)^2-m^2}n_F(p'-k) \right].
	\end{split}
\end{equation*}
We change $k$ to $p-k$ in the first term and $k\to p'-k$ in the second one:
\begin{equation*}
	\begin{split}
		\text{Re}^\beta M^V_{\mu\nu}(p,p')=&e^2\int\frac{\D^4 k}{(2\pi)^3} \bar{u}(p')\gamma_\alpha
		\left[\frac{(\slashed{p}'-\slashed{p}+\slashed{k}+m)V_{\mu\nu}(k,p'-p+k)(\slashed{k}+m)}{[(p'-p+k)^2-m^2](p-k)^2}\right.\\
		&\left.+\frac{(\slashed{k}+m)V_{\mu\nu}(p-p'+k,k)(\slashed{p}-\slashed{p}'+\slashed{k}+m)}{[(p-p'+k)^2-m^2](p'-k)^2}\right]\gamma^\alpha u(p)
		\, n_F(k)\delta(k^2-m^2)\\
		\equiv& e^2\int\frac{\D^4 k}{(2\pi)^3} \bar{u}(p')\left[\frac{N^V_1}{D_1}+\frac{N^V_2}{D_2} \right] u(p) n_F(k)\delta(k^2-m^2).
	\end{split}
\end{equation*}
After simplification the relevant parts for numerators are
\begin{equation*}
	\begin{split}
		\bar{u}(p')N^V_1 u(p)=&\bar{u}(p')\Big\{ -4\slashed{k}\, k_\mu k_\nu-\slashed{k}(k_\mu P_\nu+k_\nu P_\mu)
		+\left(k\cdot P + k\cdot q\right)(k_\mu\gamma_\nu+k_\nu\gamma_\mu)\Big\}u(p);\\
		\bar{u}(p')N^V_2 u(p)=&\bar{u}(p')\Big\{ -4\slashed{k}\, k_\mu k_\nu-\slashed{k}(k_\mu P_\nu+k_\nu P_\mu)
		+\left(k\cdot P - k\cdot q\right)(k_\mu\gamma_\nu+k_\nu\gamma_\mu)\Big\}u(p).
	\end{split}
\end{equation*}
After decomposing $k$ we only have to consider:
\begin{equation*}
	\begin{split}
		\bar{u}(p')N^V_1 u(p)=\bar{u}(p')\Big\{& -4A_u A_l^2 \slashed{\hat{l}}(u_\mu\hat{l}_\nu+u_\nu\hat{l}_\mu)
		-(4A_P+1)A_l^2\slashed{\hat{l}}(P_\mu\hat{l}_\nu+P_\nu\hat{l}_\mu)\\
		&+\left(k\cdot P + k\cdot q\right)A_P(P_\mu\gamma_\nu+P_\nu\gamma_\mu)
		+\left(k\cdot P + k\cdot q\right)A_u(u_\mu\gamma_\nu+u_\nu\gamma_\mu)\Big\}u(p);\\
		\bar{u}(p')N^V_2 u(p)=\bar{u}(p')\Big\{& -4A_u A_l^2 \slashed{\hat{l}}(u_\mu\hat{l}_\nu+u_\nu\hat{l}_\mu)
		-(4A_P+1)A_l^2\slashed{\hat{l}}(P_\mu\hat{l}_\nu+P_\nu\hat{l}_\mu)\\
		&+\left(k\cdot P - k\cdot q\right)A_P(P_\mu\gamma_\nu+P_\nu\gamma_\mu)
		+\left(k\cdot P - k\cdot q\right)A_u(u_\mu\gamma_\nu+u_\nu\gamma_\mu)\Big\}u(p).
	\end{split}
\end{equation*}
The denominators are
\begin{equation*}
	\begin{split}
		D^V_1=&(4m^2-2k\cdot P)(k\cdot q)+2(k\cdot q)^2+q^2(2m^2-k\cdot P+k\cdot q)\\
		=& 2k_z\epsilon\left(\omega_P k_0 - a\omega_P k_x -2 m^2\right)+\epsilon ^2 \left(\omega_P (k_0-a k_x)+2 k_z^2-2 m^2\right)+k_z \epsilon ^3;\\
		D^V_2=&-(4m^2-2k\cdot P)(k\cdot q)+2(k\cdot q)^2+q^2(2m^2-k\cdot P-k\cdot q)\\
		=& -2k_z\epsilon\left(\omega_P k_0 - a\omega_P k_x -2 m^2\right)+\epsilon ^2 \left(\omega_P (k_0-a k_x)+2 k_z^2-2 m^2\right)-k_z \epsilon ^3.
	\end{split}
\end{equation*}

Consider the term in  $P_\mu \gamma_\nu+P_\nu \gamma_\mu$:
\begin{equation*}
\begin{split}
I^V_{P\gamma}=& e^2\int\frac{\D^4k}{(2\pi)^3}\left[
\frac{\omega_P(k_0-k_x a)-k_z\epsilon}{D^V_1}+\frac{\omega_P(k_0-k_x a)+k_z\epsilon}{D^V_2}
\right]\frac{k_x}{a\omega_P}\delta\left(k^2-m^2\right)n_F(k),
\end{split}
\end{equation*}
after expanding at first order in $\epsilon$ and in high temperature, we find
\begin{equation*}
\begin{split}
I^V_{P\gamma}\simeq&-\frac{e^2 T^2}{96\pi\omega_P^2} I_\Omega(2,0,0,1) \\
	=&-\frac{e^2 T^2}{36\omega_P^2}\left[\frac{3}{2a^3}\log\left(\frac{1+a}{1-a}\right)-\frac{3}{a^2}\right].
\end{split}
\end{equation*}

Similarly the term in $u_\mu \gamma_\nu+u_\nu \gamma_\mu$ is
\begin{equation*}
\begin{split}
I^V_{u\gamma}\simeq& e^2\int\frac{\D^4k}{(2\pi)^3}\left[
\frac{\omega_P(k_0-k_x a)-k_z\epsilon}{D^V_1}+\frac{\omega_P(k_0-k_x a)+k_z\epsilon}{D^V_2}
\right](k_0-k_x/a)\delta\left(k^2-m^2\right)n_F(k),
\end{split}
\end{equation*}
and at high temperature it becomes
\begin{equation*}
\begin{split}
I^V_{u\gamma}\simeq&-\frac{e^2 T^2}{96\pi\omega_P}\left( I_\Omega(2,0,0,1)-I_\Omega(0,0,0,1)\right) \\
	=&-\frac{e^2 T^2}{18\omega_P}\left[\frac{3}{2a^2}-\frac{3(1-a^2)}{4a^3}\log\left(\frac{1+a}{1-a}\right) \right].
\end{split}
\end{equation*}

The term in $\slashed{l}(P_\mu \hat{l}_\nu + P_\nu \hat{l}_\mu)$:
\begin{equation*}
\begin{split}
I^V_{Pl}\simeq& -e^2\int\frac{\D^4k}{(2\pi)^3}\left[\frac{1}{D^V_1}+\frac{1}{D^V_2}\right]
\frac{(4 k_x + a \omega_P)k_y^2}{a \omega_P}\delta\left(k^2-m^2\right)n_F(k)
\end{split}
\end{equation*}
at first order in $\epsilon$ and at high temperature is
\begin{equation*}
\begin{split}
I^V_{Pl}\simeq&\frac{e^2 T^2}{192\pi\omega_P^4}\left( 16 m^2\left(3I_\Omega(2,2,0,3)+a^2I_\Omega(4,2,0,3)\right)
	+\omega_P^2\left(I_\Omega(0,2,0,3)+2I_\Omega(2,2,-2,3)-4a^2 I_\Omega(4,2,-2,3)+\right.\right.\\
	&\left.\left.-a^4 I_\Omega(4,2,0,3)	+2a^4 I_\Omega(6,2,-2,3)\right)\right) \\
=&\frac{2e^2 m^2 T^2}{15\omega_P^4}\left[\frac{5(3-2a^2)}{2a^4(1-a^2)}-\frac{15}{4a^5}\log\left(\frac{1+a}{1-a}\right) \right]
	-\frac{e^2 T^2}{72\omega_P^2}\left[\frac{3}{2a^3}\log\left(\frac{1+a}{1-a}\right) -\frac{3}{a^2} \right]\\
=&\frac{7}{360}\frac{e^2 T^2}{\omega_P^2}\left[\frac{12}{7}\frac{4m^2}{\omega_P^2}\left(\frac{5(3-2a^2)}{2a^4(1-a^2)}-\frac{15}{4a^5}\log\left(\frac{1+a}{1-a}\right) \right)
	-\frac{5}{7}\left(\frac{3}{2a^3}\log\left(\frac{1+a}{1-a}\right) -\frac{3}{a^2}\right) \right].
\end{split}
\end{equation*}

Lastly, the term in $\slashed{l}(u_\mu \hat{l}_\nu + u_\nu \hat{l}_\mu)$:
\begin{equation*}
\begin{split}
I^V_{ul}\simeq&-e^2\int\frac{\D^4k}{(2\pi)^3}\left[\frac{1}{D^V_1}+\frac{1}{D^V_2}\right]
	4 (k_0 - k_x/a)k_y^2\delta\left(k^2-m^2\right)n_F(k)
\end{split}
\end{equation*}
gives
\begin{equation*}
\begin{split}
I^V_{ul}\simeq&\frac{e^2 T^2}{96\pi\omega_P^3}
	\left( 8 m^2\left(I_\Omega(0,2,0,3)+3(a^2-1)I_\Omega(2,2,0,3)-a^2I_\Omega(4,2,0,3)\right)
	+\omega_P^2\left(I_\Omega(0,2,-2,3)+\right.\right.\\
	&\left.\left.-(1+2a^2)I_\Omega(2,2,-2,3)+2a^2 I_\Omega(4,2,-2,3)+a^4 I_\Omega(4,2,-2,3)
	-a^4 I_\Omega(6,2,-2,3)\right)\right) \\
=&\frac{4e^2 m^2 T^2}{45\omega_P^3}
	\left[\frac{15(3-2a^2)}{8a^5}\log\left(\frac{1+a}{1-a}\right)-\frac{45}{4a^4} \right]
	-\frac{e^2 T^2}{18\omega_P}
	\left[\frac{3}{2a^2}-\frac{3(1-a^2)}{4a^3}\log\left(\frac{1+a}{1-a}\right) \right]\\
=&-\frac{1}{30}\frac{e^2 T^2}{\omega_P}	\left[-\frac{2}{3}\frac{4m^2}{\omega_P^2}
	\left(\frac{15(3-2a^2)}{8a^5}\log\left(\frac{1+a}{1-a}\right)-\frac{45}{4a^4} \right)
	+\frac{5}{3}\left(\frac{3}{2a^2}
	-\frac{3(1-a^2)}{4a^3}\log\left(\frac{1+a}{1-a}\right)\right) \right].
\end{split}
\end{equation*}
%

%------------------------------------------------------------------------------------
\subsection{Contact term}
%------------------------------------------------------------------------------------
The contact diagram, Fig.~\ref{fig:SETRen}(e), is 
\begin{equation*}
	M_{\mu\nu}^C=\int\frac{\D^4 k}{(2\pi)^4} e\, a_{\mu\nu\rho\kappa}\bar{u}(p')\left[\gamma^\kappa \I S_F(p-k)(- \I e\gamma^\alpha) \I D^\rho_\alpha(k)
	+(p'\leftrightarrow p)\right]u(p),
\end{equation*}
where the contact vertex is:
\begin{equation*}
	a_{\mu\nu}^{\beta\rho}=\eta_{\mu\nu}\eta^{\beta\rho}-\frac{1}{2}\left(\eta^\beta_\mu\eta^\rho_\nu+\eta^\beta_\nu\eta^\rho_\mu\right).
\end{equation*}
The temperature part given by the Bose distribution is:
\begin{equation*}
	\begin{split}
		^\beta M_{\mu\nu}^C(p',p)=&-e^2  a_{\mu\nu\rho\kappa}\int\frac{\D^4 k}{(2\pi)^3}\delta(k^2)n_B(k)
		\bar{u}(p')\left[\frac{\gamma^\kappa(\slashed{p}-\slashed{k}+m)\gamma^\rho}{(p-k)^2-m^2}
		+(p'\leftrightarrow p)\right]u(p)\\
		\equiv&- e^2\int\frac{\D^4 k}{(2\pi)^3} \bar{u}(p')\left[\frac{N_1^C}{D_1^C}+\frac{N_2^C}{D_2^C} \right] u(p) n_B(k)\delta(k^2).
	\end{split}
\end{equation*}
We refer to ``1'' as the first term and with ``2'' to the term with $p'\leftrightarrow p$:
The numerator of the Bose part:
\begin{equation*}
	\begin{split}
		\bar{u}(p')N_1^C u(p)=&\bar{u}(p')\Big\{ -\frac{1}{2}\left(P_\mu\gamma_\nu+P_\nu\gamma_\mu\right)+(k_\mu\gamma_\nu+k_\nu\gamma_\mu)
		+\frac{1}{2}\left(q_\mu\gamma_\nu+q_\nu\gamma_\mu\right)+\left(2m+\slashed{k}\right)\eta_{\mu\nu} \Big\} u(p),\\
		\bar{u}(p')N_2^C u(p)=&\bar{u}(p')\Big\{-\frac{1}{2}\left(P_\mu\gamma_\nu+P_\nu\gamma_\mu\right)+(k_\mu\gamma_\nu+k_\nu\gamma_\mu)
		-\frac{1}{2}\left(q_\mu\gamma_\nu+q_\nu\gamma_\mu\right)+\left(2m+\slashed{k}\right)\eta_{\mu\nu}\Big\} u(p),
	\end{split}
\end{equation*}
therefore the contribution to AGM is the same for the terms ``1'' and ``2'':
\begin{equation*}
	\bar{u}(p')N_1^C u(p)=\bar{u}(p')N_2 u(p)=\bar{u}(p')\Big\{-\frac{1}{2}\left(P_\mu\gamma_\nu+P_\nu\gamma_\mu\right)+(k_\mu\gamma_\nu+k_\nu\gamma_\mu)\Big\} u(p).
\end{equation*}
Decomposing $k$ in the numerators we have
\begin{equation*}
	\bar{u}(p')N_1^C u(p)=\bar{u}(p')N_2 u(p)=\bar{u}(p')\Big\{\left(A_P-\frac{1}{2}\right)\left(P_\mu\gamma_\nu+P_\nu\gamma_\mu\right)
	+A_u(u_\mu\gamma_\nu+u_\nu\gamma_\mu)\Big\} u(p)
\end{equation*}
and the denominators are
\begin{equation*}
	\begin{split}
		D_1^C=&-\left[(k\cdot P)-(k\cdot q)\right]=-\omega_P(k_0-a k_x)-k_z\epsilon,\\
		D_2^C=&-\left[(k\cdot P)+(k\cdot q)\right]=-\omega_P(k_0-a k_x)+k_z\epsilon.
	\end{split}
\end{equation*}

The part in $\left(P_\mu\gamma_\nu+P_\nu\gamma_\mu\right)$ at leading order in $\epsilon$ is
\begin{equation*}
	\begin{split}
		I_{P\gamma}^C=&-e^2\int\frac{\D^4 k}{(2\pi)^3} \left[\frac{N_1^C}{D_1^C}+\frac{N_2^C}{D_2^C} \right] n_B(k)\delta(k^2)
		\simeq -\frac{e^2}{\omega_P^2} \int\frac{\D^4 k}{(2\pi)^3}\delta(k^2)n_B(k) \frac{\omega_P-2k_x/a}{k_0-ak_x}\\
		=&-\frac{e^2}{16\pi^3\omega_P^2} \int\D k\,k\,n_B(k)\sum_{s=\pm}\int\D\Omega_k \left[\frac{\omega_P}{k}\frac{1}{s-a\hat{k}_x}
		-\frac{2}{a}\frac{\hat{k}_x}{s-a\hat{k}_x}\right]\\
		=&-\frac{e^2}{16\pi^3\omega_P^2} \int\D k\,k\,n_B(k)\int\D\Omega_k \left[\frac{\omega_P}{k}\frac{2a\hat{k}_x}{1-a^2\hat{k}_x^2}
		-2\frac{2\hat{k}_x^2}{1-a^2\hat{k}_x^2}\right].
	\end{split}
\end{equation*}
The first term in square bracket is vanishing and the second one gives
\begin{equation*}
	\begin{split}
		I_{P\gamma}^C=&+\frac{e^2}{16\pi^3\omega_P^2}2\frac{8\pi}{3}\left(\frac{3}{2a^3}\log\left(\frac{1+a}{1-a}\right)-\frac{3}{a^2}\right) \int\D k\,k\,n_B(k)\\
		=&\frac{e^2 T^2}{18 \omega_P^2}\left(\frac{3}{2a^3}\log\left(\frac{1+a}{1-a}\right)-\frac{3}{a^2}\right).
	\end{split}
\end{equation*}
The part in $\left(u_\mu\gamma_\nu+u_\nu\gamma_\mu\right)$ at leading order in $\epsilon$ is
\begin{equation*}
	\begin{split}
		I_{u\gamma}^C\simeq&\frac{2e^2}{\omega_P} \int\frac{\D^4 k}{(2\pi)^3}\delta(k^2)n_B(k) \frac{k_0-k_x/a}{k_0-ak_x}
		=\frac{e^2}{8\pi^3 \omega_P} \int\D k\,k\,n_B(k)\sum_{s=\pm}\int\D\Omega_k \frac{s-\hat{k}_x/a}{s-a\hat{k}_x}\\
		=&\frac{e^2}{8\pi^3 \omega_P} \int\D k\,k\,n_B(k)\int\D\Omega_k \frac{2(1-\hat{k}_x^2)}{1-a^2\hat{k}_x^2}
		=\frac{e^2}{8\pi^3 \omega_P}\frac{\pi^2 T^2}{6}\frac{16\pi}{3}\left(\frac{3}{2a^2}-\frac{3(1-a^2)}{4a^3}\log\left(\frac{1+a}{1-a}\right) \right)\\
		=&\frac{e^2T^2}{9\omega_P}\left(\frac{3}{2a^2}-\frac{3(1-a^2)}{4a^3}\log\left(\frac{1+a}{1-a}\right) \right).
	\end{split}
\end{equation*}
%

%------------------------------------------------------------------------------------
\subsubsection{Fermionic part: High temperature}
%------------------------------------------------------------------------------------
%
The temperature part given by the Fermi distribution is:
\begin{equation*}
	\begin{split}
		^\beta M_{\mu\nu}^C(p',p)=&e^2  a_{\mu\nu\rho\kappa}\int\frac{\D^4 k}{(2\pi)^3}\delta(k^2-m^2)n_F(k)
		\bar{u}(p')\left[\frac{\gamma^\kappa(\slashed{k}+m)\gamma^\rho }{(p-k)^2}+(p'\leftrightarrow p)\right]u(p)\\
		=&e^2\int\frac{\D^4 k}{(2\pi)^3}\delta(k^2-m^2)n_F(k)\bar{u}(p')\gamma^\kappa(\slashed{k}+m)\gamma^\rho\left[\frac{1}{(p-k)^2}+\frac{1}{(p'-k)^2}\right]u(p)\\
		\equiv&e^2\int\frac{\D^4 k}{(2\pi)^3}\delta(k^2-m^2)n_F(k)\bar{u}(p')\left[\frac{N^C}{D_1^C}+\frac{N^C}{D_2^C}\right]u(p).
	\end{split}
\end{equation*}
The numerator is
\begin{equation*}
	N^C=-\left(\gamma_\mu k_\nu + \gamma_\nu k_\mu\right)\to
	N^C\subs{AGM}=-A_u\left(u_\mu\gamma_\nu+u_\nu\gamma_\mu\right)-A_P\left(P_\mu\gamma_\nu+P_\nu\gamma_\mu\right).
\end{equation*}
The denominators are
\begin{equation*}
	\begin{split}
		D_1^C=&\left[2m^2-(k\cdot P)+(k\cdot q)\right]=2m^2-\omega_P(k_0-a k_x)-k_z\epsilon,\\
		D_2^C=&\left[2m^2-(k\cdot P)-(k\cdot q)\right]=2m^2-\omega_P(k_0-a k_x)+k_z\epsilon.
	\end{split}
\end{equation*}

The part in $\left(P_\mu\gamma_\nu+P_\nu\gamma_\mu\right)$ at leading order in $\epsilon$ is
\begin{equation*}
	\begin{split}
		I_{P\gamma,f}^C\simeq&\frac{2e^2}{\omega_P^2 a} \int\frac{\D^4 k}{(2\pi)^3}\delta(k^2-m^2)n_F(k)\frac{k_x}{k_0-ak_x-2m^2/\omega_P}\\
		=&\frac{e^2}{8\pi^3 \omega_P^2}\frac{1}{a} \int\D k\,\frac{k^2}{E_k}\,n_F(k)\sum_{s=\pm}\int\D\Omega_k \frac{\hat{k}_x}{s\frac{E_k}{k}-a\hat{k}_x-\frac{2m^2}{\omega_P k}}.
	\end{split}
\end{equation*}
At high temperature the leading term is
\begin{equation*}
	I_{P\gamma,f}^C\simeq  \frac{e^2 T^2}{96 \pi \omega_P^2} I_\Omega(2,0,0,1)
	=\frac{1}{36}\frac{e^2 T^2}{\omega_P^2}\left[\frac{3}{2a^3}\log\left(\frac{1+a}{1-a}\right)-\frac{3}{a^2} \right].
\end{equation*}

The part in $\left(u_\mu\gamma_\nu+u_\nu\gamma_\mu\right)$ at leading order in $\epsilon$ and expanding for high temperatures is
\begin{equation*}
	\begin{split}
		I_{u\gamma,f}^C\simeq&\frac{2e^2}{\omega_P} \int\frac{\D^4 k}{(2\pi)^3}\delta(k^2-m^2)n_F(k)\frac{k_0-k_x/a}{k_0-a k_x-2m^2/\omega_P}\\
		=&\frac{e^2}{8\pi^3 \omega_P}\int\D k\,n_F(k)\sum_{s=\pm}\int\D\Omega_k\,\frac{k^2}{E_k} \frac{s E_k/k-\hat{k}_x/a}{s E_k/k-a\hat{k}_x-\frac{2m^2}{\omega_P k}}\\
		\simeq&\frac{e^2 T^2}{96\pi\omega_P}\left( I_\Omega(0,0,0,1)-I_\Omega(2,0,0,1) \right)\\
		=&\frac{1}{18}\frac{e^2 T^2}{\omega_P}\left[\frac{3}{2a^2}-\frac{3(1-a^2)}{4a^3}\log\left(\frac{1+a}{1-a}\right) \right].
	\end{split}
\end{equation*}
Summing all the relevant terms we found that the diagram can be written as:
\begin{equation*}
	M^C_{\mu\nu}= \left(I^C_{P\gamma}+\theta\subs{HT}I^C_{P\gamma,f}\right)\left( P_\mu\gamma_\nu + P_\nu \gamma_\mu\right)
	+\left(I_{u\gamma}^C+\theta\subs{HT}I^C_{u\gamma,f}\right) \left(u_\mu\gamma_\nu+u_\nu\gamma_\mu\right).
\end{equation*}
%

%------------------------------------------------------------------------------------
\subsection{Angular integrals}
%------------------------------------------------------------------------------------
\label{sec:AngularIntegrals}
Here, we report the results of the angular integrals we used to evaluate the diagrams:
\begin{equation*}
	\begin{split}
		\int\D\Omega_k \frac{2}{1-a^2 \hat{k}_x^2}=&8\pi\left[\frac{1}{2a}\log\left(\frac{1+a}{1-a}\right)\right];\quad
		\int\D\Omega_k \frac{2\hat{k}_x^2}{1-a^2 \hat{k}_x^2}=\frac{8\pi}{3}\left[\frac{3}{2a^3}\log\left(\frac{1+a}{1-a}\right)-\frac{3}{a^2}\right];\\
		\int\D\Omega_k \frac{2}{(1-a^2 \hat{k}_x^2)^2}=&8\pi\left[\frac{1}{4a}\log\left(\frac{1+a}{1-a}\right)+\frac{1}{2(1-a^2)}\right];\\
		\int\D\Omega_k \frac{2\hat{k}_x^2}{(1-a^2 \hat{k}_x^2)^2}=&\frac{8\pi}{3}\left[\frac{6}{4(1-a^2)a^2}-\frac{3}{4a^3}\log\left(\frac{1+a}{1-a}\right)\right];
	\end{split}
\end{equation*}
and the integrals involving $\hat{k}_y$ are
\begin{equation*}
	\begin{split}
		\int\D\Omega_k \frac{2\hat{k}_y^2}{(1-a^2 \hat{k}_x^2)^2}=& \frac{8 \pi}{3}\left[\frac{3(1+a^2)}{8 a^3}\log\left(\frac{1+a}{1-a}\right)-\frac{3}{4a^2}\right],\\
		\int\D\Omega_k \frac{2\hat{k}_x^2\hat{k}_y^2}{(1-a^2 \hat{k}_x^2)^2}=& \frac{8 \pi}{15} \left[\frac{15(3-a^2)}{8 a^5} \log\left(\frac{1+a}{1-a}\right)-\frac{45}{4 a^4}\right],\\
		\int\D\Omega_k \frac{2\hat{k}_y^2}{(1-a^2 \hat{k}_x^2)^3}=& \frac{8 \pi}{3}\left[\frac{3(3 a^2-1)}{16a^2(1-a^2)}+\frac{3(1+3 a^2)}{32 a^3}\log\left(\frac{1+a}{1-a}\right)\right],\\
		\int\D\Omega_k \frac{2\hat{k}_x^2\hat{k}_y^2}{(1-a^2 \hat{k}_x^2)^3}=& \frac{8 \pi}{15} \left[\frac{15(3-a^2)}{16 a^4(1-a^2)}-\frac{15(3+a^2)}{32 a^5} \log\left(\frac{1+a}{1-a}\right)\right],\\
		\int\D\Omega_k \frac{2\hat{k}_x^4\hat{k}_y^2}{(1-a^2 \hat{k}_x^2)^3}=& \frac{8\pi}{35}\left[\frac{35 (15-13 a^2)}{16 a^6(1-a^2)}+\frac{105(a^2-5)}{32a^7}\log\left(\frac{1+a}{1-a}\right)\right],
	\end{split}
\end{equation*}
and those involving $\hat{k}_z$  in the principal value sense are given by
\begin{equation*}
\begin{split}
\int\D\Omega_k \frac{2 \hat{k}_y^2}{(1-a^2\hat{k}_x^2)\hat{k}_z^2}
	=&\lim_{\tau\to 0}\int\D\Omega_k \frac{2 \hat{k}_y^2}{(1-a^2\hat{k}_x^2)(\hat{k}_z^2-\tau^2)}
	=-8\pi\left[\frac{1}{2a}\log\left(\frac{1+a}{1-a}\right)\right];\\
\int\D\Omega_k \frac{2 \hat{k}_x^2 \hat{k}_y^2}{(1-a^2\hat{k}_x^2)\hat{k}_z^2}
	=&\lim_{\tau\to 0}\int\D\Omega_k \frac{2 \hat{k}_x^2 \hat{k}_y^2}{(1-a^2\hat{k}_x^2)(\hat{k}_z^2-\tau^2)}
	=-\frac{8\pi}{3}\left[\frac{3}{2a^3}\log\left(\frac{1+a}{1-a}\right)-\frac{3}{a^2}\right];\\
\int\D\Omega_k \frac{2 \hat{k}_y^2}{(1-a^2\hat{k}_x^2)^3\hat{k}_z^2}=& -8 \pi\left[\frac{3}{16 a}\log\left(\frac{1+a}{1-a}\right)-\frac{3a^2-5}{8(1-a^2)^2}\right],\\
\int\D\Omega_k \frac{2 \hat{k}_x^2 \hat{k}_y^2}{(1-a^2\hat{k}_x^2)^3\hat{k}_z^2}=& -\frac{8 \pi}{3} \left[\frac{3(1+a^2)}{8 a^2(1-a^2)^2}
		-\frac{3}{16 a^3}\log\left(\frac{1+a}{1-a}\right)\right],\\
\int\D\Omega_k \frac{2 \hat{k}_x^4 \hat{k}_y^2}{(1-a^2\hat{k}_x^2)^3\hat{k}_z^2}=& -\frac{8 \pi}{5}\left[\frac{15}{16a^5}\log\left(\frac{1+a}{1-a}\right)
		-\frac{5(3a-5 a^3)}{8 a^5(1-a^2)^2}\right],\\
\int\D\Omega_k \frac{2 \hat{k}_x^6 \hat{k}_y^2}{(1-a^2\hat{k}_x^2)^3\hat{k}_z^2}=& -\frac{8 \pi}{7} \left[\frac{105}{16 a^7}\log\left(\frac{1+a}{1-a}\right)
	-\frac{7(15a-25a^3+8a^5)}{8 a^7(1-a^2)^2}\right].
\end{split}
\end{equation*}
%

%*********************************************************************
%\bibliographystyle{apsrev4-1}
%\bibliography{Biblio}

\end{document}